\documentclass[12pt]{article}
\pdfoutput=1
\usepackage{jheppub, bm}
\numberwithin{equation}{section}
\renewcommand\afterTocRuleSpace{\bigskip}

\graphicspath{{./figures/}}

\newcommand{\V}{{v}}
\newcommand{\HAT}{{}}
\newcommand{\RHO}{\rho}

\newcommand{\bea}{\begin{eqnarray}}
\newcommand{\eea}{\end{eqnarray}}
\newcommand{\be}{\begin{equation}}
\newcommand{\ee}{\end{equation}}
\newcommand{\bse}{\begin{subequations}}
\newcommand{\ese}{\end{subequations}}
\newcommand{\mb}{\mathbf}

\newcommand{\wt}{\widetilde}

\newcommand{\ol}{\overline}

\newcommand{\eg}{\emph{e.g.}}
\newcommand{\ie}{\emph{i.e.}}
\newcommand{\cf}{\emph{cf.}}

\newcommand{\Z}{{\mathbb Z}}
\newcommand{\R}{{\mathbb R}}
\newcommand{\C}{{\mathbb C}}

\newcommand{\cp}{{\mathbb{CP}}}

\newcommand{\Tr}{{\rm Tr \,}}
\renewcommand{\Re}{{\rm Re}}
\renewcommand{\Im}{{\rm Im}}
\newcommand{\bs}{\backslash}
\newcommand{\pd}{\partial}

\newcommand{\CA}{\mathcal{A}}

\newcommand{\CC}{\mathcal{C}}
\newcommand{\CD}{\mathcal{D}}

\newcommand{\CF}{\mathcal{F}}
\newcommand{\CG}{\mathcal{G}}
\newcommand{\CH}{\mathcal{H}}

\newcommand{\CL}{\mathcal{L}}
\newcommand{\CM}{\mathcal{M}}
\newcommand{\CN}{\mathcal{N}}
\newcommand{\CO}{\mathcal{O}}

\newcommand{\CR}{\mathcal{R}}
\newcommand{\CS}{\mathcal{S}}

\newcommand{\CZ}{\mathcal{Z}}

\title{The Coulomb Branch of 3d $\CN=4$ Theories}

\author[1,2]{Mathew Bullimore}
\author[1]{Tudor Dimofte}
\author[2]{Davide Gaiotto}

\affiliation[1]{School of Natural Sciences, Institute for Advanced Study, Princeton, NJ 08540, USA}
\affiliation[2]{Perimeter Institute for Theoretical Physics, Waterloo, Ontario, Canada N2L 2Y5}

\abstract{We propose a construction for the quantum-corrected Coulomb branch of a general 3d gauge theory with $\CN=4$ supersymmetry, in terms of local coordinates associated with an abelianized theory.
In a fixed complex structure, the holomorphic functions on the Coulomb branch are given by expectation values of chiral monopole operators. 
We construct the chiral ring of such operators, using equivariant integration over BPS moduli spaces.
We also quantize the chiral ring, which corresponds to placing the 3d theory in a 2d Omega background. 
Then, by unifying all complex structures in a twistor space, we encode the full hyperk\"ahler metric on the Coulomb branch.
We verify our proposals in a multitude of examples, including SQCD and linear quiver gauge theories, whose Coulomb branches have alternative descriptions as solutions to Bogomolnyi and/or Nahm equations.
}

\begin{document}

\maketitle

\section{Introduction}

Three-dimensional gauge theories with eight supercharges ($\CN=4$ supersymmetry) generically have a moduli space of supersymmetric vacua 
parameterized by the expectation values of a triplet of vectormultiplet scalar fields. This branch of vacua is conventionally called the Coulomb branch $\CM_C$. 
Classically, the expectation values of the scalars are diagonal, and generically break the gauge group $G$ to a maximal abelian subgroup. The low-energy abelian gauge fields can then be dualized to periodic scalars (the ``dual photons"), which parametrize additional directions in the moduli space, giving the Coulomb branch a classical description
\be \CM_C \approx (\R^3\times S^1)^{\text{rank}(G)}\big/\raisebox{-.05cm}{{\small Weyl($G$)}}\,. \label{MCclass} \ee
Extended supersymmetry requires that the moduli-space metric be hyperk\"ahler.

The naive classical geometry of the Coulomb branch receives quantum corrections, 
both perturbative and non-perturbative \cite{Seiberg-3dbranes, SW-3d}. The quantum-corrected geometry 
can be derived through a direct calculation for abelian gauge theories \cite{dBHOY, KS-mirror}. For nonabelian gauge theories
that admit a brane construction, the infrared Coulomb branch geometry can be derived through S-duality. The basic example of A-type quivers of unitary groups 
was first analyzed in \cite{HananyWitten}, and admits several extensions to a variety of quivers and gauge groups, see \eg\ \cite{FH-O3, HZ-orientifolds}. 
The brane constructions can be extended further and systematized by applying S-duality to compactifications of four-dimensional $\CN=4$ gauge theory \cite{GW-Sduality}. 

Perhaps surprisingly, this large set of well-understood examples has not yet yielded a general description of the Coulomb branch, 
valid for a generic $\CN=4$ gauge theory. The purpose of this paper fill that gap. 

Standard local operators such as gauge-invariant polynomials of the vectormultiplet scalars are insufficient to parameterize the Coulomb branch,
because they fail to capture the expectation values of the dual photons. In order to fully parametrize the Coulomb branch, one needs to 
study the vacuum expectation values of BPS monopole operators, a three-dimensional analogue of 
't Hooft line operators in four dimensions \cite{BKW-monopoles, Borokhov-monopoles}. 

The chiral operators built out of monopole operators dressed by vectormultiplet scalar fields form a chiral ring $\C[\CM_C]$, and their expectation values are expected to give a complete set of holomorphic functions on the Coulomb branch, seen as a complex symplectic manifold. 
Monopole operators are labelled by the GNO charge $A$, which specifies a way to embed a $U(1)$ monopole singularity into 
the full gauge group $G$. The monopole charge breaks the gauge group to a subgroup $G_A$. A monopole of charge $A$ 
can be dressed by a general $G_A$-invariant polynomial $p$ in the vectormultipet scalar fields restricted to $G_A$
to produce a chiral operator $M_{A,p}$. 
 
As observed by the authors of \cite{CHZ-Hilbert}, one can gain information about the Coulomb branch as a complex manifold by 
studying its Hilbert series. This counts all dressed monopole operators in order to derive the quantum numbers of the generators and relations 
of the corresponding chiral ring $\C[\CM_C]$. In complicated examples, though, one stills has to guess the precise form of the ring relations
and of the Poisson brackets between generators. One of our objectives is to determine the full Poisson algebra structure 
of the chiral operators/holomorphic functions $M_{A,p}$.

Our strategy is to define an ``abelianization map'', which embeds the Poisson algebra of holomorphic functions $\C[\CM_C]$
on the Coulomb branch into a larger algebra $\C[\CM_C^{\mathrm{abel}}]$ of holomorphic functions on an ``abelian patch'' of the Coulomb branch,
which is roughly described as the complement of the locus where nonabelian gauge symmetry would be classically restored.

The abelianization map has a transparent physical meaning: it maps the vev of a monopole operator of the full theory to a linear combination of abelian monopole 
operator vevs $v_B$ in the low-energy abelian gauge theory, with coefficients that are meromorphic functions of the abelian vectormultiplet scalars $\varphi_a$:
\begin{equation}
M_{A,p} \to \sum_{B \prec A} c^B_{A,p}[\varphi_a] v_B\,.
\end{equation}
The coefficients $c^B_{A,p}[\varphi_a]$ capture the microscopic physics which converts the nonabelian monopole singularity of charge $A$ into the low-energy
abelian charge $B$. Localization calculations such as \cite{GOP-tHooft,IOT-Hitchin} suggest that the $c^B_{A,p}[\varphi_a]$ coefficients should only receive contributions from 
BPS ``bubbling monopole'' geometries, \ie\ should be computed by a path integral localized on BPS solutions of the equations of motion in the presence of the monopole singularity, with given abelian magnetic charge at infinity.

Schematically, we expect the relation to take the form 
\begin{equation}\label{equiv-intro}
c^B_{A,p}[\varphi_a] = \int^{G\mathrm{-equivariant}}_{\CM_A^B} c_p[E_A] e(\CD_m)\,,
\end{equation}
\ie\ an equivariant integral over the moduli space $\CM_A^B$ of bubbling solutions of the Bogomolnyi equations,
with an integrand assembled from the Euler class of the bundle $\CD_m$ of Dirac zero modes for the matter fields 
and an appropriate characteristic class of the universal $G_A$-bundle $E_A$ associated to the singularity. The abelian 
vectormultiplet scalars $\varphi_a$ should play the role of equivariant parameters for the action of the gauge group $G$.%
\footnote{The moduli space $\CM_A^B$ has singularities labelled by lower magnetic charges $A'$. The equivariant path integral is expected to have 
regularization ambiguities proportional to full monopoles $M_{A',p'}$ of lower charge}

In this paper, we will mainly focus on theories whose Poisson algebra is generated by 
monopole operators such that the moduli spaces $\CM_A^B$ are a point. This includes all quivers built from unitary gauge groups. In particular, for linear quivers of this type, brane constructions predict two alternative descriptions of the Coulomb branch: as a moduli space of solutions to the Bogomolnyi equations with singularities, and as a moduli space of solutions to the Nahm equations on an interval. Both constructions involve auxiliary gauge groups associated to the quiver. We will compare the predictions of the abelianization map with the results of these alternative descriptions, finding exact agreement in all cases.
We leave the detailed analysis of more general gauge theories to future work. 

It is also possible to extend the abelianization map to gain a description of the Coulomb branch of four-dimensional gauge theories 
compactified on a circle, or five-dimensional gauge theories compactified on a torus. 

The abelianization map can be extended in a straightforward way to give a canonical quantization of the Poisson algebra of holomorphic functions,
simply by working equivariantly under space-time rotations. Physically, the quantization is associated to a (twisted) Omega deformation of the three-dimensional gauge theory. The quantized monopole operators $\hat M_{A,p}$ can be directly compared with the expressions found in localization of supersymmetric 
correlation functions. We leave the comparison to a companion paper.

Having described Coulomb branches as complex symplectic manifolds, we will also conjecture how to extend the abelianization map to construct their twistor spaces.  The twistor space unifies all complex structure, and captures the full hyperk\"ahler geometry of a Coulomb branch.

In Section~\ref{sec:gen} we will review the properties of $\CN=4$ gauge theories. In Section~\ref{sec:abelian} we will review the geometry 
of the Coulomb branch of abelian gauge theories, in a language which is suitable for the construction of the abelianization map, and propose 
a natural quantization for the ring of holomorphic functions on the Coulomb branch. In Section~\ref{sec:NA} we describe the abelianization map.
In Sections~\ref{sec:SQCD} and \ref{sec:quiver} we discuss several examples. In Appendix \ref{app:monopoles}. we develop some tools for analyzing singular monopole moduli spaces as holomorphic symplectic manifolds; and in Appendix \ref{app:equivariant} we describe some of the simplest equivariant integrals of the type \eqref{equiv-intro}.

\section{Generalities} \label{sec:gen}

A renormalizable 3d $\CN=4$  gauge theory is defined by a choice of gauge group, a compact Lie group $G$,
and a choice of matter content, \ie\ a quaternionic representation $\CR$ of $G$.%
\footnote{For further background material on 3d $\CN=4$ theories, see \eg\ \cite{SW-3d, IS}.} %
The fields of the theory consist of a vectormultiplet transforming in the adjoint representation of $G$, and hypermultiplets transforming in the representation $\CR$. The vectormultiplet contains a triplet of real scalars $\vec\phi = (\phi_1,\phi_2,\phi_3)\in (\mathfrak g)^3$. 
The hypermultiplets contain $4N$ real scalars (for some $N\geq 0$), which parametrize $\R^{4N}$ with its standard hyperk\"ahler structure. 
The representation $\CR$ can be understood as mapping $G$ to a subgroup of the hyperk\"ahler isometry group $USp(N) = U(2N)\cap Sp(2N,\C)$ of $\R^{4N}$.
The Lagrangian of the gauge theory is fully determined by the choice of $(G,\CR)$, together with (dimensionful) gauge couplings for every factor in $G$ and a set of canonical deformation parameters (masses and FI parameters) that we review below.

3d $\CN=4$ gauge theories always have an R-symmetry group $SU(2)_C \times SU(2)_H$.
The three scalar fields in each vectormultiplet transform as a triplet of $SU(2)_C$, while 
hypermultiplets transform as complex doublets of $SU(2)_H$. 

In addition, there is a global symmetry group $G_C\times G_H$ that commutes with $SU(2)_C \times SU(2)_H$ (and the supersymmetry algebra).
The hypermultiplets transform under a ``Higgs-branch'' global symmetry group $G_H$, which, formally,  is the normalizer of the gauge group inside $USp(N)$, modulo the action of the gauge group itself
\be  G_H = N_{USp(N)}(\CR(G))\,\big/\,\CR(G)\,, \label{GH} \ee
with `$\CR$' denoting the map from $G$ to $USp(N)$. 
Thus, loosely speaking, the hypermultiplet scalars transform in a quaternionic representation of $G \times G_H$. If the gauge group $G$ contains abelian factors, 
the theory will also have ``Coulomb-branch'' global symmetries whose conserved currents are simply the abelian field strengths. Monopole operators with magnetic charges in the abelian factors,
by construction, are charged under Coulomb-branch global symmetries. Altogether, in the ultraviolet gauge theory,
\be G_C = U(1)^{\text{$\#$ $U(1)$ factors in $G$}}\,, \label{GC} \ee
though in the infrared $G_C$ may be enhanced to a nonabelian group whose maximal torus is \eqref{GC}.

In the absence of mass deformations, 
a typical $\CN=4$ gauge theory has a rich moduli space of vacua, a union of ``branches'' of the form $\CC_a \times \CH_a$,
where $\CC_a$ is a hyperk\"ahler manifold parameterized by the expectation values of gauge-invariant combinations of
vectormultiplet scalars and monopole operators and $\CH_a$ is a hyperk\"ahler manifold parameterized by the expectation values gauge-invariant 
combinations of hypermultiplets.
We will generically refer to the ``Coulomb branch'' $\CM_C$ as a branch of vacua where the $\CH_a$ factor is trivial, and 
to the ``Higgs branch'' $\CM_H$ as a branch of vacua where $\CC_a$ is trivial. Other branches of the moduli space of 
vacua are usually referred to as mixed branches, and consist of a product of hyperk\"ahler sub-manifolds of 
$\CM_C$ and $\CM_H$.

The Higgs branch $\CM_H$ is not affected by quantum corrections, and is simply the hyperk\"ahler quotient
\be \CM_H = \R^{4N}/\!/\!/\,G\,,\qquad \dim_\C\CM_H = 2N-2\dim(\CR(G))\,. \label{MH-intro} \ee
In contrast, the Coulomb branch does suffer quantum corrections. Classically,
\be \CM_C  \approx (\R^3\times S^1)^{\text{rank}\,G}/W_G\,,\qquad \dim_\C\CM_C = 2\,\text{rank}\,G\,, \label{MC-intro} \ee
where $W_G$ is the Weyl group of $G$; but quantum corrections modify the topology and geometry of $\CM_C$.

The global symmetries $G_C\times G_H$ act tri-holomorphically on the corresponding branches of vacua and are associated to a triplet of protected moment map operators. 
On the other hand, the R-symmetries rotate among themselves the hyperk\"ahler forms of $\CM_C$ and $\CM_H$ respectively. 
In particular, in the absence of mass deformations all choices of complex structure on $\CM_C$ and $\CM_H$ are equivalent.

The $\CN=4$ gauge theories admit two classes of deformation parameters, masses and FI parameters, each associated to 
a Cartan generator of the global symmetry of the theory. Masses $\vec m \in (\mathfrak t_H)^3$ transform as a triplet of $SU(2)_C$, while 
FI parameters $\vec t\in (\mathfrak t_C)^3$ transform as a triplet of $SU(2)_H$. Masses deform/resolve the geometry of the Coulomb factors $\CC_a$ 
and restrict the Higgs factors $\CH_a$ to the fixed point of the corresponding isometries. FI parameters do the opposite. 

The geometry of the Higgs and Coulomb branches of vacua is captured by the expectation values of two types of
half-BPS local operators. Higgs-branch operators parameterize $\CM_H$ and transform in irreducible representations of $SU(2)_H$:
a spin $n/2$ Higgs branch operator is the projection to spin $n/2$ of a gauge-invariant polynomial of $n$ elementary hypermultiplets.  

Coulomb-branch operators parameterize $\CM_C$ and transform in irreducible 
representations of $SU(2)_C$. A generic Coulomb-branch operator can be described as a ``dressed monopole operator'':
a BPS monopole operator combined with some polynomial in the vectormultiplet scalars which is invariant under the 
subgroup of the gauge group preserved by the monopole singularity.

It is useful to pick an $\CN=2$ subalgebra of the $\CN=4$ superalgebra and look at the Higgs and Coulomb branch operators which are chiral 
under the $\CN=2$ subalgebra. These operators form a chiral ring and map to holomorphic functions on $\CM_C$ and $\CM_H$.
Indeed, the choice of $\CN=2$ subalgebra is equivalent to a choice of complex structures $\zeta$ and $\zeta'$ on the two branches of vacua;
and the rings of holomorphic functions $\C[\CM_C]_\zeta$, $\C[\CM_H]_{\zeta'}$ in these complex structures are subrings of the $\CN=2$ chiral ring.
(We often drop the explicit dependence on $\zeta,\zeta'$.)
The choice of $\CN=2$ subalgebra is preserved by a Cartan subalgebra of $SU(2)_C \times SU(2)_H$, and the corresponding abelian R-charge 
of BPS monopole operators in $\C[\CM_C]_\zeta$ can be computed by a standard formula \cite{GW-Sduality, CHZ-Hilbert}. 

We will denote the chiral combinations of the hypermultiplet scalars as pairs $(X,Y)\in \C^{2N}$ (or more precisely $(X_{\zeta'},Y_{\zeta'})$), 
implicitly assuming that the matter representation $\CR$ is the sum of two conjugate complex representations.
If $\CR$ is truly pseudo-real, $X$ and $Y$ should be unified into a single set of fields. 
We will denote the chiral combination of vectormultiplet scalars, containing two of the three real adjoint-valued fields, as $\varphi\in \mathfrak g_\C$ (or more precisely $\varphi_\zeta$). Chiral BPS monopole operators require a singular vev for 
the remaining real scalar field $\sigma\in \mathfrak g$, matching the singularity in the gauge fields. 

The imprint of the hyperk\"ahler geometry on the ring of holomorphic functions on $\CM_C$ and $\CM_H$
is a holomorphic Poisson bracket, associated to the complex symplectic forms built out of the appropriate linear combination 
of hyperk\"ahler forms (See Eqn. \eqref{Omegazeta} for an explicit formula). For hypermultiplets, we simply have
$\{X,Y\}=1$. 

The data of the chiral rings and Poisson brackets is captured by two topologically twisted theories: the 
Rozansky-Witten theory \cite{RW} combines space-time rotations and $SU(2)_C$ to select a supercharge 
whose cohomology captures the complex geometry of the Higgs branch; while a twisted Rozansky-Witten theory
 combines space-time rotations and $SU(2)_H$ to select a supercharge 
whose cohomology captures the complex geometry of the Coulomb branch.
It is likely that the results of this paper could be verified by studying monopole operators 
in the language of twisted Rozansky-Witten theory.

The description of $\CM_C$ and $\CM_H$ as complex symplectic manifolds does not 
capture the hyperk\"ahler metric on the moduli spaces. The metric can be captured, though, 
by a twistor construction. We refer to Sections \ref{sec:twistor-ab}, \ref{sec:twistor-SQCD}, \ref{sec:twistor-quiver} for details. Our results strongly suggest that 
it should be possible to extend the language of the (twisted) Rozansky-Witten theory to capture the 
full twistor geometry, perhaps in a manner similar to projective superspace constructions 
in physics, see \eg\ \cite{IvanovRocek}.

\subsection{The chiral ring is independent of gauge couplings}
\label{sec:ind-g}

A key ingredient in many our constructions is a simple variation of a non-renormalization theorem. Consider a 3d $\CN=4$ gauge theory, with Coulomb and Higgs branches $\CM_C$, $\CM_H$. Let us choose an $\CN=2$ subalgebra of  the  $\CN=4$ superalgebra, as above,
corresponding to a choice of complex structures on the Coulomb and Higgs branches. Then the chiral rings $\C[\CM_C]_\zeta$ and $\C[\CM_H]_{\zeta'}$ are subrings of the $\CN=2$ chiral ring.

The gauge coupling constants $g_i$ of our theory are real parameters with no natural complexification. They can be promoted to background $\CN=2$ superfields in several ways: either as the real scalar components of linear multiplets $\Sigma_i$, as the scalar components of real (vector) multiplets $V_i$, or as the scalar components of chiral multiplets $\Phi_i$ that only ever enter the theory in the combination $\Phi_i+\Phi_i^\dagger$. None of these multiplets can ever occur in an effective superpotential of the $\CN=2$ theory, or in the ${\CN=2}$ chiral ring. Indeed, this was the basis behind the non-renormalization theorems of \cite{AHISS}.

In the present case, we conclude that for any fixed choice of complex structure, the chiral rings $\C[\CM_C]_\zeta$, $\C[\CM_H]_{\zeta'}$ do not depend on gauge couplings $g_i$. In particular, this means that there must exist a set of chiral operators $\{\CO_a\}$ that generate each chiral ring, in such a way that the ring relations (structure constants, etc.) are independent of the $g_i$. The $\CO_a$ do include monopole operators, whose ultraviolet definition $V_\pm\sim \exp\big(\pm\frac{1}{g^2}(\sigma+i\gamma)\big)$ does implicitly involve gauge couplings. However, once such operators are correctly identified, the relations among them never contain the $g_i$.

Note that the status of gauge couplings in 3d $\CN=2$ or $\CN=4$ theories is fundamentally different from that in 4d $\CN=2$ theories. In 4d, the real gauge couplings do have a natural complexification by the theta-angle, and they do enter the complex geometry of the Coulomb branch \cite{SW-I,SW-II}.

In a nonabelian 3d $\CN=4$ gauge theory at a generic point in the Coulomb branch, instanton corrections to the metric are controlled by the instanton action, which is proportional to $\frac{1}{g^2}$
but goes to zero as one approaches the locus where nonabelian gauge symmetry is classically restored. Thus the non-renormalization theorem protects the 
complex geometry on the Coulomb branch from the effect of instantons, allowing corrections only at the classical nonabelian locus. This is the the motivation for our abelianization map. 

\section{Abelian Coulomb branches}
\label{sec:abelian}

We review the structure of Coulomb (and Higgs) branches of abelian gauge theories, building up gradually from SQED to a general theory. Our main goal is to describe the chiral ring of the Coulomb branch $\C[\CM_C] = \C[\CM_C]_\zeta$ (for any fixed $\zeta$) intrinsically, in a way that will generalize to nonabelian theories. We also describe quantization of the chiral ring in the presence of an Omega background, as well as the twistor construction that unifies all complex structures $\zeta$ and reproduces the hyperk\"ahler metric on the Coulomb branch.

\subsection{SQED}

SQED with $N$ hypermultiplets is a gauge theory with $G=U(1)$ and $\CR \simeq \R^{4N} \simeq T^*\C^{N}$, in the notation of Section \ref{sec:gen}. Given a complex structure on the Higgs branch, the $2N$ complex hypermultiplet scalars $(X^i,Y^i)_{i=1}^N$ carry charges $(+1,-1)$ under the $U(1)$ gauge symmetry. This theory has a Higgs-branch symmetry $G_H = PSU(N)$ that rotates the $N$ hypermultiplets, and a Coulomb-branch symmetry $G_C = U(1)_t$ that rotates the dual photon.

The Higgs branch is protected from quantum corrections and may be constructed as a hyperk\"ahler quotient. The complex and real moment maps of the $U(1)$ gauge group action are
\be \mu = \vec X \cdot \vec Y\,,\qquad \mu_\R = |\vec X|^2 - |\vec Y|^2\,,\ee
and the Higgs branch is the hyperk\"ahler quotient
\be  \CM_H  =  ( \mu = t\,,\; \mu_\R = t_\R ) / U(1)\,. \ee
When the complex FI parameter $t$ is zero but the real FI parameter $t_\R$ is nonzero, $\CM_H$ is isomorphic to $T^*\cp^{N-1}$ as a complex manifold. If both FI parameters vanish, $\CM_H$ becomes a singular hyperk\"ahler cone. In terms of representation theory, this cone can be identified with the orbit of the minimal nilpotent element inside $\mathfrak{sl}(N,\C)$, and its resolution at finite $t_\R$ is the Springer resolution of the orbit. When $t,t_\R$ are both nonzero, $\CM_H$ still has the topology of $T^*\cp^{N-1}$, but is no longer isomorphic to $T^*\cp^{N-1}$ as a complex manifold (in particular, the base $\cp^{N-1}$ is no longer a holomorphic submanifold).

The Higgs-branch chiral ring is obtained by starting with the free polynomial ring $\C[X,Y]$ generated by the $X_i$ and $Y_i$, taking invariant functions, and imposing the complex moment map condition $\mu=t$. Abstractly,
\be \C[\CM_H] = \C[X,Y]^{U(1)}/(\mu-t)\,. \label{SQED-OH} \ee
This can also be thought of as a holomorphic symplectic reduction of $\C[X,Y]$. Concretely, $\mathbb{C}[\CM_H]$ is generated by an $N\times N$ matrix of functions $z_{ij} = X^iY^j$, subject to $\Tr |\!|z_{ij}|\!| = t$ and the condition that the determinant of any $2\times 2$ minor vanishes, \ie\ $\text{rank}\, |\!|z_{ij}|\!| = 1$. 

Classically, the Coulomb branch is parametrized by the vectormultiplet scalar fields $\vec\phi = (\phi_1,\phi_2,\phi_3)\in\R^3$,
together with the dual photon $\gamma$, which obeys $d\gamma = *dA$. In our conventions, the dual photon is a periodic scalar with $\gamma \sim \gamma + 2\pi g^2$, where $g^2$ is the tree-level gauge coupling. Therefore, topologically, the classical Coulomb branch is simply $\CM_C^{\rm class} = \R^3\times S^1$. The symmetry $G_C = U(1)_t$ shifts the dual photon and so rotates the $S^1$ factor. 

The Coulomb branch of an abelian theory does not receive non-perturbative quantum corrections as there are no dynamical monopoles. Furthermore, the classical Coulomb branch only receives a 1-loop quantum correction, which can be explicitly computed \cite{Seiberg-3dbranes, SW-3d, IS}. Topologically, this correction has the effect of changing the topology of $\CM_C^{\rm class}$ at infinity from a product $S^2\times S^1$ to a nontrivial fibration of Euler number $N$; and correspondingly shrinking the dual photon circle $S^1$ at any values of $(\varphi,\sigma)$ where a hypermultiplet becomes massless.

To analyze this we introduce $N$ hyperk\"ahler triplets of masses $\vec m_i\in (\mathfrak t_H)^3$, $i=1,...,N$,
valued in a Cartan subalgebra of the $PSU(N)$ flavor symmetry of the Higgs branch. (These masses are constant vevs for a background $PSU(N)$ vectormultiplet, and are defined up to an overall shift which can be absorbed in $\vec\phi$.) The effective mass of the $i$-th hypermultiplet is $|\vec\phi+\vec m_i|$, so the $S^1$ circle of $\CM_C^{\rm clas}$ shrinks at the $N$ points
\be  \vec \phi = -\vec m_i\,. 
\label{SQED-shrink} \ee 

How can we describe the quantum-corrected Coulomb branch $\CM_C$ as a complex symplectic manifold?
Fixing a complex structure, we form chiral combinations (say) $\varphi = \phi_2+i\phi_3$ and $\sigma+i\gamma = \phi_1+i\gamma$, and correspondingly $m_i :=m_{i,2}+i\,m_{i,3}$, $m_i^\R := m_{i,1}$.
Classically, the holomorphic functions on $\CM_C^{\rm class}$ are given by the vevs of the complex scalar $\varphi$ and the monopole operators $\V^\pm = e^{\pm\frac1{g^2}(\sigma+i\gamma)}$.\footnote{See \cite{AHISS, BKW-monopoles, GW-Sduality, IS-aspects} for some further discussions of monopole operators and their properties.} The monopole operators simply satisfy $\V^+\V^-=1$ and the holomorphic symplectic form is
\be \Omega_C = d\varphi\wedge d \log \V^+ =- d\varphi \wedge d\log \V^-\,.\label{SQED-Omega} \ee
This identifies $\mathcal{M}_C^{\rm class} = \mathbb{C} \times \mathbb{C}^* $ as a complex symplectic manifold. A natural guess for the quantum-corrected Coulomb branch is that the vevs of monopole operators become $\C$ (rather than $\C^*$) valued functions, and satisfy a modified relation (\cf\ \cite{BKW-monopoles})
\be \CM_C\,:\quad \V^+\V^- = \prod_{i=1}^N (\varphi + m_i)\,, \label{SQED-MC} \ee
with the same holomorphic symplectic form $\Omega_C$. The modified relation beautifully accounts for the shrinking of the $S^1$ at points \eqref{SQED-shrink}. It identifies the Coulomb branch with a deformation of the $A_{N-1}$ singularity $\C^2/\Z_N$.

The relation \eqref{SQED-MC} is consistent with transformations of $\V^\pm$ and $\varphi$ under the topological symmetry $U(1)_t$ and the R-symmetry $U(1)_C\subset SU(2)_C$. Indeed, the monopole operators have charge $\pm1$ under $U(1)_t$, while $\varphi$ is neutral. On the other hand, $\varphi$ has R-charge $+2$ while $V_\pm$ both have R-charge $N$ \cite{SW-3d, AHISS, BKW-monopoles}. (Note that $U(1)_C$ is broken unless $m_i \equiv 0$, in which case the RHS becomes homogeneous, transforming with charge~$2N$.)

The exact hyperk\"ahler metric on the Coulomb branch of an abelian gauge theory can be determined from a 1-loop calculation, as discussed in \cite{Seiberg-3dbranes, SW-3d, IS}. 
For SQED with $N$ flavors, this calculation reproduces the hyperk\"ahler metric on the $N$-centered Taub-NUT space. In Gibbons-Hawking coordinates, the hyperk\"ahler metric is
\be ds^2 = U(\phi)\, d\vec\phi\cdot d\vec\phi + U(\phi)^{-1} (g^{-2}d\gamma+\vec\omega(\phi)\cdot d\vec\phi)^2\,. \ee
This metric describes an $S^1$-fibration over $\mathbb{R}^3$ with fiber and base coordinates given by $\gamma$ and $\vec\phi$, respectively. The function $U(\phi)$ encodes the 1-loop correction to the tree-level gauge coupling,
\be U(\phi) = \frac1{g^2} + \sum_{i=1}^N\frac{1}{\big|\vec\phi+\vec m_i\big|}\,,\qquad \vec\nabla U = \vec\nabla\times\vec\omega\,.  \label{metric-U1} \ee
At the $N$ points $\vec\phi=-\vec m_i$ a fundamental hypermultiplet becomes massless and the 1-loop corrections force $U(\phi) \to \infty$, shrinking the $S^1$ fiber at that point. 
In addition, the Dirac connection $\vec\omega$ modifies the topological structure of $S^1$ bundle on the sphere at infinity to a fibration of Euler number $N$.

In the infrared, as $g\to\infty$, this metric describes the deformation and/or resolution of the $A_{N-1}$ singularity $\C^2/\Z_{N}$, precisely agreeing with \eqref{SQED-MC}. The non-renormalization argument of Section \ref{sec:ind-g}, however, guarantees that the chiral-ring relation \eqref{SQED-MC} holds for all values of $g^2$ --- and thus also describes the chiral ring of the $N$-centered Taub-NUT space.

\subsection{General charges}
\label{sec:abel-gen}

The analysis of the Coulomb branch for SQED can be upgraded to a general abelian gauge theory. The construction is local, in the sense that it can be performed separately for each $U(1)$ gauge group in a general theory, where the SQED results apply.

Consider, then, a theory with gauge group $G=\prod_{a=1}^rU(1)_a$ and representation $\CR = T^*\C^N$ for $N$ hypermultiplets, such that the $i$-th hypermultiplet $(X^i,Y^i)$ has charges $(Q^a{}_i,-Q^a{}_i)$ under $U(1)_a$.  The flavor symmetry group $G_H\times G_C$ of the theory includes a subgroup  $G_H = U(1)^{N-r}=\prod_{\alpha=1}^{N-r} U(1)_\alpha$ that rotates the hypermultiplets with some charges $(q^\alpha{}_i,-q^\alpha{}_i)$ and a subgroup $G_C = U(1)^r= \prod_{a=1}^rU(1)^{(a)}$ of topological symmetries shifting the dual photons.%
\footnote{In special cases, such as SQED, the flavor symmetry may have a nonabelian enhancement. For the current general discussion, we are just considering maximal tori of the flavor groups.} %
The $N-r$ vectors $\vec q^{\,\,\alpha}\in \Z^N$ are only well defined modulo the $\vec Q^a\in \Z^N$, and together with the $\vec Q^a$ form a basis for $\R^N$. It is therefore convenient to combine the vectors $\vec Q^a,\vec q^{\,\,\alpha}$ into a square $N\times N$ matrix
\be 
{\mb Q} = \begin{pmatrix} Q \\ q \end{pmatrix} \label{abel-charges} \,,
\ee
\ie\ such that $\mb Q^a{}_i = Q^a{}_i$ for $\alpha =1,\ldots,r$ and $\mb Q^{N-a+1}{}_i = q^{a}{}_i$ for $a=1,\ldots,N-r$.
Without loss of generality, we can take $|\!\det \mb Q|  = 1$ to insure that our basis of gauge and flavor generators is minimal.%
\footnote{That condition is equivalent to the requirement that $\mb Q^{-1}$ is integral, so that the matter fields can only be coupled to integrally quantized background flavor and gauge fluxes.}

Classically, the Coulomb branch has the form $\CM_C^{\rm class} = (\R^3\times S^1)^{r}$, parametrized by the scalars $(\sigma_a,\varphi_a)$ and the dual photon $\gamma_a$ for each $U(1)_a$ gauge group. As in the case of SQED, we expect this picture to be modified by 1-loop quantum corrections. The effective real and complex masses of the $i$-th hypermultiplet are
\be M_i = \sum_a Q^a{}_i \varphi_a + \sum_\alpha q^\alpha{}_i m_\alpha
 \,, \qquad
 M_i^\R = \sum_a Q^a{}_i \sigma_a + \sum_\alpha q^\alpha{}_i m^\R_\alpha\,, \label{abel-defM} \ee
where $(m^\R_\alpha,m_\alpha)$ are a triplet of mass parameters for each $U(1)_\alpha$ factor in the flavor group $G_H$. When the $i$-th hypermultiplet becomes massless, $M^\R_i=M_i=0$, 1-loop quantum corrections will cause the circle parametrized by the associated dual photon $\sum_a Q^a{}_i\gamma_a$ to shrink.

In order to describe the quantum-corrected Coulomb branch $\CM_C$ as a complex symplectic manifold, we use the expectation values of the complex fields $\varphi_a$ along with monopole operators. For each $U(1)_a$ factor in $G$, there is a pair of monopole operators $\V_a^\pm \sim e^{\pm\frac1{g_a^2}(\sigma_a+i\gamma_a)}$. The shrinking of $S^1$'s at the location of massless hypermultiplets is then captured by the relations
\be  \V_a^+ \V_a^- = \prod_{i=1}^N (M_i)^{|Q^a{}_i|} =: P_a(\varphi,m)\,. \label{abel-simple} \ee
This is simply a copy of the SQED formula \eqref{SQED-MC} for every $U(1)_a$ factor, slightly modified to allow for more general gauge and flavor charges. The relations transform homogeneously under the topological symmetry group $G_C = \prod_{a=1}^r U(1)^{(a)}$, whose factors act on $\V_a^\pm$ with charge $\pm 1$, and act trivially on $\varphi$. When $\vec m_\alpha = 0$, they also transform homogeneously under the R-symmetry $U(1)_C$, acting on $\varphi$ with charge 2 and on the monopole operators $\V_a^\pm$ with charge $\sum_i |Q^a{}_i|$. The nontrivial R-charges of the monopoles are determined just as they were for SQED \cite{SW-3d, AHISS, BKW-monopoles}.

The operators $(\V_a^+,\V_a^-,\varphi_a)_{a=1}^r$, subject to relations \eqref{abel-simple}, don't quite generate all the holomorphic functions on the Coulomb branch. A few additional generators are required.
For every cocharacter $A \in \text{Hom}(U(1),G) \simeq \Z^r$, there exists a pair of monopole operators $\V_A^+$, $\V_A^-=\V_{-A}^+$ constructed from the dual photon for the corresponding $U(1)$ subgroup of $G$. They have charges $\pm A_a$ (where $A = (A_1,...,A_r)$) under the topological symmetry $U(1)^{(a)}$. These more general monopole operators can always be expressed as rational functions of the $(\V_a^+,\V_a^-,\varphi_a)_{a=1}^r$, but not necessarily as polynomials. 

For general monopole operators labelled by $A,B\in\mathbb{Z}^r$ we propose that
\be 
\V^-_A =\V^+_{-A} \qquad  \V^+_A \V^+_B  = \V^+_{A+B} P_{A,B}(\varphi,m)\,, 
\label{abel-gen}
\ee
where
\be
P_{A,B}(\varphi,m) = \prod_{i=1}^N (M_i)^{(Q^T  A)^i_+ +(Q^TB)^i_+-(Q^T(A+B))^i_+}\,,
\ee
and $(x)_+ := {\rm max}(x,0)$. The vectors $(Q^T  A)\in\mathbb{Z}^N$ are defined using the charge matrix as a map $Q^T:\Z^r \to \Z^N$. When $B = -A$ it is straightforward to verify that equation \eqref{abel-gen} becomes
\be
\V^+_A \V^-_A = \prod_{i=1}^N (M_i)^{|(Q^TA)^i|} \,, \label{abel-nonsimple} 
\ee
which is an immediate generalization of the formula~\eqref{abel-simple} for simple monopole operators. Another consequence of equation \eqref{abel-gen} is that $\V^+_{nA}=(\V^+_A)^n$ when $n\geq 0$. The formula \eqref{abel-gen} implies that a general monopole operator $\V^+_A$ can be written as a rational function of $(\V_a^+,\V_a^-,\varphi_a)_{a=1}^r$.

We propose that the full Coulomb branch chiral ring is generated as
\be \boxed{ \C[\CM_C] = \C\big[ \{\V^\pm_A\}_{A\in \Z^r},\{\varphi_a\}_{a=1}^r \big]\big/ \eqref{abel-gen}}\,. \label{abel-O}
\ee
We will prove this result in Section \ref{sec:abelMS} using 3d abelian mirror symmetry, assuming that the charge matrix $\mb Q$ is unimodular. Note that due to $\V^+_{nA}=(\V^+_A)^n$ it is sufficient to take a finite set of primitive monopole operators as the generators of $\C[\CM_C]$. Further properties of the Coulomb branch chiral ring of abelian gauge theories, including a concise combinatorial description of its basis, will appear in joint work with Justin Hilburn \cite{BDGH}.

Mirror symmetry will also show that the holomorphic symplectic form on $\CM_C$ is given by
\be \Omega_C = \sum_{a=1}^r d\varphi_a\wedge d\log \frac{\V_a^+}{P_a(\varphi,m)^{1/2}}\,. \label{abel-Omega} \ee
The induced Poisson brackets include
\be \{\varphi_a,\varphi_b\} = 0\,,\qquad \{\varphi_a,\V^\pm_b\} = \pm \delta_{ab} \V^\pm_b\,,\qquad \{\V_a^+,\V_a^-\} = - \frac{\pd}{\pd \varphi_a}P_a(\varphi,m)\,.\ee
The non-vanishing brackets $\{\V^+_A,\V^+_B\}$ among monopole operators can most easily be derived by taking a classical limit of the quantum relations in Section \ref{sec:quant-ab}.

Finally, the hyperk\"ahler metric generalizing \eqref{metric-U1} takes the form (\cf\ \cite{dBHOY})
\be ds^2 = U^{a{b}}\,d\vec\phi_a \cdot d\vec\phi_{b} + (U^{-1})_{a{b}}(g_a^{-2}d\gamma_a+\vec\omega^{ac}\cdot d\vec\phi_{c})(g_a^{-2}d\gamma_{b}+\vec\omega^{bc'}\cdot d\vec\phi_{c'})\,, \label{metric-gen} \ee
\be U^{ab} = \frac{1}{g_a^2}\delta^{ab} + \sum_{i=1}^N \frac{Q^a{}_iQ^{b}{}_i}{\big|\vec M_i\big|} \,,\qquad   \vec\nabla U^{ab} = \vec\nabla\times \vec\omega^{ab}\,,
\label{U-abel} \ee
where $\vec M_i=(M_i^\R,\Re\,M_i,\Im\,M_i)$ is the triplet of effective masses for the $i$-th hypermultiplet.
This metric describes an $(S^1)^r$ fibration over $(\mathbb{R}^3)^r$.
The metric receives corrections only at one loop, which appear in the functions $U^{ab}$ and the Dirac connection $\vec\omega^{ab}$. This metric can be used directly to justify the chiral ring relations \eqref{abel-gen} and to derive the holomorphic symplectic form --- most easily, by studying the limit $g_a^2\to \infty$ and then using the non-renormalization argument to ensure that the result is independent of $g_a$. Instead, we will prove formulas \eqref{abel-gen} and \eqref{abel-Omega} using 3d abelian mirror symmetry (together with non-renormalization).

\subsection{Derivation via mirror symmetry}
\label{sec:abelMS}

The mirror of an abelian theory $T$ with $N$ hypermultiplets and gauge group $G = \prod_{a=1}^rU(1)_a$ is another abelian theory $\wt T$ with $N$ hypermultiplets $(X_i,Y_i)$ and gauge group $\wt G = \prod_{\alpha=1}^{N-r} U(1)_\alpha$. The gauge charges $\wt Q_\alpha{}^i$ and flavor charges $\tilde q_a{}^i$ in the mirror theory may be combined into an $N\times N$ matrix $\wt {\mb Q}$, which is related to the matrix $\mb Q$ defined in \eqref{abel-charges} by \cite{dBHOY}
\be \wt {\mb Q} = \begin{pmatrix} \tilde q\\ \wt Q \end{pmatrix} = \mb Q^{-1,T}\,. \label{tQq}\ee
As this matrix is unimodular, $|\!\det \mb Q|  = 1$, the mirror charge matrix  $\wt Q_\alpha{}^i$ has integer entries.
Mirror symmetry interchanges the Coulomb and Higgs branch symmetries of these theories so that $\wt G_H = G_C = \prod_{a=1}^r U(1)_a$ and $\wt G_C = G_H = \prod_{\alpha=1}^{N-r} U(1)_\alpha$. In particular, the $N-r$ FI parameters of $\wt T$ are related to the masses of $T$: $\tilde t_\alpha = m_\alpha$. Subject to these relations, the Higgs branch chiral ring $\C[\wt \CM_H]$ of $\wt T$ should be identical to the Coulomb branch chiral ring $\C[\CM_C]$ of $T$.

Since the Higgs branch of $\wt T$ receives no quantum corrections, we can describe its chiral ring explicitly as a holomorphic symplectic quotient of the ring of functions $\C[X_i,Y_i]$ built from hypermultiplets,
\be \textstyle \C[\wt \CM_H] = \C[X_i,Y_i]^{\wt G}\big/(\tilde \mu_\alpha := \sum_i \wt Q_\alpha{}^iX_i Y_i = \tilde t_\alpha)\,.\ee
In other words, just as in \eqref{SQED-OH} for SQED, we take polynomials in $X_i,Y_i$ that are invariant under the $\wt G$ gauge action and impose complex moment map constraints.

Clearly the $N$ functions $Z_i := X_iY_i$ are gauge invariant. They are not all independent, since $\sum_i \wt Q_\alpha{}^i Z_i = \tilde t_\alpha$. As independent elements we can take the $N-r$ complex moment maps for the flavor symmetry group, $\nu_a:=\sum_i \tilde q_a{}^i Z_i$. Then since $\wt{\mb Q} = \mb Q^{-1,T}$ we have
\be Z_i = \sum_a Q^a{}_i\nu_a + \sum_\alpha q^\alpha{}_i \tilde t_\alpha\,, \label{Znu}\ee
analogous to \eqref{abel-defM}. The remaining gauge-invariant monomials are all of the form
\be \textstyle W^w := X^{w_+}Y^{w_-} = \prod_i (X_i)^{w^i_+}(Y_i)^{w^i_-}  \,,\qquad w = w_+-w_-\in \Z^N\,,\ee
for vectors $w$ that are in the kernel of $\wt Q:\Z^{N}\to \Z^{N-r}$, with positive and negative parts $w_\pm \in \Z^N_{\geq 0}$. In particular, the monomials $W^{Q_a}$ are gauge invariant, where $Q_a = (Q_a{}^1,...,Q_a{}^N)$ denotes a charge vector of $T$. Indeed, assuming that $|\!\det\mb Q|=1$, the kernel of the map $\wt Q: \mathbb{Z}^N \to \mathbb{Z}^{N-r}$ simply equals the image of $Q^T :\Z^r\to \Z^N$.

It is now completely straightforward to calculate that
\be W^{Q^a} W^{-Q^a} = \prod_{i=1}^N (Z_i)^{|Q^a{}_i|} =: P_a(\nu,\tilde t)\,,\ee
and more generally
\be W^w W^v = W^{w+v} P^{w_++v_+-(w+v)_+}(\nu,\tilde t)\,, \ee
with $P^w(\nu,\tilde t) := \prod_i (Z_i)^{w^i}$ and $w,v\in \text{im}\,Q^T$. The Higgs-branch chiral ring is then generated as
\be \C[\wt\CM_H] = \big\langle \{\nu_a\}_{a=1}^r,\, \{W^w\}_{w\in {\rm Im}\,Q^T}\,\big|\,
  W^w W^v = W^{w+v} P^{w_++v_+-(w+v)_+}(\nu,\tilde t)\,\big\rangle\,.
\ee
If we replace
\be \begin{array}{c@{\;\;\to\;\;}c}
 \nu_a & \varphi_a \\
 W^{\pm Q^T\cdot A} & \V^\pm_A \\
 Z_i & M_i \\
 \tilde t_\alpha & m_\alpha\,,
\end{array} \ee
then we recover the presentation \eqref{abel-O} for $\C[\CM_C]$ in the original theory.

Finally, we can check that the holomorphic symplectic form \eqref{abel-Omega} is correct. On the Higgs branch of the mirror theory $\wt T$, the holomorphic symplectic form $\wt\Omega_H$ descends from the  form $\sum_i dY_i\wedge dX_i$ on $T^*\C^N$ upon symplectic reduction. Observe that
\begin{align}
 \sum_{a}  d\nu_a\wedge d\log\frac{W^{Q^a}}{P_a(\nu,\tilde t)^{1/2}} &=
  \sum_{a,i} d(\tilde q_a{}^i X_i Y_i)\wedge d\log(X^{\frac12 Q^a}Y^{-\frac12 Q^a}) \label{Omega-der1} \\
  &= \sum_{a,\alpha;i} d(\wt{\mb Q}_{\alpha,a}{}^i X_i Y_i) \wedge d\log( X^{\frac12 \mb Q^{\alpha,a}}Y^{-\frac12\mb Q^{a,\alpha}}) \label{Omega-der2} \\
  &\hspace{-1.3in}= \frac12\sum_{i,j}  (\wt{\mb Q}^T\mb Q)^{ij} \Big( \frac{Y_i}{X_j}dX_i\wedge dX_j - \frac{X_i}{Y_j}dY_i\wedge dY_j + \Big(\frac{X_i}{X_j}+\frac{Y_i}{Y_j}\Big)dY_i\wedge dX_j\Big)  \notag\\
  &= \sum_i dY_i\wedge dX_i\,, \notag
\end{align}
where in \eqref{Omega-der2} we used the constraints $\sum_i \wt Q_\alpha{}^iX_iY_i=t_\alpha$ to promote the sum over $\alpha=1,...,r$ to a sum over both $\alpha$ and $a$ covering all indices of $\wt{\mb Q}$. The calculation shows that the symplectic reduction of $\sum_i dY_i\wedge dX_i$ can be expressed as the LHS of \eqref{Omega-der1}, which translates on the Coulomb branch of $T$ to \eqref{abel-Omega}.

\subsubsection{Example: SQED from mirror symmetry}
For SQED we have the following matrix of charges: 
\be 
{\mb Q} = \begin{pmatrix} 1 & 1 & 1 & \cdots & 1 \\ 0 & 1 & 0 & \cdots  & 0\\ 0 & 0 & 1 & \cdots & 0 \\ \cdots & \cdots & \cdots & \cdots & \cdots \\ 0 & 0 & 0 &\cdots & 1 \end{pmatrix} \label{SQED-abel-charges} \,,
\ee
and thus the charges of the mirror theory can be presented as
\be \wt {\mb Q} = \mb Q^{-1,T}= \begin{pmatrix} 1 & 0 & 0 & \cdots & 0 \\ -1 & 1 & 0 & \cdots  & 0\\ -1 & 0 & 1 & \cdots & 0 \\ \cdots & \cdots & \cdots & \cdots & \cdots \\ -1 & 0 & 0 &\cdots & 1 \end{pmatrix} \,. \label{SQED-tQq}\ee
An equivalent presentation of the mirror theory, related by a redefinition of the gauge charges, is as a linear quiver of $N-1$ $U(1)$ gauge groups with a single flavor at either end (whose Coulomb branch we return to in Section \ref{sec:quiver}).

We have a single basic gauge-invariant bilinear $\nu = Z_1= X_1 Y_1$ on the Higgs branch of the mirror theory, since $Z_i = \nu + \tilde t_{i-1}$ for $i>1$. The mirrors of the basic monopole operators are
$W^{1, \cdots,1} = \prod_i X_i$ and $W^{-1, \cdots,-1} = \prod_i Y_i$, which satisfy the expected relation
\be 
W^{1, \cdots,1}  W^{-1, \cdots,-1} = \nu \prod_{\alpha=1}^{N-1} (\nu + \tilde t_\alpha )\,.
\ee

\subsection{Quantization}
\label{sec:quant-ab}

Since the Higgs and Coulomb branches of a 3d $\CN=4$ theory are complex symplectic manifolds, it is natural to ask whether they admit a quantization, and whether this quantization plays a physical role. The answer to both questions turns out to be positive.

\subsubsection{Quantization via Omega background}
\label{sec:quant-gen}

Physically, quantization can be achieved by placing a 3d $\CN=4$ theory in a two-dimensional Omega background. The details were recently presented in \cite{Yagi-quantization}. Conceptually, the idea is to reduce the 3d theory to a 1d quantum mechanics, so that operators become fixed to a line and their product (potentially) becomes non-commutative. 
This same basic idea was used in \cite{GW-surface} and later \cite{GMNIII} to quantize algebras of line operators in four-dimensional theories, by forcing the line operators to lie in a common plane (see also the recent review \cite{Gukov-surface}).%
\footnote{If one additionally puts a 4d theory on a half-space, the (quantized) algebra of line operators in the bulk acts on the boundary condition. This leads to Ward identities for line operators that have appeared in numerous recent works, \eg\ related to the 3d-3d correspondence \cite{DGG} and to integrable systems \cite{GK-3d, GGP-spectral}.} %
A direct reduction of the 4d constructions leads to the Omega background that quantizes 3d chiral rings.

Another example of quantization of an operator algebra appeared in \cite{Kapustin-Witten, GW-branes}. Namely, it was found that in a two-dimensional A-model, a ``canonical coisotropic brane'' boundary condition \cite{KapustinOrlov} induces a deformation quantization of the algebra of operators on the boundary. This quantization is also related to the 3d quantization discussed here, since the reduction of a 3d theory along the isometries of an Omega background is precisely expected to produce a 2d A-model with canonical coisotropic boundary \cite{NekWitten}.

To describe the background that quantizes the chiral ring of a 3d $\CN=4$ theory $T$, we first rewrite the 3d theory on $\R^3=\R^2\times \R_t$ as a two dimensional $\CN=(2,2)$ theory on $\R^2$ whose fields are valued in functions of the third direction $\R_t$. In general, there is some freedom in choosing an $\CN=(2,2)$ subalgebra of 3d $\CN=4$ to make manifest. The choice is parameterized by the coset $R_{[\text{3d $\CN=4$]}}/R_{[\text{3d $\CN=(2,2)$]}}$, where the numerator and denominator are the R-symmetries of the respective superalgebras. More concretely, the R-symmetry $U(1)_A\times U(1)_V$ of 2d $\CN=(2,2)$ embeds as a maximal torus of $SU(2)_C\times SU(2)_H$, so there is a $\cp^1\times \cp^1$ of choices. This, however, is the same as the choice of complex structure on 3d Higgs and Coulomb branches. After fixing complex structures, we will select the unique 2d $\CN=(2,2)$ subalgebra whose R-symmetry $U(1)_A\times U(1)_V$ leaves our distinguished complex structures invariant, \ie\ $U(1)_A = U(1)_C$ and $U(1)_V=U(1)_H$.

Now, ``turning on'' an Omega background in a 2d $(2,2)$ theory involves choosing a nilpotent supercharge $Q$ and deforming both the supersymmetry algebra and the Lagrangian so that \cite{LNS-Omega, LNS-SW, NekSW, Shadchin-2d}
\be Q^2 = \epsilon\,\CL_V\,,\label{Q2V}\ee
where $V$ is the vector field that generates rotations of $\R^2$, and $\CL_V$ the corresponding Lie derivative. There are two standard candidates for a nilpotent $Q$: the A-type supercharge $Q_C = Q_-+\ol Q_+$ and the B-type supercharge $Q_H = \ol Q_-+\ol Q_+$. Note that $Q_C$ is invariant under $U(1)_H$ and transforms with charge $+1$ under $U(1)_C$, whereas the opposite is true for $Q_H$.%
\footnote{From a 3-dimensional perspective, $Q_C$ and $Q_H$ coincide with the supercharges of Rozansky-Witten theory \cite{RW}. For example, if we turn on FI parameters (but not masses) and flow to the infrared, so that the gauge theory is well approximated by a sigma model to the Higgs branch, $Q_H$ becomes the Rozansky-Witten supercharge for the sigma-model. Similarly, $Q_C$ is the Rozansky-Witten supercharge for a sigma-model to the Coulomb branch.} %
We will loosely denote both types of Omega-deformed spacetimes as $\R^2_\epsilon\times \R_t$ or simply $\R^2_\epsilon\times \R$.

It was argued in \cite{Yagi-quantization} that the Omega background using $Q_H$ quantizes the Higgs-branch chiral ring.%
\footnote{Strictly speaking, \cite{Yagi-quantization} considered 3d $\CN=4$ sigma-models, but the results extend easily to gauge theories. A detailed description of the Omega background for gauge theories appears in \cite{LTYZ-5d}.} %
In the presence of the Omega background, $Q_H$-closed operators (which include elements of the Higgs-branch chiral ring) are restricted to lie at the origin of $\R^2_\epsilon$. The position of these operators in the third direction $\R_t$ then determines an ordering, and their operator product is no longer required to be commutative. Similarly, the Omega background with $Q_C$ quantizes the Coulomb-branch chiral ring.

The quantizations $\CA_H$ and $\CA_C$ of the Higgs and Coulomb branch chiral rings that are produced by Omega backgrounds should be unique, or almost so: they depend only on the complex FI and mass parameters (respectively) that deform the chiral rings $\C[\CM_H]$ and $\C[\CM_C]$. Notice that the set of complex FI's $t$ corresponds to a class in $H^2(\CM_H,\C)$, namely the class of the complex symplectic form $\Omega_H$. Similarly, $m$ corresponds to the class of $\Omega_C$ in $H^2(\CM_C,\C)$.

In the mathematical theory of deformation quantization, the quantization of the ring of functions on a complex symplectic manifold $\CM$ (with certain ``nice'' properties) is uniquely characterized by an element of $H^2(\CM,\C)\otimes \C[\![\epsilon]\!]$, \ie\ a formal power series in $\epsilon$ with coefficients in $H^2(\CM,\C)$, called the \emph{period} of the quantization. The types of spaces that arise as Higgs and Coulomb branches of a 3d $\CN=4$ gauge theory possess the required ``nice'' properties as long the branches can be fully resolved by turning on mass and FI parameters \cite{BK-quantization}, \cf\ \cite[Sec. 3]{BPW}. (This is equivalent to requiring that mass and FI parameters can make the theory fully massive.) More so, if one requires that quantization be equivariant with respect to the $U(1)_H$, $U(1)_C$ actions on the rings of functions
then the period must simply lie in $H^2(\CM,\C)$ itself \cite{Losev-quantizations}. The quantizations produced physically by the Omega background are precisely such equivariant quantizations, uniquely characterized by the choice of $m \in H^2(\CM_C,\C)$ and $t\in H^2(\CM_H,\C)$.

\subsubsection{Explicit presentation}

The quantization $\CA_H$ of the Higgs branch chiral ring in any 3d $\CN=4$ gauge theory can easily be described by virtue of the fact that $\C[\CM_H]$ is a complex symplectic quotient. The result is very explicit for an abelian theory. Let us use the notation of Section \ref{sec:abelMS}, considering a theory $\wt T$ (mirror to $T$) with $N$ hypermultiplets, gauge group $\wt G = \prod_{\alpha=1}^{N-r}U(1)_\alpha$, and charge matrix \eqref{tQq}.

The quantization $\ol \CA_H$ of the ``ungauged'' ring $\C[T^*\C^N]$ is canonical, due to its affine structure. Namely, the generators $X_i,Y_i$ are promoted to operators $\hat X_i,\hat Y_i$ with commutation relations
\be [\hat Y_i,\hat X_j] = \delta_{ij}\epsilon\,. \ee
Thus the quantization $\ol \CA_H$ is just $N$ copies of a Heisenberg algebra. The ring $\CA_H$ is obtained by a quantum symplectic reduction, \cf\ \cite{CBEG, BLPW-hyp}%
\footnote{Such a quantum symplectic reduction also appeared in \cite{Dimofte-QRS}, in a rather different context.}%
: first taking a subring of gauge-invariant operators in $\ol\CA_H$, and then imposing the moment-map constraints
\be  \hat{\tilde\mu}^\alpha \,=\, \sum_i \wt Q_\alpha{}^i \hat Z_i \;= \tilde t_\alpha\,, \label{hatmu} \ee
where $\hat Z_i=\;:\!\hat X_i\hat Y_i\!: \,= \hat X_i\hat Y_i+\frac\epsilon2 = \hat Y_i\hat X_i-\frac\epsilon2$ is the normal-ordered product.%
\footnote{Other operator orderings could also be used, but the resulting ambiguities can be absorbed into the complex FI parameters $\tilde t_\alpha$.} %
Note that the gauge-invariant operators in $\ol\CA_H$ are precisely those that commute with the moment maps.

As in Section \ref{sec:abelMS}, we define quantum moment maps for the flavor symmetry, $\hat \nu_a  = \sum_i \tilde q_a{}^i\hat Z_i$, as well as monomials $\hat W^w  = \hat X^{w_+}\hat Y^{w_-}$ that suffer no ordering ambiguities since for every $i$ either $(w_+)_i=0$ or $(w_-)_i=0$. After imposing \eqref{hatmu} we still have $\hat Z_i = \sum_a Q^a{}_i\hat\nu_a+\sum_\alpha q^\alpha{}_i\tilde t_\alpha$ just as in \eqref{Znu}. After some straightforward calculations, we find that $\CA_H$ is generated by $\{\hat W^w\}_{w\in\text{im} Q}$ and $\{\hat\nu_a\}_{a=1}^r$, subject to the relations
\be [\hat\nu_a,\hat\nu_b]=0\,,\qquad [\hat \nu_a,\hat W^w] = (\tilde q_a\cdot w)\epsilon\, \hat W^w\,,\ee
and the deformed product
\be  \hat W^v\hat W^w = \bigg(\prod_{\substack{\text{$i$ s.t. $|v_i|\leq |w_i|$,}\\[.05cm] v_i w_i<0}} [\hat Z_i]^{-v_i} \bigg) \hat W^{v+w} \bigg(\prod_{\substack{\text{$i$ s.t. $|v_i|> |w_i|$,}\\[.05cm] v_i w_i<0}} [\hat Z_i]^{w_i} \bigg)
\label{abel-qproduct-H} \ee
where
\be [a]^b := \begin{cases} \prod_{i=1}^{b} (a+(i-\frac12)\epsilon) & b>0 \\
 \prod_{i=1}^{|b|} (a-(i-\frac12)\epsilon) & b <0 \\
 1 & b=0\,. \end{cases}
\ee

By using abelian mirror symmetry as in Section \ref{sec:abelMS}, we can translate the quantization of $\CA_H$ for theory $\wt T$ to a quantization of the Coulomb branch $\CA_C$ for theory $T$. Namely, $\CA_C$ for theory $T$ is generated by $\{\hat \V^A_\pm\}_{A\in \Z^r}$ and $\{\hat\varphi_a\}_{a=1}^r$ subject to the commutation relations
\be  [\hat\varphi_a,\hat\varphi_b]=0\,,\qquad [\hat\varphi_a,\hat\V^\pm_A] = \pm \epsilon\, A_a\, \hat\V^\pm_A\,, \label{abel-comm} \ee
together with $\hat\V^+_A = \hat\V^-_{-A}$ and a product formula
\be \hat \V^+_A\hat\V^+_B = \bigg(\prod_{\substack{\text{$i$ s.t. $|(Q^T A)_i|\leq |(Q^T B)_i|$,}\\[.05cm] (Q^TA)_i(Q^TB)_i<0}} [\hat M_i]^{-(Q^TA)_i} \bigg) \hat \V_{A+B}^+ \bigg(\prod_{\substack{\text{$i$ s.t. $|(Q^T A)_i|> |(Q^T B)_i|$,}\\[.05cm] (Q^TA)_i(Q^TB)_i<0}} [\hat M_i]^{(Q^TB)_i} \bigg)\,, \label{abel-qproduct}
\ee
which has as a special case
\be \hat\V_a^+\hat\V_a^- = \prod_{i=1}^N [\hat M_i]^{-Q^a{}_i} =:\, \hat P_a(\hat\varphi,m)\,. \label{abel-Vproduct}\ee
Here $\hat M_i = \sum_a Q^a{}_i\hat\varphi_a+ \sum_\alpha q^\alpha{}_im_\alpha$, as in \eqref{abel-defM}.

Note that in (say) a Coulomb-branch Omega background, the R-symmetry $U(1)_C = U(1)_A$ is broken explicitly by the RHS of \eqref{Q2V}. It could be restored by giving $\epsilon$ charge $+2$, which is precisely the charge of the holomorphic symplectic form $\Omega_C$. Correspondingly, the quantum operator products \eqref{abel-comm}, \eqref{abel-Vproduct} break $U(1)_C$, even at zero complex mass, unless $\epsilon$ is given a charge $+2$. 

\subsubsection{Example: SQED from mirror symmetry, quantized}

In the mirror of SQED, quantization of the basic bilinear is $\hat \nu = \hat Z_1= \hat X_1 \hat Y_1+ \frac{\epsilon}{2}$, with $\hat Z_i = \hat \nu + \tilde t^{i-1}$. Moreover,
$\hat W^{1, \cdots,1} = \prod_i \hat X_i$ and $\hat W^{-1, \cdots,-1} = \prod_i \hat Y_i$. We have 
\be
\begin{array}{rl}
\hat W^{1, \cdots,1}  \hat W^{-1, \cdots,-1} &= \big(\hat \nu - \tfrac{\epsilon}{2}\big) \prod_{\alpha=1}^{N-1} \big(\hat \nu + \tilde t_\alpha - \tfrac{\epsilon}{2}\big)\,, \\[.2cm]  \hat W^{-1, \cdots,-1} \hat W^{1, \cdots,1} &= \big(\hat \nu +\tfrac{\epsilon}{2}\big) \prod_{\alpha=1}^{N-1} \big(\hat \nu + \tilde t_\alpha + \tfrac{\epsilon}{2}\big)\,.
\end{array}
\ee
In SQED itself, these translate to
\be \hat\V^+\hat\V^- = P\big(\hat\varphi-\tfrac\epsilon2,m\big) = \prod_{i=1}^N\big(\hat\varphi-\tfrac\epsilon2+m_i\big)\,,\qquad \hat\V^-\hat\V^+ = P\big(\hat\varphi+\tfrac\epsilon2,m\big)\,,\ee
along with $[\hat\varphi,\hat\V^\pm] = \pm \epsilon \hat\V^\pm$. For $N=2$ these are the relations for a central quotient of the universal enveloping algebra of $\mathfrak{sl}_2$; while for general $N$ the operators $\hat\V^\pm,\hat \varphi$ generate a spherical Cherednik algebra (\cf\ \cite{Gordon-Cherednik, EGGO}).

\subsection{Twistor space} 
\label{sec:twistor-ab}

So far, we have focused on the Coulomb branch $\CM_C$ as a complex symplectic manifold. In abelian gauge theories, we could go further and write down the hyperk\"ahler metric, as it receives only 1-loop quantum corrections. As a warm-up for theories with nonabelian gauge groups, where such an explicit construction of the metric is not possible, we will now describe the hyperk\"ahler structure using the twistor construction. In general, a hyperk\"ahler manifold $M$ defines a twistor space $\CZ\simeq M\times \cp^1$ with certain properties, and vice versa \cite{HKLR-HK}. A review of the construction appears (\eg) in \cite[Sec.~3]{GMN}. We recall a few relevant facts.

A hyperk\"ahler manifold $M$ has an $S^2$ worth of complex structures, parametrized as $aI+b J +cK$, with $a^2+b^2+c^2=1$, where $I,J,K$ are complex structures satisfying $I^2=J^2=K^2 = IJK = -1$. One identifies $S^2$ with $\cp^1$, with its standard complex structure $\tilde I$. Let $\zeta$ be an affine coordinate on $\cp^1$, and denote by $I_\zeta$ the corresponding complex structure on $M$, so that, for example, $I_0=I$ and $I_\infty = - I$. The twistor space $\CZ := M\times \cp^1$ is a complex manifold, whose complex structure at a point $(m,\zeta)$ is $(I_\zeta, \tilde I)$. One then verifies that:
\begin{enumerate}
\item The projection $p:\CZ\to \cp^1$ is holomorphic, so that $p^{-1}(\zeta)$ is a copy of $M$ with complex structure $I_\zeta$.
\item The antipodal map $\tau_0:\cp^1\to \cp^1$ ($\tau_0\zeta=-\ol\zeta^{-1}$)
lifts to an antiholomorphic involution $\tau:\CZ\to \CZ$, providing a real structure on $\CZ$.
\item There is a section $\Omega_\zeta$  of $\bigwedge^2T^*_{\CZ/\cp^1}\otimes \CO(2)$ satisfying $\tau^*\Omega_\zeta = \Omega_\zeta$, which in the fiber $p^{-1}(\zeta)$ becomes the holomorphic symplectic form on $M$ in complex structure $\zeta$. Explicitly, let $\omega_I,\omega_J,\omega_K$ denote the K\"ahler forms on $M$ in complex structures $I,J,K$. Then
\be  \Omega_\zeta = \omega_J+i\omega_K + 2\zeta \omega_I - \zeta^2(\omega_J-i\omega_K) = \Omega + 2\zeta\omega + \zeta^2\ol\Omega\,, \label{Omegazeta} \ee
where $\omega:=\omega_I$ and $\Omega:=\Omega_0=\Omega_I$. (The involution $\tau^*$ acts as a composition of the antipodal map $\Omega_\zeta\mapsto \ol\zeta^2\Omega_{-\ol\zeta^{-1}}$ and complex conjugation in the fibers $\Omega_\zeta\mapsto\ol{\Omega_\zeta} = \ol\Omega-2\bar\zeta\ol\omega+\bar\zeta^2\Omega$, which together preserve $\Omega_\zeta$.) 
\item For all points $m\in M$, the section $\{m\}\times \cp^1$ of $p:\CZ \to M$ is holomorphic and real with respect to $\tau$. The normal bundle to any such section is isomorphic to $\C^{\dim_\C M}\otimes_\C \CO(1)$.
\end{enumerate}
Conversely, given a $2d+1$ dimensional complex manifold $\CZ$ with a projection $p:\CZ\to\cp^1$, satisfying the first three properties above, the moduli space of real sections as in (4) parametrizes a hyperk\"ahler manifold.

The sections in (4), restricted to a fixed $\zeta\in \cp^1$, provide (locally) holomorphic functions on $M$ in complex structure $\zeta$. In our main case of interest, where $M$ is the Higgs or Coulomb branch of a 3d $\CN=4$ theory, these functions should arise as the expectation values of chiral operators -- \ie\ operators obeying a BPS condition with respect to a combination of supercharges that is \emph{also} labelled by the twistor parameter~$\zeta$. The R-symmetry $SU(2)_H$ or $SU(2)_C$, as appropriate, rotates the $\cp^1$ twistor sphere.

Some of the holomorphic functions we have encountered arise naturally as values of a complex moment map.
If a compact group $G$ acts on $M$ via hyperk\"ahler isometries, then it preserves all three forms $\omega_I,\omega_J,\omega_K$, and gives rise to three $\mathfrak g^*$-valued moment maps $\mu_I,\mu_J,\mu_K$. Letting $\mu^\R=\mu_I$ and $\mu = \mu_J + i\mu_K$ be the real and complex moment map in complex structure $I$, \eqref{Omegazeta} implies that $\mu_\zeta = \mu + 2\zeta\mu^\R - \zeta^2\ol\mu$ is the complex moment map in complex structure $I_\zeta$. In other words, moments maps are real sections of $\CO(2)$.

\subsubsection{Twistor space for $\R^4$}
\label{sec:XYtwistor}

A basic example of a hyperk\"ahler manifold is $M= \R^4$. This occurs as the Higgs branch of a theory with a single hypermultiplet (with trivial gauge group $G$). Using property (4) and identifying $M$ with its own cotangent bundle, we find that the twistor space $\CZ$ is the total space of the bundle $\CO(1)\oplus \CO(1)\to \cp^1$.

Suppose that in complex structure $I$ the holomorphic coordinates on $M\simeq \C^2$ are $(X,Y)$. Then, using $\zeta$ as an affine parameter centered at the ``north pole'' $I\in \cp^1$, the sections in (4) are
\be X_\zeta = X - \zeta \ol Y\,,\qquad Y_\zeta = Y + \zeta \ol X\,. \label{XYnorth} \ee
Physically, these can be obtained by applying an $SU(2)_H$ rotation to a hypermultiplet. Letting $\wt \zeta = 1/\zeta$ be an affine parameter centered at the south pole of $\cp^1$, and recalling that local coordinates $(u,\zeta)$ on $\CO(k)\to\cp^1$ transform as $(u,\zeta)\mapsto(u/\zeta^k,1/\zeta) = (\wt \zeta^k u,\wt \zeta)$, we also find the continuation of \eqref{XYnorth} to a neighborhood of the south pole,
\be \wt X_{\tilde \zeta} = \wt\zeta X - \ol Y\,,\qquad \wt Y_{\tilde \zeta} = \wt\zeta Y + \ol X\,.\ee
The holomorphic symplectic form is
\begin{align} \Omega_\zeta &= dX_\zeta\wedge dY_\zeta = dX\wedge dY + \zeta(dX\wedge d\ol X + dY\wedge d\ol Y) + \zeta^2d\ol X\wedge d\ol Y \\
& = \Omega + 2\zeta\omega + \zeta^2\ol \Omega \notag
\end{align}
around the north pole, as desired; and around the south pole we have $\wt \Omega_{\wt \zeta} = \wt \zeta^2\Omega_{1/\wt \zeta}$, as appropriate for a section of $\CO(2)$. Also observe that
\be \ol {X_\zeta} = \ol X - \ol\zeta Y = \wt Y_{-\ol\zeta}\,,\qquad \ol {Y_\zeta} = \ol Y + \ol \zeta X = -\wt X_{-\ol \zeta}\,; \label{XYreal} \ee
thus $X_\zeta,Y_\zeta$ are real sections with respect to an involution $\tau$ that combines the antipodal map on $\cp^1$ with a twisted conjugation $(X_\zeta, Y_\zeta)\mapsto (\ol{Y_\zeta},-\ol{X_\zeta})$ in the fibers.

The space $M$ admits a $G_H=U(1)$ hyperk\"ahler isometry, whose complex moment map is $\mu_\zeta = X_\zeta Y_\zeta = XY + \zeta(|X|^2-|Y|^2) - \zeta^2 \ol{XY}$. We recognize in the middle term the real moment map in complex structure $I$, $\mu^\R = \frac12(|X|^2-|Y|^2)$.

\subsubsection{SQED in the IR}
\label{sec:IRtwistor}

Now consider the Coulomb branch $\CM_C$ of SQED with $N$ hypermultiplets. The complex field $\varphi$ is the complex moment map for the topological $U(1)_t$ isometry, and so must define a real section of $\CO(2)$ in twistor space,
\be \varphi_\zeta = \varphi + 2\zeta\sigma - \zeta^2\ol\varphi\,,\qquad \wt\varphi_{\wt \zeta} = -\ol\varphi+2\wt\zeta\sigma+\wt\zeta^2\varphi \,; \qquad \ol{\varphi_\zeta} = -\wt\varphi_{-\ol\zeta}\,, \label{real-phi} \ee
where ``$\sigma$'' denotes the real moment map for $U(1)_t$.

In the infrared, \ie\ at infinite gauge coupling $g^2\to\infty$, the twistor description of monopole operators can be obtained by using mirror symmetry. (Alternatively, classic references such as \cite{Hitchin-HK} provide a twistor description of the resolved  $\C^2/\Z_N$ singularity.)
In the mirror of SQED, the mirrors of monopole operators $V_\pm$ are products $X_1X_2\cdots X_N$ and $Y_1Y_2\cdots Y_N$ of chiral fields. Since the $X_i$ and $Y_i$ are promoted to sections of $\CO(1)$ in twistor space, the monopole operators $\V_\zeta^\pm$ must be sections of $\CO(N)$. Around the north pole,  $\V_\zeta^\pm$  are degree-$N$ polynomials in $\zeta$, and around the south pole $\wt \V_{\wt \zeta}^\pm = (\wt\zeta)^N \V_{1/\wt\zeta}^\pm$. The real structure is inherited from \eqref{XYreal}:
\be \ol {\V_\zeta^+} = \wt \V_{-\ol \zeta}^-\,,\qquad \ol {\V_\zeta}^i = (-1)^N \wt \V_{-\ol\zeta}^+\,. \label{real-v} \ee

The full twistor space of the Coulomb branch can be obtained by starting with the vector bundle $\CO(N)\oplus \CO(N) \oplus \CO(2) \to \cp^1$ and imposing the equation
\be \V_\zeta^+ \V_\zeta^-  = \varphi_\zeta^N \label{Vzeta1} \ee
among respective sections. This may be deformed by choosing $N$ fixed sections $m_{\zeta,i} = m_i + 2\zeta m_i^\R - \zeta^2\ol m_i$ of $\CO(2)$, encoding the real and complex masses of the theory:
\be \V_\zeta^+ \V_\zeta^-  =  \prod_{i=1}^N(\varphi_\zeta+m_{\zeta,i})\,. \label{Vzeta2} \ee
The holomorphic symplectic form is as in \eqref{SQED-Omega}, $\Omega_\zeta = d\varphi_\zeta\wedge d\log \V_\zeta^+$.

\subsubsection{SQED at finite gauge coupling}
\label{sec:gtwistor}

At finite gauge coupling, the Coulomb branch of SQED has the same form as a complex manifold, but is modified to a multi-centered Taub-NUT space as a hyperk\"ahler manifold. Such a modification was described mathematically in \cite{Hitchin-HK}.

Physically, we may understand the modification by considering the semi-classical expressions for the chiral monopole operators in complex structure $I$, $\V^\pm \sim e^{\pm\frac{1}{g^2}(\sigma+i\gamma)}$. If we rotate such expression naively with $SO(3)_C$ we obtain a monopole operator which is 
chiral in complex structure $\zeta$, involving a rotated real combination $\sigma_\zeta$ of the three vectormultiplet scalar fields. 
The scalar field $\sigma_\zeta$, though, is not holomorphic in $\zeta$ and it is thus unsuitable for the purpose of describing the twistor space. 
We can ameliorate that problem by multiplying the rotated monopole operator by an appropriate function of the complex scalar $\varphi_\zeta$,
to obtain a dressed monopole operator that is holomorphic in $\zeta$: $\V_\zeta^\pm \sim e^{\pm\frac1{g^2}(\sigma+i\gamma - \zeta\ol\varphi)}$.
\footnote{In detail, $\sigma_\zeta = \frac{1-|\zeta|^2}{1+|\zeta|^2} \sigma - \frac{\bar \zeta}{1+|\zeta|^2}\varphi- \frac{\zeta}{1+|\zeta|^2}\ol \varphi$ and thus $\sigma_\zeta = \sigma - \zeta\ol\varphi - \frac{\bar \zeta}{1+|\zeta|^2}\varphi_\zeta$} 

Similarly, in complex structure $-I$, the monopole operators are $\V^\pm\sim e^{\mp\frac1{g^2}(\sigma-i\gamma)}$, which become $\wt \V_{\wt\zeta}^\pm\sim e^{\mp\frac1{g^2}(\sigma-i\gamma+\wt\zeta\varphi)}$ in a neighborhood of the south pole. The transformation from the north to the south poles is multiplication by
\be  \exp\Big( \mp \frac1{g^2}\frac{\varphi_\zeta}{\zeta}\Big)\,, \label{Ltrans} \ee
where, notably, $\varphi_\zeta$ is the complex moment map for $U(1)_t$.

Combining this observation with the ``topological'' quantum correction of Section \ref{sec:IRtwistor}, which made the monopole operators sections of $\CO(N)$, we might guess the following description for the twistor space of the Coulomb branch. We introduce a complex line bundle $L$ over the total space of $\CO(2)\to\cp^1$, with transition function $\exp\big( \frac1{g^2}\frac{\varphi_\zeta}{\zeta}\big)$ on the intersection of affine charts, and its dual $L^*$ with transition function $\exp\big(- \frac1{g^2}\frac{\varphi_\zeta}{\zeta}\big)$. Then we view the vector bundle $\CO(N)\oplus \CO(2)\to \cp^1$ as a line bundle $\hat\CO(N)\to (\CO(2)\to \cp^1)$, and twist it by $L\oplus L^*$ to obtain
\be E = \hat\CO(N)(L\oplus L^*)\,. \ee
The twistor space $\CZ$ is the subvariety of $E$ defined by choosing $N$ sections $m_{\zeta,i}$ of $\CO(2)$ as in \eqref{Vzeta2}, and then imposing 
\be \V_\zeta^+ \V_\zeta^- = \prod_{i=1}^N(\varphi_\zeta + m_{\zeta,i}) \ee
among (local) sections $\V_\zeta^+$, $\V_\zeta^-$, and $\varphi_\zeta$  of $\hat\CO(N)L$, $\hat\CO(N)L^*$, and $\CO(2)$, respectively. This description coincides with that of \cite{Hitchin-HK} for $N=2$.

The real structure and the holomorphic symplectic form are unchanged from Section \ref{sec:IRtwistor}. The only difference is that now the monopole operators, \ie\ the real sections of $\hat\CO(N)L$ or $\hat\CO(N)L^*$, take the form of a degree-$N$  polynomial in $\zeta$ multiplied by the exponential factors $e^{\pm \frac1{g^2}(\sigma+i\gamma-\zeta\ol\varphi)}$. For example, when $N=1$,
\be \begin{array}{ll} \V_\zeta^+ = (a-\zeta\ol b) e^{\frac1{g^2}(\sigma+i\gamma-\zeta\ol\varphi)}\,, \qquad & \wt \V_{\wt \zeta,+} = (\wt\zeta a-\ol b) e^{-\frac1{g^2}(\sigma-i\gamma+\wt \zeta\varphi)}\,, \\
\V_\zeta^- = (b+\zeta\ol a) e^{-\frac1{g^2}(\sigma+i\gamma-\zeta\ol\varphi)}\,,\qquad &
\wt \V_{\wt \zeta,-} = (\wt \zeta b+\ol a)e^{\frac1{g^2}(\sigma-i\gamma+\wt \zeta\varphi)}\,,
\end{array}
\ee
with $v_\zeta^+v_\zeta^-= \varphi_\zeta$, or equivalently $ab = \varphi$ and $|a|^2-|b|^2 = 2\sigma$.

\subsubsection{The general case}

The twistor space of the Coulomb branch of a general abelian theory is a straightforward generalization of SQED. Suppose the gauge group is $G=\prod_{a=1}^r U(1)_a$ and that there are $N$ hypermultiplets with gauge charges $Q^a{}_i$ and flavor charges $q^\alpha{}_i$ as in Section \ref{sec:abel-gen}.
The holomorphic scalars $\varphi^a$ in complex structure $I$ are promoted to sections $\varphi_\zeta^a$ of $\CO(2)$, together defining a section of $\CO(2)^{\oplus r}\to\cp^1$. Each monopole operator $\V_A^\pm$ is promoted to a section $\V_{A,\zeta}^\pm$ of the bundle $\CO(\sum_i|(Q^TA)_i|)\to \CO(2)^{\oplus r}$, twisted by a line bundle $L^{\pm A}$ with transition function $\exp\big(\pm\sum_a \frac1{g_a^2}A_a\frac{\varphi_{a,\zeta}}\zeta\big)$. Let $\{\V_A\}_{A\in \mb A}$ be any finite set of monopole operators that together with the $\varphi_a$ generate the Coulomb-branch chiral ring $\C[\CM_C]$. Then the twistor space of the Coulomb branch is the subvariety of
\be E = \bigoplus_{A\in\mb A} \Big(\CO(\sum_i|(Q^TA)_i|)\otimes L^A\Big) \to \CO(2)^{\oplus r} \label{E-abel} \ee
cut out by the straightforward $\zeta$-dependent generalization of \eqref{abel-gen}. The complex symplectic form is the straightforward $\zeta$-dependent generalization of \eqref{abel-Omega} and the real structure is $\ol{\varphi_\zeta^a} = -\wt\varphi_{-\ol\zeta}^a$ and $\ol{\V_{A,\zeta}^+} = \epsilon_{A,\pm}\wt \V_{-A,-\ol\zeta}^+$, with a choice of signs $\epsilon_{A}\in\{\pm1\}$ consistent with the chiral ring relations.

\section{Nonabelian gauge theories}
\label{sec:NA}

In the remainder of the paper, we aim to describe the Coulomb branch of nonabelian gauge theories. Here we present a set of properties that should be valid in any gauge theory. Part of the key to our analysis is the non-renormalization result from the Section \ref{sec:gen}. Another is the expected structure of the metric on the Coulomb branch. We will argue that the only corrections to the classical metric that can effect the complex structure of the Coulomb branch are one-loop corrections; and moreover that these one-loop corrections are determined by an abelianized version of the theory. This allows us to upgrade many results from Section \ref{sec:abelian} to arbitrary gauge theories.

\subsection{The metric on the Coulomb branch}

Consider a 3d $\CN=4$ theory with gauge group $G$ of rank $r$. In general, $G$ can be a product of abelian and simple factors, or a central quotient thereof.
As discussed in Section \ref{sec:gen}, the vectormultiplet has a triplet of scalar fields $\vec\phi \in (\mathfrak g)^3$ valued in the real Lie algebra of $G$. On the Coulomb branch, the scalars $\vec\phi$ take expectation values in a Cartan subalgebra $\mathfrak t\subset \mathfrak g$. In particular, the scalar potential contains terms of the form $|[\phi_i,\phi_j]|^2$, which guarantee that all three components of $\vec\phi$ belong to the \emph{same} Cartan subalgebra. One also typically requires that the expectation values of $\vec\phi$ are sufficiently generic to break the gauge group to a maximal torus $\mathbb T\subset G$. The massless abelian gauge fields for the $r$ $U(1)$ factors in $\mathbb T$ can be dualized to $r$ periodic dual photons $\gamma\in \mathfrak t/\Lambda_G$, where $\Lambda_G = \text{Hom}(U(1),\mathbb T) \subset \mathfrak t$ is the cocharacter lattice. The classical Coulomb branch $\CM_C^{\rm clas}$ then takes the form
\be \CM_C^{\rm clas} \simeq \big[(\R^{3r}-\Delta)\times (S^1)^{r}\big]/{W_G}\,, \ee
where $W_G$ is the Weyl group of $G$ (the residual gauge transformations acting on the Cartan-valued $\vec\phi$ and $\gamma$); and $\Delta$ is the discriminant locus: the set of $\vec\phi\in (\mathfrak t)^3 \simeq \R^{3r}$ that do not fully break $G$ to the maximal torus $\mathbb T$. For example, if $G=U(r)$, $\Delta$ is the set where eigenvalues $\vec\phi_a$ of the three components of $\vec\phi$ simultaneously coincide, $\Delta = \{\prod_{a< b} \big|\vec\phi_a-\vec\phi_b\big|=0\}$.

The classical metric on the Coulomb branch takes the same form as the classical metric for an abelian theory with gauge group $\mathbb T$. For concreteness, we choose a factorization $\mathbb T\simeq  \prod_{a=1}^r U(1)_a$ and a corresponding basis $\{\chi^a\}_{a=1}^r$ for the cocharacter lattice $\Lambda_G$, such that $\chi_a: U(1)\overset\sim\to U(1)_a$. We expand $\vec\phi = \sum_a \vec\phi_a\chi^a$ and $\gamma = \sum_a \vec\gamma_a\chi^a$, and let $\kappa^{ab}$ denote the Cartan-Killing form in this basis. The classical metric on the Coulomb branch takes the form
\be ds^2_{\rm clas} =\frac{1}{g_a^2}\kappa^{ab}\big(d\vec\phi_a\cdot d\vec\phi_b + d\gamma_a d\gamma_b\big)\,, \ee
where the $g_a$ are couplings for the abelianized gauge group.%
\footnote{Recall that our normalization for the dual photons is such that $\gamma_a\sim \gamma_a+2\pi g_a^2$.}

The classical Coulomb branch has both perturbative corrections at one loop, and nonperturbative corrections due to BPS monopoles. Perturbative corrections come from hypermultiplets and from W-bosons, and are almost identical to the corrections in a purely abelian theory. Suppose that our theory has $N$ hypermultiplets $(X^i,Y^i)$, transforming in a quaternionic representation of $G$ with weights $\mu_i \in \mathfrak t^*$ for
$1\leq i \leq N$. (The component $\mu_i{}^a = \langle\mu_i,\chi^a\rangle$ can be understood as the charge of $X_i$ under the factor $U(1)_a$ in the abelianized gauge group.) Similarly, let $\alpha_j$ be the roots of $G$, \ie\ the (nonzero) weights of the adjoint representation, with components $\alpha_j{}^a = \langle \alpha_j,\chi^a\rangle$. Then the perturbative metric at (essentially) one loop is \cite{ChalmersHanany, DKMTV-monopoles, DTV-matter, Tong-ADE, GibbonsManton} 
\be ds^2_{\rm pert} = U^{ab}\,d\vec\phi_a \cdot d\vec\phi_b + (U^{-1})_{ab}(g_a^{-2}d\gamma_a+\vec\omega^{ac}\cdot d\vec\phi_c)(g_a^{-2}d\gamma_b+\vec\omega^{bc'}\cdot d\vec\phi_{c'})\,, \label{gpert} \ee
with
\be  U^{ab} = \frac{\kappa^{ab}}{g_a^2} + \sum_{i=1}^N \frac{\mu_i{}^a\mu_i{}^b}{\big|\vec M_i\big|} - \sum_{j=1}^{\text{dim}(G)-r} \frac{\alpha_j{}^a\alpha_j{}^b}{\big|\vec M^{\rm W}_j\big|}\,,\qquad \vec\nabla U^{ab} = \vec\nabla\times\vec\omega^{ab}\,. \label{U-nonabel}  \ee
Here $\vec M_i = \langle \mu_i,\vec\phi\rangle+... = \sum_a \mu_i{}^a\vec\phi_a+...$ is the effective masses of each hypermultiplet (where additional mass terms from flavor symmetries can enter in the `$...$' terms, just like in the abelian case); and $\vec M^{\rm W}_j = \langle \alpha_j,\vec\phi\rangle = \sum_a\alpha_j{}^a\vec\phi_a$ is the effective masses of each W-boson. Comparing \eqref{U-nonabel} to \eqref{U-abel}, we see that the only difference between this metric and that of an abelian theory are the W-boson corrections, entering with opposite sign to the hypermultiplets.

The nonperturbative corrections to the metric on the Coulomb branch come from monopoles, and are notoriously difficult to compute. Direct computations for gauge group $G=SU(n)$ were carried out  explicitly in \cite{DKMTV-monopoles, DTV-matter, FraserTong}. In the case of pure $G=SU(2)$ theory, symmetry and smoothness of the moduli space uniquely identifies the Coulomb branch as the Atiyah-Hitchin manifold \cite{SW-3d}, whose exact hyperk\"ahler metric was described in \cite{AH}; but for most gauge groups and matter content, the full set of nonperturbative corrections are unknown.%
\footnote{In principle, they may be obtained by using the methods of \cite{GMN} for 4d $\CN=2$ theory on a circle of finite radius, then taking the radius to zero size.}

The only fact we need to know about non-perturbative corrections is that they are proportional to the instanton action
\be \exp\bigg( -C \frac{\big|\vec M^{\rm W}_j\big|}{g_a^2}\bigg)\,. \ee
They are exponentially suppressed by the inverse gauge couplings and by the W-boson masses, which measure the distance from the discriminant locus $\Delta = \{\prod_j |\vec M^{\rm W}_j\big| = 0\}$.

\subsection{Chiral ring via abelianization}
\label{sec:abelianize}

Due to the non-renormalization argument of Section \ref{sec:ind-g}, we can analyze the chiral ring of the Coulomb branch in the limit $g_a\to 0$ (or more precisely $\min_j |\vec M^{\rm W}_j| \gg \max_a g_a$), and obtain a result that should hold for all $g_a$. As long as all the W-boson masses are nonzero, the nonperturbative corrections to the metric disappear in this limit. Thus, in the complement of the discriminant locus, it suffices to look at the perturbative metric \eqref{gpert}. Let us call the hyperk\"ahler manifold with this metric $\CM_C^{\rm abel}$, since it is essentially the Coulomb branch of an abelianized theory.

Fixing a complex structure, we may split the abelian vevs $\vec\phi$ into real and complex parts, say $\sigma=\phi_1$ and $\varphi = \phi_2+i\phi_3$, with components $\sigma_a$ and $\varphi_a$ with respect to the basis of the cocharacter lattice. For each $U(1)_a$ factor in the maximal torus of the gauge group, we construct monopole operators $v_a^\pm \sim e^{\pm\frac1{g_a^2}(\sigma_a+i\gamma_a)}$. More generally, for every cocharacter $A$ of $G$ there are abelian monopole operators $v_A^\pm$. The analysis of Section \ref{sec:abel-gen}, adapted to the geometry \eqref{gpert}, suggests that the ring of functions on $\CM_C^{\rm abel}$ should be generated by the (vevs of) $v_A^\pm$ and the complex scalars $\varphi_a$, subject to constraints of the form
\be v_a^+ v_a^- = \frac{P_a^{\rm hypers}(\varphi,m)}{P_a^{\text{W}}(\varphi)} \label{vva} \ee
with $P_a^{\rm hypers}(\varphi,m) = \prod_{i=1}^N (M_i)^{|\mu_i{}^a|}$ the product of effective complex masses of the hypermultiplets and $P_a^{\text{W}}(\varphi) = \prod_{j} (M^{\rm W}_j)^{|\alpha_j{}^a|}$ the product of effective complex masses of the W-bosons.
Explicitly, $M^{\rm W}_j = \langle \alpha_j,\varphi\rangle = \sum_a\alpha_j{}^a\varphi_a$ just as above; while $M_i = \langle \mu_i,\varphi\rangle+\langle \mu^F_i,m\rangle$, where $\mu_i\in \mathfrak t^*$  is the weight of the $i$-th hypermultiplet under the gauge group, and $\mu^F_i \in \mathfrak t_H^*$ is the weight of the $i$-th hypermultiplet under the Higgs-branch flavor symmetry group $G_H$. (Recall that $G_H$ is the normalizer of $G$ in $USp(N)$, and we can turn on complex masses $m\in \mathfrak t_H^\C$ valued in a Cartan of $G_H$.)

The appearance of W-boson masses in the denominator of \eqref{vva} is a direct consequence of the sign of the W-boson contribution to \eqref{U-nonabel}. It is consistent with the formula for the R-charge of monopole operators proposed in \cite{BKW-monopoles, GW-Sduality}. Namely, $\varphi$ has charge $+2$ under the $U(1)_C$ symmetry that preserves our choice of complex structure, while $v_a^\pm$ should have charge $\sum_i |\mu_i{}^a|-\sum_j|\alpha_j{}^a|$.

Following \eqref{abel-gen}, the relations \eqref{vva} can be generalized to 
\be v_A^+ = v_{-A}^-\,,\qquad v_A^+v_B^+ = v^+_{A+B} \frac{P^{\text{hypers}}_{A,B}(\varphi,m)}{P^{\text{W}}_{A,B}(\varphi)}\,,\label{rel-nonab}\ee 
for general cocharacters $A,B\in \mathfrak t$, with
\begin{subequations} \label{P-nonab}
\begin{align} P^{\text{hypers}}_{A,B}(\varphi,m) &= \prod_{\text{hypers $i$}}( \langle \mu_i,\varphi\rangle + \langle \mu^F_i,m\rangle)^{\langle \mu_i,A\rangle_+ + \langle \mu_i,B\rangle_+ - \langle \mu_i,A+B\rangle_+}\,, \\
 P^{\text{W}}_{A,B}(\varphi,m) &= \prod_{\text{$\alpha_j\in$roots}} \langle \alpha_j,\varphi\rangle^{\langle \alpha_j,A\rangle_+ + \langle \alpha_j,B\rangle_+ - \langle \alpha_j,A+B\rangle_+}\,, 
\end{align}
\end{subequations}
where $(x)_+ = \max\{x,0\}$. Altogether, the ring of functions on $\CM_C^{\rm abel}$ takes the form
\be \boxed{\C[\CM_C^{\rm abel}] = \Big(\C[\{v_A^\pm\}_{A\in\text{cocharacters}},\{\varphi_a\}, \{(M^{\rm W}_j)^{-1}\}_{j\in\text{roots}}]/\eqref{rel-nonab}\Big)^{W_G}\,.} \label{C-abel}\ee
In addition to the standard generators $v_A^\pm$ and $\varphi_a$, we have added the inverses of W-boson masses $M^{\rm W}_j = \langle \alpha_j,\varphi\rangle$. This is because our current description of $\CM_C^{\rm abel}$ is valid only in the complement of the discriminant locus. (The necessity of inverting the $M^{\rm W}_j$ can be seen immediately from expressions like \eqref{rel-nonab}.)
Moreover, due to residual gauge symmetry, we must only consider the part of the abelianized chiral ring that is invariant under the Weyl group $W_G$, denoted by $(\;\;)^{W_G}$.

The holomorphic symplectic form on $\CM_C^{\rm abel}$, as well as the Poisson structure on the ring $\C[\CM_C^{\rm abel}]$ and its quantization, all follow immediately from Section \ref{sec:abelian}. For example, the holomorphic symplectic form is
\be \Omega_C^{\rm abel} = \sum_{a=1}^r d\varphi_a\wedge d\,\log \bigg( v_a^+ \frac{P^{\rm W}_a(\varphi)^{\frac12}}{P^{\rm hypers}_a(\varphi,m)^{\frac12}}\bigg)\,. \label{O-abel} \ee

The ring \eqref{C-abel} must be equivalent to the ring of functions on the true Coulomb branch, in the complement of the discriminant locus:
\be  \C[\CM_C^{\rm abel}] = \C[\CM_C \backslash \Delta]\,. \label{compD} \ee
Thus, the true chiral ring is a subring
\be \C[\CM_C]\subset\C[\CM_C^{\rm abel}]\,. \label{sub} \ee
It must satisfy several properties:
\begin{enumerate} \label{Cproperties}

\item The ring $\C[\CM_C]$ has a Poisson structure, compatible with the Poisson structure on $\C[\CM_C^{\rm abel}]$. Thus $\C[\CM_C]\subset\C[\CM_C^{\rm abel}]$ is closed under the Poisson bracket.

\item The relation \eqref{compD} implies that if we invert all functions in $\C[\CM_C]$ that vanish on the discriminant locus, adjoin their inverses to $\C[\CM_C]$, we will recover $\C[\CM_C^{\rm abel}]$.

\item In many theories, the true Coulomb branch is expected to be smooth. This happens, in particular, when there is no Higgs branch, either because there are no hypermultiplets or because, in the presence of a generic mass deformation, all hypermultiplets are massive.%
\footnote{The existence of a suitable mass deformation is equivalent to the existence of a $U(1)\subset G_H$ in the Higgs-branch flavor group whose action on the Higgs branch has isolated fixed points.} %
Then $\C[\CM_C]$ is the ring of functions on a smooth variety. It cannot contain any of the functions $(M^{\rm W}_j)^{-k}$ (suitably Weyl-symmetrized) for $k\geq 1$.

\item The ring $\C[\CM_C]$ must contain dressed non-abelian monopole operators $M_{A,p}$, described in the next section, for all cocharacters $A\in \Lambda_G$ and dressing polynomials~$p$; and the embedding $\C[\CM_C]\subset \C[\CM_C^{\rm abel}]$ must preserve $U(1)_C$ R-symmetry and $G_C$ flavor symmetries.

\end{enumerate}
The predictions of our `abelianization map' will be consistent with these expectations. 

\subsection{Non-abelian monopole operators: an educated guess}
\label{sec:ops}

Some of the operators in the chiral ring $\C[\CM_C]$ are gauge-invariant functions of the complex scalar $\varphi\in \mathfrak g$. Under the abelianization map \eqref{sub}, such functions map Weyl-invariant polynomials in the $\varphi_a$, \ie\ to functions in the symmetric algebra $S[\mathfrak t]^W \subset \C[\CM_C^{\rm abel}]$. This algebra is generated (say) by Weyl-averages of monomials
\be p_{n}(\varphi) = \sum_{w\in W} w\cdot \varphi^{n} := \sum_{w\in W} w\cdot (\varphi_1^{n_1}\varphi_2^{n_2}\cdots \varphi_r^{n_r})\,,\qquad n\in \mathbb N^r\,. \ee
For example, for $G=U(r)$, the nonabelian operators $\Tr(\varphi^k)$ map to $p_{(k,0,...,0)}(\varphi) = \sum_{a=1}^r (\varphi_a)^k$\,. Formally, the Harish-Chandra isomorphism guarantees that gauge-invariant functions of $\varphi$ are in 1-1 correspondence with Weyl-invariant polynomials in the $\varphi_a$; we will frequently invert the isomorphism to label gauge-invariant functions of $\varphi$ by ``$p_n$'' or simply ``$n$''.

Of course, we expect many more functions in $\C[\CM_C]$ than the $p_n(\varphi)$. In particular, the chiral ring $\C[\CM_C]$ must include non-abelian monopole operators $M_A$. These are disorder operators, defined by specifying the singularity in the classical fields near an insertion point. Specifically, they are obtained by embedding an abelian Dirac singularity in the gauge group $G$, and thus are labelled by an element of the cocharacter lattice $\Lambda_G = \mathrm{Hom}(U(1),G)$ modulo Weyl transformations, or equivalently a dominant cocharacter $A$ of $G$. (See \cite[Sec. 10]{Kapustin-Witten} and references therein for an extended discussion of monopole singularities.) A monopole operator can also be \emph{dressed} by a function of the field $\varphi'\in \mathfrak g_A$, where $\mathfrak g_A$ is the Lie algebra of the Levi subgroup $G_A\subset G$ left unbroken by the embedding $A:U(1)\hookrightarrow G$. The dressing factor is required to be invariant only under $G_A$, rather than all of $G$. We can denote the resulting dressed operator as $M_{A,p}$, where $p$ is the $G_A$-invariant polynomial and the pair $(A,p)$ is defined up to the action of the Weyl group.   

For example, when $G=U(r)$, the embeddings $U(1)\hookrightarrow G$ are specified by an $r$-tuple of integers $A\in \Z^r$ modulo the standard action of the permutation group $S_r$. If $A$ has $r_i$ entries equal to $i\in \Z$, with $\sum_i r_i=r$, then the Levi subgroup is $G_A = \prod_i U(r_i)$. The dressed monopoles take the form $M_A \prod_i \Tr(\varphi_{(i)}{}^{k_i})$, where each $\varphi_{(i)}\in \mathfrak u(r_i)$ and $k_i\geq 0$.

It is believed that the chiral ring $\C[\CM_C]$ is completely generated by $G$-invariant functions of $\varphi$ and dressed monopole operators. (In fact, the $G$-invariant functions of $\varphi$ may themselves be considered dressings for the trivial monopole operator $M_{0,p}$.) This idea was recently tested in computations of Hilbert series for the Coulomb branch chiral ring \cite{CHZ-Hilbert}. The dressed monopole operators, however, constitute a vastly overcomplete set of generators, and in general satisfy nontrivial relations.

The relations would be automatic if we knew the abelian images of dressed monopole operators under the inclusion \eqref{sub}. This comes from computing the expectation value of 
a generic dressed monopole operator $M_{A,p}$ in a vacuum with a generic expectation value for $\varphi$. We expect the path integral to localize 
on BPS configurations which preserve the same supersymmetry as the BPS monopole itself. The BPS equations set the matter fields to zero, require $\varphi$ to be constant and
$\sigma$ and the gauge connection to both satisfy the Bogomolnyi equations and commute with $\varphi$. 

Naively, as $\varphi$ is generic, that restricts the path integral to abelian solutions of the Bogomolnyi equations with Dirac singularity 
$w \cdot A$ for some element $w$ of the Weyl group. Thus the naive expectation value of a monopole operator $M_{A,p}$, dressed by a polynomial 
\be p_{A,n}(\varphi) :=
 \sum_{w'\in W_A} w'\cdot (\varphi_1^{n_1}\cdots \varphi_r^{n_r})\, \ee
where $W_A$ is the Weyl group of $G_A$,
would be given by 
Weyl averages
of the type
\be V_{A,n} = \sum_{w\in W} v^+_{w\cdot A}\, p_{w\cdot A,w\cdot n}(\varphi) =|W_A|  \sum_{w\in W} v^+_{w\cdot A}\, w\cdot (\varphi_1^{n_1}\cdots \varphi_r^{n_r})
\, , \label{V2ab} \ee
labelled by a dominant cocharacter $A$ and a vector $n\in \mathbb N^r$.
These are Weyl averages of \emph{abelian} monopole operators labelled by the cocharacter $A$, whose dressing factors are $G_A$-invariant polynomials of the abelian fields $\varphi_a$.

This naive expectation is countered by the existence of {\it monopole bubbling}: the moduli space of solutions of nonabelian Bogomolnyi equations
with a Dirac singularity has a singular locus corresponding to the collapse of several smooth monopoles onto the Dirac singularity. 
In the neighbourhood of the singular locus, one finds ``bubbling'' solutions which are essentially abelian outside an arbitrarily small 
neighbourhood of the Dirac singularity, which is thus partially screened. Such bubbling solutions only exist if the abelian monopole charge measured at infinity 
is a dominant cocharacter $B$ that is smaller than the charge $A$ at the Dirac singularity, such that $A-B$ is a positive coroot.

Monopole bubbling solutions have the potential to contribute to the expectation value of a general monopole operator $M_{A,p}$. 
Based on other examples of localization computations, it is clear that the contribution to the path integral of a sector of abelian charge $B$ 
will involve an equivariant integral over the moduli space $\CM_A^B$ of solutions of Bogomolnyi equations with a Dirac singularity $A$ and abelian charge $B$ 
at infinity, with the expectation values at infinity of $\varphi_i$ playing the role of equivariant parameters for the gauge group action. 

The integrand should account for the one-loop determinants of the matter fields around the monopole configuration 
and for the insertion of $p(\varphi)$ at the origin. As the gauge group is broken to $G_A$ at the origin, the moduli space 
$\CM_A^B$ supports a universal $G_A$-bundle $E_A$.%
\footnote{To construct $E_A$, recall (as in Appendix \ref{app:monopoles}) that $\CM_A^B$ is realized as an infinite-dimensional hyperk\"ahler quotient of the affine space of gauge connections (and scalar fields) on $\R^3$, $\CM_A^B= \CA/\!/\!/\CG = \vec\mu^{-1}(0)/\CG$, where $\CG$ is group of gauge transformations that preserve the boundary conditions at the origin and at infinity. Letting $\CG'\subset \CG$ denote the subgroup of gauge transformations acting trivially at the origin, we have $\CG/\CG' \simeq G_A$ and $E_A = \vec\mu^{-1}(0)/\CG'$.} %
We can interpret the insertion of 
$p(\varphi)$ as a characteristic class $c_p[E_A] = p(x)$, where $x$ are the Chern classes of the lines in $E_A$. 
Thus we arrive to the expression proposed in the introduction, repeated here for convenience: 
\begin{equation}
M_{A,p} \to \sum_{B \prec A}  v_B^+ \int^{\varphi_a-\mathrm{equivariant}}_{\CM_A^B} c_p[E_A] e(\CD_m)\,,
\end{equation}
with an implicit average over the Weyl group.

When $A$ is a minuscule cocharacter, there is no monopole bubbling, and we can simply take $M_{A,p_n} = V_{A,n}$. Recall that a (dominant) minuscule cocharacter is the highest weight of a representation of the Langlands dual group $G^\vee$ that contains no other dominant weights. We will see that in some theories --- particularly those where $G$ is a product of $U(r)$ or $PSU(r)$ factors --- the minuscule monopole operators generate the entire chiral ring.

The ring relations between the monopole operator vevs should be compatible with a direct localization computation in the presence of multiple monopole singularities. 
In particular, we would expect to find a direct calculation to give an integral over the moduli space of Bogomolnyi solutions with two Dirac singularities:
\begin{equation}
M_{A,p} M_{A',p'} \to \sum_{B \prec A+A'}  v_B^+ \int^{\varphi_a-\mathrm{equivariant}}_{\CM_{A,A'}^B} c_p[E_A] c_{p'}[E_{A'}]e(\CD_m)\,.
\end{equation}
Such an expression is compatible with the expressions for the individual $M_{A,p}$ and $M_{A',p'}$ because the fixed points of the 
$\varphi_a$ action correspond to the collapse of smooth monopoles on either Dirac singularity and thus 
the integral over $\CM_{A,A'}^B$ can be recast as a sum over $B'$ of integrals over $\CM_A^{B-B'}$ and $\CM_{A'}^{B'}$.
The expressions will be fully consistent if the ratio $\frac{v_{B'}^+ v_{B-B'}^+}{v_B^+}$ 
accounts precisely for the contribution to one-loop determinants of the normal directions to $\CM_A^{B-B'} \times \CM_{A'}^{B'}$
in $\CM_{A,A'}^B$. It is straightforward to recognize the appropriate contributions in 
\be \frac{v_{B'}^+ v_{B-B'}^+}{v_B^+} = \frac{P_{\text{hypers}}^{B',B-B'}(\varphi,m)}{P_{\text{W}}^{B',B-B'}(\varphi)}\,.
\ee

\subsection{Higher-dimensional generalizations}

We comment here briefly on the extension of our abelianization formula to gauge theories in higher dimension, compactified on tori. 

The simplest example is the computation of the expectation value of 't Hooft-Wilson loop operators in four-dimensional $\CN=2$ 
gauge theories compactified on a circle of finite size $\beta$. We expect the localization formula to hold, up to the obvious replacement of 
rational characteristic classes with trigonometric ones, say 
\begin{equation}
M_{A,R} \to \sum_{B \prec A}  v_B \int^{\varphi_a-\mathrm{equivariant}}_{\CM_A^B} ch_R[E_A] ch(\CD_m)
\end{equation}
where we labelled the electric charge of the line defect by a representation $R$ of $G_A$. 
If we introduce a rotation twist $q$ in the circle compactification, we can make the answer equivariant with respect to rotations. 
This localization formula coincides with the results of  \cite{GOP-tHooft,IOT-Hitchin}.

In five-dimensional gauge theories, one can consider 't Hooft surfaces dressed by two-dimensional 
degrees of freedom coupled to $G_A$. This should result in localization formulae involving 
elliptic characteristic classes.

\subsection{Quantization}
\label{sec:quant-nonabel}

The chiral ring $\C[\CM_C]$ of a nonabelian theory can be quantized to a non-commutative algebra $\CA_C$, whose physical meaning was discussed in Section \ref{sec:quant-ab}.
We expect the algebra $\CA_C$ to be a subalgebra of a canonical quantization $\CA_C^{\rm abel}$ of the abelianized ring $\C[\CM_C^{\rm abel}]$, with coefficients of the abelianization map given by equivariant integrals. Namely, if we consider our gauge theory on $\R^2_\epsilon\times \R$, we should find the same integrals over moduli spaces of Bogomolnyi solutions, but made equivariant under rotations in space.

The quantization $\CA_C^{\rm abel}$ is straightforward to describe by generalizing Section \ref{sec:quant-ab}. It is generated by operators
\be \{\hat v_A^\pm\}_{A\,\in\,\text{cocharacters}}\,,\quad \{\hat \varphi_a\}_{a=1}^r\,, \label{quant-generators} \ee
and the inverses of the W-boson masses $\hat M^W_j = \langle \alpha_j, \hat \varphi\rangle$ for roots $\alpha_j$.
Motivated by preliminary localization results, we propose that the contribution of W-bosons to products of monopole operators should be the inverse of an adjoint hypermultiplet of complex mass $-\epsilon/2$. (Such a shift of the mass is also familiar in 4d localization, \cf\ \cite{OkudaPestun}.) Altogether, the relations among generators \eqref{quant-generators} take the form
\be \label{quant-gen}
\begin{array}{c}
\hat v_A^+ = \hat v_{-A}^-\,,\qquad [\hat \varphi_a, \hat v_A^\pm] = \pm\epsilon\langle a,A\rangle v_A^\pm\,,\qquad [\hat \varphi_a,\hat\varphi_b] = 0\,,\\[.4cm]
\displaystyle \hat v_A^+ \hat v^+_B = \frac{ P_{A,B;-}^{\rm hypers}(\hat\varphi,m)}{P^{\rm W}_{A,B;-}(\hat\varphi)}\, v^+_{A+B} \, \frac{ P_{A,B;+}^{\rm hypers}(\hat\varphi,m)}{P^{\rm W}_{A,B;+}(\hat\varphi)}\,,
\end{array}\ee
where
\begin{subequations}
\be P_{A,B;-}^{\rm hypers}(\hat\varphi,m)\;= \hspace{-.5cm}\prod_{\substack{\text{hypers $i$ s.t.}\\|\langle \mu_i,A\rangle|\,\leq \,|\langle \mu_i,B\rangle| \\  \langle \mu_i,A\rangle\langle \mu_i,B\rangle \,<\, 0}}
 \hspace{-.5cm} [\hat M_i]^{-\langle \mu_i,A\rangle}\,,\qquad
 P_{A,B;+}^{\rm hypers}(\hat\varphi,m)\;= \hspace{-.5cm}\prod_{\substack{\text{hypers $i$ s.t.}\\|\langle \mu_i,A\rangle|\,>\, |\langle \mu_i,B\rangle| \\  \langle \mu_i,A\rangle\langle \mu_i,B\rangle \,<\, 0}}
 \hspace{-.5cm} [\hat M_i]^{\langle \mu_i,B\rangle}\,,
\ee
\be P_{A,B;-}^{\rm W}(\hat\varphi) \;=  \hspace{-.7cm}\prod_{\substack{\text{roots $\alpha_j$ s.t.}\\|\langle \alpha_j,A\rangle|\,\leq \,|\langle \alpha_j,B\rangle| \\  \langle \alpha_j,A\rangle\langle \alpha_j,B\rangle \,<\, 0}}
 \hspace{-.5cm} [\hat M^W_j-\tfrac\epsilon2]^{-\langle \alpha_j,A\rangle}\,,\qquad
 P_{A,B;+}^{\rm W}(\hat\varphi)\;= \hspace{-.7cm}\prod_{\substack{\text{roots $\alpha_j$ s.t.}\\|\langle \alpha_j,A\rangle|\,>\, |\langle \alpha_j,B\rangle| \\  \langle \alpha_j,A\rangle\langle \alpha_j,B\rangle \,<\, 0}}
 \hspace{-.5cm} [\hat M^W_j-\tfrac\epsilon2]^{\langle \alpha_j,B\rangle}\,,\ee
\end{subequations}
and $\hat M_i = \langle\mu_i,\hat\varphi\rangle+\langle\mu^F_i,m\rangle$ are the effective complex masses of the hypermultiplets. The notation here is the same as in \eqref{abel-qproduct}. 

The quantization of the abelianization map that embeds $\CA_C\subset\CA_C^{\rm abel}$ is sometimes possible to describe without a direct localization calculation.
In the examples of Sections \ref{sec:SQCD} and \ref{sec:quiver}, we will find that $\C[\CM_C]\subset \C[\CM_C^{\rm abel}]$ is generated by monopole operators in minuscule representations. Moreover, in these examples, it will suffice to consider dressing factors that commute with the monopole operator itself. Such dressed operators will have an unambiguous quantization.

\subsection{Twistor space}
\label{sec:twist-gen}
In order to define the twistor space for the Coulomb branch of nonabelian gauge theories, we cannot invoke the simple non-renormalization theorem which 
lead us to the abelianization map. More precisely, we can invoke it for every fixed value of $\zeta$, but we need to do some extra work in order to find how to
glue the whole twistor space together. 

Recall that the full Coulomb branch $\CM_C$ contains $\CM_C^{\rm abel}$ (which is covered by abelian coordiantes $\varphi_a,v_A^\pm$) as an open subset. Correspondingly, we propose that the twistor space $Z$ for $\CM_C$ contains a twistor space $Z^{\rm abel}$ for $\CM_C^{\rm abel}$ as an open subset, where the inclusion map $Z^{\rm abel}\hookrightarrow Z$ is holomorphic and preserves the real structure $\tau$.
The abelianized twistor space $Z^{\rm abel}$ is easy to construct by generalizing the results of Section \ref{sec:twistor-ab}. Then the transition functions and the real structure on $Z$ will then be fully determined by those on $Z^{\rm abel}$.

Concretely, we construct the abelianized twistor space $Z^{\rm abel}$ as in \eqref{E-abel}. We take $\{v_A:=v_A^+\}_{A\in \mb A}$ be any finite set of abelian monopole operators that together with $\{\varphi_a\}_{a=1}^r$ generate $\C[\CM_C^{\rm abel}]$, and construct a bundle
\be E = \bigoplus_{A\in\mb A} \Big(\CO(\Delta(A))\otimes L^A\Big) \to \CO(2)^{\oplus r}  \ee
over the twistor sphere $\cp^1$.
The scalars $\varphi_a\to \varphi_{a,\zeta}$ are promoted to a section of $\CO(2)^{\oplus r}$, while each monopole operator $v_A \to v_{A,\zeta}$ is promoted to a section of $\CO(\Delta(A))\otimes L^A$, where
\be \Delta(A) := \sum_{\text{hypers $i$}} \langle \mu_i,A\rangle - \sum_{\text{W-bosons $j$}} \langle \alpha_j,A\rangle \ee
is the $U(1)_C$ charge of the monopole operator and the transition function for $L^A$ is $\exp \frac{1}{g^2\zeta}(A,\varphi_\zeta)$, with $(-,-)$ the Cartan-Killing form, scaled by the appropriate gauge coupling(s) $g^2$. Altogether, the $E$ is covered by two coordinate charts, with
\be \wt\varphi_{a,\wt \zeta} = \zeta^{-2}\varphi_{a,\zeta}\,,\qquad \wt v_{A,\wt \zeta} = \zeta^{-\Delta(A)}e^{\frac{1}{g^2}\frac{(A,\varphi_\zeta)}{\zeta}}v_{A,\zeta}\,.\ee
The real structure is as in \eqref{real-phi}--\eqref{real-v}. The twistor space $Z^{\rm abel}$ itself is the sub-bundle of $E$ cut out by the chiral-ring equations $v_{A,\zeta}v_{B,\zeta} = v_{A+B,\zeta} P^{\rm hypers}_{A,B}(\varphi_\zeta,m_\zeta)/P^{\rm W}_{A,B}(\varphi_\zeta)$.

Similarly, the full twistor space $Z$ is covered by two coordinate charts. The real sections of $Z$ are monopole operators $M_{A,p}^\zeta$ in complex structure $\zeta$. On $Z^{\rm abel}\subset Z$, the monopole operators $M_{A,p}^\zeta$ are related to $\varphi_{a,\zeta}, v_{A,\zeta}$ by the (trivial) $\zeta$-dependent generalization of the abelianization map. This fully determines their transition functions and their real structure. 
It is useful to note that the transition function for $v_{A,\zeta}$ involves the exponential of a canonical vector field, namely the vector field generated by the Hamiltonian $\frac{1}{2g^2}(\varphi_\zeta,\varphi_\zeta)$. Correspondingly, in the nonabelian setting, we must have
\be  \wt M_{A,p}^{\wt \zeta} = \zeta^{-\Delta(A,p)} e^{\big\{\frac{1}{2g^2}(\varphi_\zeta,\varphi_\zeta),-\big\}} M_{A,p}^\zeta\,, \label{trans-M} \ee
where $\Delta(A,p_n) = \Delta(A) + |n|$ is the $U(1)_C$ charge of $M_{A,p}^\zeta$.
In practice, the transformation \eqref{trans-M} will relate $\wt M_{A,p}^{\wt \zeta}$ to an infinite sum of monopole operators $M_{A,p'}^{\zeta}$ with other dressing factors.

It is natural to expect that the Coulomb branch of a gauge theory defined by a more general prepotential $\CF(\varphi)$ 
will be associated to a twistor space glued together by the exponential of the vector field $\{ \CF(\varphi), - \}$. 

\section{SQCD}
\label{sec:SQCD}

In this section, we analyze the Coulomb branch of the simplest nonabelian theories, in terms of the abelianization map of Section \ref{sec:NA}. We focus on SQED with gauge group $G=U(N_c)$ and $N_f$ `flavors' of hypermultiplets in the fundamental representation, \ie\ $\CR\simeq (T^*\C^{N_c})^{\oplus N_f} \simeq T^*\C^{N_cN_f}$.
In this case, we argue that the chiral ring $\C[\CM_C]$ is generated by dressed monopole operators of fundamental weight (cocharacter). We also consider some examples of $PSU(N_c)$ theories, and theories with an adjoint hypermultiplet.

The Coulomb branches of all these theories are smooth manifolds when generic mass parameters are turned on. Various descriptions of them are already known.
For $PSU(2)\simeq SO(3)$ theory with $N_f=0$, the Coulomb branch was identified in \cite{SW-3d} as the Atiyah-Hitchin hyperk\"ahler manifold. More generally, brane constructions \cite{HananyWitten} predict the Coulomb branch of  $U(N_c)$ SQCD with $N_f$ hypermultiplets to be equivalent to a moduli space of $N_c$ smooth $PSU(2)$ monopoles in the background of $N_f$ Dirac singularities. (The Coulomb branch of $PSU(N_c)$ theories is also such a monopole moduli space, with center-of-mass degrees of freedom factored out.) We apply classic scattering methods \cite{Donaldson-scattering, Hurtubise-scattering} to explicitly describe these monopole moduli spaces as complex manifolds, and to directly verify our abelianization construction for Coulomb branches.%
\footnote{Our description of Coulomb branches via scattering matrices was partially inspired by the recent works \cite{NP-quiver, NPS-quiver}, which considered Coulomb branches of 4d $\CN=2$ theories on a circle of finite radius.}

\subsection{SU(2) theory: the Atiyah-Hitchin manifold}
\label{sec:AH}

As a warmup, let's consider pure $PSU(2)\simeq SO(3)$ gauge theory. Both flavor groups $G_C,G_H$ are trivial. The abelianized ring $\C[\CM_C^{\rm abel}]$ is generated by the eigenvalue $\varphi$ of the adjoint complex scalar in the vectormultiplet, its inverse $\varphi^{-1}$ (since we remove the discriminant locus), and by abelian monopole operators $v^\pm$, all subject to the relation
\be v^+v^- = \frac1{-\varphi^2}\,, \label{AH-abel} \ee
where $\pm\varphi$ are the complex masses of the two W-bosons.
The $\Z_2$ Weyl symmetry acts as $(\varphi,v^+,v^-)\mapsto (-\varphi,v^-,v^+)$. The Poisson brackets are
\be \{\varphi,v^\pm\} = \pm v^\pm\,,\qquad \{v^+,v^-\} = -\frac{\partial}{\partial\varphi}\Big(\frac{1}{-\varphi^2}\Big) = -\frac{2}{\varphi^3}\,.\ee

The simplest Weyl-invariant functions of $\varphi,v^\pm$ are
\be \Phi := \varphi^2\,,\qquad  Y := v^+ + v^-\,,\qquad Z := \varphi(v^+-v^-)\,. \label{ops-AH}\ee
$Y$ and $Z$ are easily identified as the expectation value of the non-abelian monopole operator labelled by the fundamental cocharacter of $SO(3)$ (\ie\ the fundamental weight of the dual group $SU(2)$), and its dressed version. In the notation of Section \ref{sec:ops}, we would write $\Phi,Y,Z$ as $M_{A,n}=V_{A,n}$ with $(A,n)=(0,2)$, $(1,0)$ and $(1,1)$, respectively. 

The operators $\Phi,Y,Z$ satisfy the relation
\be Z^2 - \Phi Y^2 = 4\,. \label{AH} \ee
We claim that the chiral ring $\C[\CM_C]$ is precisely the subring of $\C[\CM_C^{\rm abel}]$ generated by $\Phi,Y,Z$, \ie\ the ring of functions on the \emph{smooth} hypersurface described by \eqref{AH}. Indeed, \eqref{AH} is precisely the complex equation for the Atiyah-Hitchin manifold \cite{AH}, known to be the Coulomb branch of pure $SO(3)$ theory \cite{SW-3d}. The embedding $\C[\CM_C]\subset \C[\CM_C^{\rm abel}]$ automatically identifies the Poisson structure on $\C[\CM_C]$ to be
\be \{\Phi,Y\} = 2Z\,,\qquad \{\Phi,Z\} = 2\Phi Y\,,\qquad \{Y,Z\} = -Y^2\,, \ee
showing that, the Poisson bracket is closed. Thus, we see that $\C[\CM_C]$ satisfies the first three conditions from page \pageref{Cproperties}. To satisfy the fourth condition, observe that all monopole operators $M_{A,n}=V_{A,n}+...$ of higher charge can be constructed as
\be M_{A,n} = \begin{cases} \Phi^{\frac n2}Y^A & \text{$n$ even} \\
 \Phi^{\frac{n-1}{2}}Y^{A-1}Z & \text{$n$ odd}\,. \end{cases}
\ee

We can generalize this construction to include matter. Since the gauge group is $PSU(2)$, the simplest possibility is a hypermultiplet in the adjoint representation, $\CR \simeq T^*\C^3$. Now $G_H = U(1)$. Giving the hypermultiplet a complex mass $m$, the abelian relations above are modified as
\be v^+v^- = \frac{m(\varphi+m)(-\varphi+m)}{-\varphi^2} = -\frac{m^3}{\varphi^2}+m\,,\qquad \{v^+,v^-\} = -\frac{2m^3}{\varphi^2}\,.\ee
The nonabelian operators that generate $\C[\CM_C]$ are still defined by \eqref{ops-AH}, but satisfy
\be Z^2-\Phi Y^2 = 4m^3-m\Phi\,, \ee
together with the Poisson brackets
\be \{\Phi,Y\} = 2Z\,,\qquad \{\Phi,Z\} = 2\Phi Y\,,\qquad \{Y,Z\} = -Y^2+4m\,.\ee

\subsection{Pure $U(N_c)$ gauge theory}
\label{sec:pureUN}

We next consider the general case of pure $U(N_c)$ SQCD, \ie\ $N_f=0$ and $\CR=\varnothing$. Obviously $G_H$ is trivial, but the topological symmetry is $G_C=U(1)_t$, corresponding to the abelian factor in $G=U(N_c)$. We can proceed as above, but it is much more informative to perform the analysis vis-\`a-vis the scattering method for moduli spaces of monopoles. 

We recall from \cite{ChalmersHanany, HananyWitten} that the Coulomb branch of pure $U(N_c)$ SQCD is expected to be a moduli space of $N_c$ smooth monopoles on $\R^3$, for the auxiliary gauge group $G'=PSU(2)$. 
(This is best understood via a brane construction in type IIB string theory, which we review in Section \ref{sec:monopoles}.) In turn, scattering methods \cite{Donaldson-scattering, Hurtubise-scattering} can be used to describe such monopole moduli spaces as complex manifolds. 

Monopoles for a compact group $G'$ are solutions to the Bogomolnyi equations $F_A=*d_A\phi$, where $A$ is a $G'$-connection on $\R^3$ and $\phi$ an adjoint valued scalar.%
\footnote{These fields, and the $\R^3$ in question, have no direct relation to the fields and spacetime of $U(N_c)$ SQCD.} %
The basic setup of the scattering approach is to choose a complex structure by splitting $\R^3$ as $\R_t \times\C$, and for each line parallel to the $t$-axis (at some fixed point $z\in \C$) to study solutions to the ordinary differential equation $\pd_t + A_t + i\phi = 0$. When $G'=PSU(2)$, the two-dimensional space of solutions has two distinguished lines $\ell_0^+,\ell_0^-$ that decay exponentially in the two asymptotic regimes $t\to\infty$ and $-\infty$. These lines can be completed to bases $(\ell_0^\pm,\ell_1^\pm)$, and the transformation between the bases is the scattering matrix
\be  \begin{pmatrix} \ell_0^+ \\ \ell_1^+\end{pmatrix} = S(z) \begin{pmatrix} \ell_0^- \\ \ell_1^-\end{pmatrix}\,.\ee
For each $z$, scattering matrix can be normalized to lie in $G_\C'=PSL(2,\C)$, and is well defined modulo multiplication by upper-triangular $G_\C'$ matrices on the right and lower-triangular $G_\C'$ matrices on the left (corresponding to changing the choice of bases). Moreover, given a solution to the Bogomolnyi equations and passing to a complexified holomorphic gauge $A_{\bar z}=0$, the scattering matrix depends \emph{holomorphically} on $z$. 

For a configuration of $N_c$ smooth monopoles and no Dirac singularities $(N_f=0)$, the boundary condition at infinity requires that $Q(z):= S_1^1(z) \sim z^{N_c}$ as $z\to\infty$. Then, by multiplying on the left and right by triangular $G_\C'$ matrices depending holomorphically on $z$, we can fix
\be S(z) = \begin{pmatrix} Q(z) & U^+(z) \\ U^-(z) & \wt Q(z) \end{pmatrix}\,, \label{scat-SQCD} \ee
such that
\be \det S(z) = Q(z)\wt Q(z) - U^+(z)U^-(z) = 1\,, \label{detS} \ee
with $Q(z)$ a monic polynomial of degree $N_c$ and $U^\pm(z)$ polynomials of finite degree $\leq N_c-1$. It follows that $\wt Q(z)$ is also a polynomial of degree $\leq N_c-2 $. A theorem of Donaldson \cite{Donaldson-scattering} states that the $N_c$-monopole moduli space is diffeomorphic to the moduli space of matrices \eqref{scat-SQCD}.

The scattering analysis therefore implies that the chiral ring of the Coulomb branch of pure $U(N_c)$ SQCD is generated by the coefficients of the polynomials $Q(z)$, $U^\pm(z)$ and $\wt Q(z)$, subject to the relations \eqref{detS}. In fact, the coefficients of $\wt Q(z)$ are not needed: due to the relation \eqref{detS}, they are automatically elements of the ring generated by the $3N_c$ coefficients of $Q(z)$ and $U^\pm(z)$. After  solving for these coefficients, \eqref{detS} still imposes $N_c$ independent ring relations, in agreement with the expected dimension of the Coulomb branch $\dim_\C\CM_C = 2N_c$.

It is a bit trickier to compute the Poisson bracket of these generators. As the monopole moduli space is defined by an infinite-dimensional 
hyperk\"ahler quotient, one can compute the Poisson bracket of two functions by lifting the functions to functionals on the infinite-dimensional linear space of 
gauge connections and scalar fields and applying the Poisson bracket on that linear space. 
The calculation is somewhat involved, and we only briefly sketch it in Appendix \ref{app:monopoles}. The result is that 
the scattering matrix $S(z)$ satisfies the Poisson bracket 
\be
\{ S^i_j(z), S^{i'}_{j'}(w) \} = \frac{S^i_{j'}(z) S^{i'}_j(w) - S^{i'}_{j}(z) S^{i}_{j'}(w)}{z-w}  \, ,
\label{s-poisson}
\ee
but only up to the ambiguity by multiplication by triangular matrices which we gauge-fixed in \eqref{scat-SQCD}. In other words, one should be careful to only use \eqref{s-poisson} on functionals of $S(z)$ which are strictly invariant under the triangular transformations. The Poisson bracket \eqref{s-poisson} is closely related to the Yangian algebra of $SL(2)$, a point upon which we elaborate in Section~\ref{sec:quiv-quant}.

In the neighbourhood of a point where the zeroes of $Q(z)$ are distinct, the zeroes $x_a$ themselves and 
the values $y^\pm_a$ of $U^\pm(z)$ at $z=x_a$ are strictly gauge invariant and give a good local coordinate system,
with $y^+_a y^-_a = -1$. The Poisson brackets of these coordinates following from \eqref{s-poisson} are then found to be 
\be
\{x_a,x_b\} = 0 \qquad \qquad \{x_a, y^\pm_b\} = \pm \delta_{ab} \, y^\pm_a \qquad \qquad \{y^+_a,y^+_b \} =0 \, .
\ee
One can re-write $Q(z)$, $U^\pm(z)$ in terms of the $x_a$, $y^\pm_a$, compute the Poisson brackets and extend them 
trivially away from the discriminant locus of $Q$. 

Now let's try to identify the coefficients in the scattering matrix with expectation values of gauge-invariant operators in $U(N_c)$ SQCD.
We propose, in close analogy to \cite{NP-quiver, NPS-quiver}, that $Q(z)$ is the characteristic polynomial of the $U(N_c)$ adjoint scalar $\varphi$, \ie\ the generating function of invariant polynomials
\be Q(z) = \big\langle\! \det(\varphi-z)\big\rangle  = \sum_{n=0}^{N_c} (-1)^nM_{0,(
\raisebox{.06cm}{$\underbrace{{}_{1,...,1}}_n$}
,0,..,0)} z^{{N_c}-n}\,. \label{Q-na} \ee
Similarly, we propose that $U^\pm(z)$ are the generating functions for dressed monopole operators labelled by the fundamental cocharacters $A = (\pm 1,0,...,0)$, which break the gauge group to $U(1)\times U(N_c-1)$:
\be
\begin{aligned}
  U^\pm(z) & = \sum\limits_{n=0}^{N_c-1}(-1)^n M_{(\pm 1,0,..,0),(0,
\raisebox{.06cm}{$\underbrace{{}_{1,...,1}}_n$}
,0,..,0)} z^{N_c-1-n}  \\
\qquad\text{or}\qquad
\frac{U^\pm(z)}{Q(z)} & = \sum_{n=0}^\infty M_{(\pm 1,0,..,0),(n,0,\ldots,0)} z^{-n-1}\,.
\label{U-na}
\end{aligned}
\ee

On the other hand, the abelianization map of Section \ref{sec:ops} predicts that $M_{A,n}=V_{A,n}$ for fundamental $A$, since this is a minuscule charge. Therefore, the generating functions $Q(z),\,U(z)$ should be expressed in terms of abelianized coordinates $\{\varphi_a,v^\pm_a\}_{a=1}^{N_c}$ as
\be Q(z) = \prod_{a=1}^{N_c}(z-\varphi_a)\,;\qquad  U^\pm(z) = \sum_{a=1}^{N_c} u_a^\pm\prod_{b\neq a}(z-\varphi_b)\,,\qquad \frac{U^\pm(z)}{Q(z)}= \sum_{a=1}^{N_c} \frac{u_a^\pm}{z-\varphi_a}\,. \label{QU-abel} \ee
(Here and in the remainder of the section we redefine abelian monopole operators by a sign
\be u_a^+ := v_a^+\,,\quad u_a^-:= (-1)^{N_c} v_a^- \label{sign-SQCD}\ee
in order to simplify many expressions.)
We find exact agreement with the monopole scattering construction, provided we identify $\varphi_a = x_a$ and $u^\pm_a = y_\pm^a/Q'(x_a)$.
Evaluating the determinant relation \eqref{detS} at $z=\varphi_a$ (simply $y^+_ay^-_a=-1$) beautifully reproduces the abelian chiral-ring relations
\be u_a^+u_a^- = \frac{-1}{\prod_{b\neq a}(\varphi_a-\varphi_b)^2}\,.\ee
Conversely, the abelian chiral-ring relations guarantee that, given polynomials $Q(z),U^\pm(z)$ parameterized as in \eqref{QU-abel}, there exists a $\wt Q(z)$ such that the determinant relation \eqref{detS} holds. 

Furthermore, with the identification $\varphi_a = x_a$ and $u^\pm_a = y^\pm_a/Q'(x_a)$, the Poisson brackets of the monopole scattering matrix \eqref{s-poisson} give
\be
\{\varphi_a,\varphi_b\} = 0 \qquad \qquad \{\varphi_a, u^\pm_b\} = \pm \delta_{ab} u^\pm_a \qquad \qquad \{u^\pm_a,u^\pm_b \} =  \pm\frac{2u^\pm_a u^\pm_b}{\varphi_a - \varphi_b}
\ee
which nicely agrees with Poisson structure derived from the abelianized theory. Note that the operator $\Tr\varphi = \sum_a\varphi_a$, which appears as the subleading coefficient of $Q(z)$, is the complex moment map for the $G_C=U(1)_t$ topological symmetry. The Poisson bracket $\{\Tr\varphi,-\}$ measures $U(1)_t$ charge.

We have found that our abelianization map is fully compatible with the scattering analysis. More interestingly, the scattering analysis has unambiguously identified the ring generators of $\C[\CM_C]$, a subring of the abelianized $\C[\CM_C^{\rm abel}]$: they are dressed monopole operators labelled by the trivial and fundamental cocharacters.

\subsubsection{Other operators}

Since the coefficients of $Q(z)$ and $U^\pm(z)$ generate the entire chiral ring, we must be able to use them to build monopole operators of non-fundamental charge.

For example, we can propose an interpretation of the coefficients of the auxiliary polynomial $\wt Q(z)$ as generating functions of monopoles with ``adjoint'' charge $(1,0,..,0,-1)$
dressed by characteristic polynomial of $\varphi$ restricted to the $U(N_c-2)$ block of unbroken gauge symmetry:
\be \wt Q(z)  = \sum_{n=0}^{N_c-2}(-1)^n M _{(1,0,..,0,-1),(0,
\raisebox{.06cm}{$\underbrace{{}_{1,...,1}}_n$}
,0,..,0)}   z^{N_c-2-n}
\; =  \sum_{a,b | a \neq b} u_a^+ u_b^- \prod_{c \neq a, c\neq b} (z-\varphi_c)  + \cdots
\ee
The ellipsis indicates terms with abelian charge $0$, which should in principle be computed from the bubbling monopole moduli space $\CM(1,0,..,0,-1)^0$. This moduli space 
has a singularity corresponding the collapse of a smooth monopole onto the Dirac singularity. There are interesting issues concerning how to properly resolve the singularity without 
losing the $U(N_c-2)$ bundle used in dressing the monopole. We refer to appendix \ref{app:UN} for a more in-depth analysis. 

It is useful to define the polynomials 
\begin{equation}
U^\pm_{(k)}(z) := \left( z^k U_\pm(z)\, \mathrm{mod} \,Q(z) \right) = \sum_a u_\pm^a \varphi_a^k \prod_{b \neq a} (z-\varphi_b)\,.
\end{equation}
which are generating functions for monopole operators of the form $M\raisebox{-.1cm}{$\scriptstyle (\pm 1,0,..,0),(k,\raisebox{.06cm}{$\overbrace{{}_{1,...,1}}^n$},0,..,0)$}$.
These can be used, for example, to generate monopole operators in other (non-fundamental) minuscule representations of $U(N_c)$, such as the first antisymmetric tensor power
\begin{align}
\frac{U_+^{(2)}(z) U_+(z) -U_+^{(1)}(z) U_+^{(1)}(z)}{Q(z)} 
&=  \sum_{n=0}^{N_c-2}(-1)^n M_{(1,1,0,..,0),(0,0,
\raisebox{.06cm}{$\underbrace{{}_{1,...,1}}_n$}
,0,..,0)}   z^{N_c-2-n} \\
&= \sum_{a,b | a \neq b} u_+^{a,b}  \prod_{c \neq a, c\neq b} (z-\varphi_c)\,,  \notag
\end{align}
where $u_+^{a,b} =  (\varphi_a - \varphi_b)^2 u_+^a u_+^b$ is the abelian monopole operator of charge $+1$ for $a$-th and $b$-th $U(1)$'s in the maximal torus.

\subsubsection{$U(2)$ and $PSU(2)$ theories}
\label{sec:U2PSU2}

To illustrate the scattering approach more concretely, consider SQCD with gauge group $G=U(2)$. The polynomials in the scattering matrix are
\be \begin{array}{rll} Q(z) &= z^2 - (\varphi_1+\varphi_2)z + \varphi_1\varphi_2 &=:\; z^2-\Phi_1 z + \Phi_2\,, \\[.2cm]
 U^\pm(z) &=  (u_1^\pm+u_2^\pm)z - u_1^\pm\varphi_2-u_2^\pm\varphi_1 &=:\; V^\pm z - W^\pm\,,
\end{array}\ee
and the auxiliary $\wt Q(z)$ is just a constant, containing the monopole operator with adjoint cocharacter,
\be \wt Q(z) = V^+V^- = (u_1^++u_2^+)(u_1^-+u_2^-) = u_1^+u_2^-+u_2^+u_1^- -2(\varphi_1-\varphi_2)^{-2}\,.\ee
The abelian relations $u_a^+u_a^-=-1/(\varphi_1-\varphi_2)^2$ ensure that $Q(z)\wt Q(z)-U^+(z)U^-(z) = 1$. In terms of non-abelian operators, these constraints take the form
\be \Phi_2 V^+ V^- - W^+ W^- - 1 = 0\,,\qquad \Phi_1 V^+ V^- - V^+W^--V^-W^+  = 0\,. \label{U2rel} \ee
Thus, $\C[\CM_C]\subset \C[\CM_C^{\rm abel}]$ is generated by $V^\pm,W^\pm,\Phi_1,\Phi_2$ subject to \eqref{U2rel}.

The abelian Poisson brackets
\be \{\varphi_a,u_b^\pm\} = \pm\delta_{ab}u_b^\pm\,,\quad \{u_1^+,u_1^-\}=-\{u_2^+,u_2^-\} = \frac{-2}{(\varphi_1-\varphi_2)^3}\,,\quad \{u_1^\pm,u_2^\pm\} = \frac{\pm 2 u_1^+u_2^+}{\varphi_1-\varphi_2} \notag \ee
imply
\be \label{PoissonU2}
 \hspace{-.3in}\begin{array}{c}
 \{\Phi_1,V^\pm\} = \pm V^\pm\,,\quad \{\Phi_1,W^\pm\} = \pm W^\pm\,,\\[.2cm] \{\Phi_2,V^\pm\} = \pm W^\pm\,,\quad \{\Phi_2,W^\pm\} = \pm(\Phi_1 W^\pm-\Phi_2 V^\pm) \\[.2cm]
 \{V^+,V^-\} = 0\,,\quad \{V^\pm,W^\pm\} = 0\,,\quad \{V^\mp,W^\pm\} = \pm V^+V^-\,,\quad \{W^+,W^-\}  = -\Phi_1V^+V^-\,.
\end{array}\ee
The operator $\Phi_1=\Tr\varphi$ is the moment map for the $G_C=U(1)_t$ action on $\CM_C$ (which is complexified to a $\C^*$ action on the chiral ring); and the Poisson bracket $\{\Phi_1,-\}$ measure $U(1)_t$ charge.

We can recover the Coulomb branch of $PSU(2)$ theory from Section \ref{sec:AH} in two different ways. On one hand, since $U(2) \approx PSU(2)\times U(1)$, we expect the $U(2)$ Coulomb branch to be roughly a product of the $PSU(2)$ Coulomb branch (the Atiyah-Hitchin manifold) and a pure $U(1)$ Coulomb branch (equivalent to $\R^3\times S^1\simeq \C\times\C^*$). Indeed, the $U(1)$ Coulomb branch $\R^3\times S^1$ simply parametrizes the center-of-mass degrees of freedom of the moduli space of two $SU(2)$ monopoles that we just analyzed with a scattering matrix; the full two-monopole moduli space is (a finite quotient of) the \emph{metric} product of the Atiyah-Hitchin manifold and $\R^3\times S^1$.

On the other hand, since $PSU(2) = U(2)/U(1)$, we also expect to see the Atiyah-Hitchin manifold as the  hyperk\"ahler quotient (albeit a rather trivial one) of a double cover of the Coulomb branch of $U(2)$ theory. In terms of the chiral ring, it is a holomorphic symplectic quotient. The operator $\Phi_1= \varphi_1+\varphi_2$, which is set to zero in $PSU(2)$ theory, plays the role of a moment map for the quotient by the topological $U(1)_t$ symmetry. Passing to a cover is necessary because there exist extra monopole operators labelled by embeddings $U(1)\hookrightarrow \left(PSU(2) \times U(1) \right)$ that do not lift to $U(2)$. We take a moment to explain how this works.

In the abelianized ring, we set $\varphi = \varphi_1-\varphi_2$ and introduce new variables
\be v^+ = \sqrt{u_1^+u_2^-}\,,\qquad v^- = \sqrt{u_1^-u_2^+}\,. \ee
(These operators are well defined on the double cover of the original abelianized moduli space.)
The triple $(\varphi,v^\pm)$ all Poisson-commute with $\Phi_1$, and satisfy the same ring and Poisson-algebra relations as in Section \ref{sec:AH}. They generate the abelianized chiral ring of the $PSU(2)$ theory. Conversely, the operators
\be u^+ = (\varphi_1-\varphi_2)\sqrt{u_1^+u_2^+}\,,\qquad u^- = (\varphi_1-\varphi_2)\sqrt{u_1^-u_2^-}\ee
and $\Phi_1$ all commute with $(\varphi,v^\pm)$. They obey $u^+u^-=1$ and generate the chiral ring of pure $U(1)$ theory. Altogether, we find that the ring $\C[\CM_C^{\rm abel}]$ for $U(2)$ theory, slightly enlarged by taking square roots, \emph{factors} as the product of the $PSU(2)$ ring generated by $(\varphi,v^\pm)$ and the $U(1)$ ring generated by $(\Phi_1,u^\pm)$. Taking a symplectic quotient with moment map $\Phi_1$ kills the $U(1)$ part.

On the full non-abelian Coulomb branch of $U(2)$ theory, we may define operators
\be Y = v^+ + v^- = \sqrt{V^+V^-}\,,\qquad Z = \varphi(v^+-v^-) = 2i\sqrt{W^+W^-}\,. \ee
These operators, together with $\Phi:=-4\Phi_2$, Poisson-commute with the moment map $\Phi_1$ and so descend to the symplectic quotient. As desired, the triple $(\Phi,Y,Z)$ generates the true chiral ring of the Atiyah-Hitchin manifold. In particular, \eqref{U2rel} implies that $Z^2-\Phi Y^2 = 4$.

\subsection{Adding matter}
\label{sec:SQCDmatter}

Adding fundamental matter to SQCD leads to Coulomb branches that are moduli spaces of singular monopoles, while adding adjoint matter leads to moduli spaces of periodic instantons.

\subsubsection{$U(N_c)$ SQCD}

$U(N_c)$ SQCD with $N_f$ fundamental hypermultiplets has a topological symmetry $G_C=U(1)_t$ as well as a Higgs-branch flavor symmetry $G_H = SU(N_f)$.
The Coulomb branch can be identified with the moduli space of $N_c$ smooth $PSU(2)$ monopoles $\R^3$ in the presence of $N_f$ Dirac singularities \cite{CherkisKapustin-mon}. The locations of the singularities in $\R^3$ coincide with the mass parameters of the hypermultiplets. Such moduli spaces of ``singular monopoles'' have been studied (\eg) in \cite{CherkisKapustin-mon, CherkisKapustin-grav, CherkisKapustin-mongrav, CherkisDurcan} and more recently in \cite{MRB-monopole, MRB-monopole2}.

As a complex manifold, the monopole moduli space is again described by a $G_\C' = PSL(2,\C)$ valued scattering matrix $S(z)$, defined modulo holomorphic upper triangular gauge transformations acting on the right, and holomorphic lower triangular on the left.
The entries of $S(z)$, however, may now be meromorphic, with poles as $z$ approaches the positions of Dirac singularities. 
Concretely, let
\be P(z) = \prod_{\alpha=1}^{N_f}(z-m_\alpha) \ee
be the characteristic polynomial for the flavor group $G_H = SU(N_f)$, containing the masses of the hypermultiplets. Then the top-left component of the scattering matrix, which is strictly invariant under the triangular gauge transformations, should have the form
\be
S^1_1(z) = \frac{Q(z)}{P(z)^{1/2}}
\label{fixS11}
\ee
where $Q(z)$ is a monic polynomial of degree $N_c$ \footnote{The square roots are allowed since the complexified gauge group is $PSL(2,\C) = SL(2,\C)/\mathbb{Z}_2$ rather than $SL(2,\C)$. To remove them, one can use the faithful adjoint representation of $PSL(2,\C)$, or view $PSL(2,\C) = PGL(2,\C)$ as $GL(2,\C)/\C^*$, and factor out an overall root.}. The boundary condition at infinity requires that $S^1_1(z)\sim z^{N_c-\frac{N_f}2}$ as $z\to \infty$ where as the boundary condition at each monopole singularity requires $S^1_1(z) \sim (z-m_\alpha)^{-1/2}$ as $z\to m_\alpha$. 

The boundary conditions for the scattering matrix can be specified in a somewhat more formal way as follows. First, the boundary conditions at the Dirac singularities specify that the scattering matrix should have the form
\be S(z) = g(z) \begin{pmatrix} P(z)^{\frac12} & 0 \\ 0 & P(z)^{-\frac12} \end{pmatrix} g'(z)\label{gPg} \ee
for some (non-unique) polynomial matrices $g,g'\in G_\C'[z]$. Here $g,g'$ can be interpreted as scattering matrices for smooth monopoles surrounding the Dirac singularities. Second, provided we have a good or balanced theory $(N_f\geq 2N_c)$, the boundary condition at infinity requires that the scattering matrix can be brought into the form
\be
S(z) \sim M(z) 
\begin{pmatrix}
z^{N_c-\frac{N_f}{2}} & 0 \\
0 & z^{-N_c+\frac{N_f}{2}}
\end{pmatrix}\,,
\label{bc-infty}
\ee
up to the usual triangular gauge transformations, where $M(z)$ is a meromorphic matrix satisfying $M(z)\to1$ as $z\to\infty$. In section \ref{sec:Grass}, we will explain how this presentation of the boundary conditions identifies an intersection of slices in the affine grassmannian of the monopole gauge group.

Viewing the monopole gauge group $G_\C' \simeq PGL(2,\C)$ as $GL(2,\C)/GL(1,\C)$, it is convenient to introduce a rescaled scattering matrix
\be \wt S(z) = P(z)^{\frac12}S(z)  \label{tS} \ee
that has \emph{polynomial} entries and determinant $\det \wt S(z) = P(z)$. As in the case of completely smooth monopoles, the gauge invariant component of the scattering matrix $\wt S^1_1(z):=Q(z)$ is a monic polynomial of degree $N_c$. Then, by upper and lower triangular gauge transformations, $\wt S(z)$ can be brought to the familiar form
\be \wt S(z) = \begin{pmatrix} Q(z) & U^+(z) \\ U^-(z) & \wt Q(z) \end{pmatrix}\,, \ee
where $U^\pm(z)$ are polynomials of degree at most $N_c-1$, and $\wt Q(z)$ has degree at most $N_c-2$ if $N_f \leq 2N_c-2$, and degree $N_f-N_c$ otherwise. The Coulomb branch chiral ring is then generated by the coefficients of $Q(z)$ and $U^\pm(z)$, subject to the determinant relation $Q(z) \wt Q(z) - U_+(z) U_-(z) = P(z)$.

As before, we can define coordinates $\varphi_a$ as the roots of $Q(z)$ and $u_a^\pm = U^\pm(\varphi_a) / Q'(\varphi_a)$, which are identified with the abelian coordinates of $\C[\mathcal{M}_C]^{\mathrm{abel}}$.  We still have
\be Q(z) = \prod_{a=1}^{N_c} (z-\varphi_a)\,,\qquad U^\pm(z) = \sum_{a=1}^{N_c} u_a^\pm \prod_{b\neq a}(z-\varphi_a)\,, \label{QUflavor} \ee
which are generating functions for gauge-invariant polynomials in the $\varphi_a$ and for monopole operators of fundamental cocharacter. Now $Q(z) \wt Q(z) - U_+(z) U_-(z) = P(z)$ becomes equivalent to the expected abelian relation
\be u_a^+u_a^- = -\frac{P(\varphi_a)}{\prod_{b\neq a}(\varphi_a-\varphi_b)^2}\,. \ee
The auxiliary polynomial $\wt Q(z)$ still has the form
\be \wt Q(z)  = \sum_{a,b\;\text{s.t.}\, a\neq b} u_a^+u_b^- \prod_{c\neq a,\,c\neq b}(z-\varphi_c)+\cdots\,,\ee
and is a generating function for monopole operators adjoint cocharacter. If $N_f \geq 2 N_c -1$ the ellipsis has degree higher than $N_c-2$, but the coefficients 
of degree higher than $N_c-2$ are simply polynomials in the masses and $\varphi_a$.

Various explicit examples of the chiral ring $\C[\CM_C]$ for SQCD with matter appear in Section \ref{sec:slices}, where the Coulomb branch is also interpreted as the intersection of certain nilpotent orbits and transverse slices in $\mathfrak{sl}(N_f,\C)$.

\subsubsection{$U(N_c)$ SQCD with an adjoint}

We can add to SQCD an adjoint hypermultiplet, to get a theory associated to an ADHM quiver. This theory is mirror to a necklace quiver of $N_f$ $U(N_c)$ gauge groups 
with a single flavor at one node. The Coulomb branch should admit a description as a moduli space of non-commutative instantons 
on $R^3 \times S^1$, with the adjoint mass $m$ playing the role of the non-commutativity parameter. We will not attempt to match 
these descriptions to the output of our abelianization map. It would be interesting to do so.  

At the level of the abelian variables, we have a relation of the form 
\begin{equation}
u_+^a u_-^a = - P(\varphi_a)\prod_{b \neq a}\left[ 1- \frac{m^2}{(\varphi_a - \varphi_b)^2} \right]
\end{equation}
If we define generating functions $Q(z)$ and $U_\pm(z)$ as before, we can write a polynomial equation such as 
\begin{equation}
Q(z) \wt Q(z) - U_+(z) U_-(z) = P(z) \frac{Q(z+m)Q(z-m)}{m^2} 
\end{equation}
Again, inspection of a few examples shows that $\wt Q(z)$ is a generating function of monopoles 
$M^{(1,-1,0,\cdots, 0), (0,0,1,\cdot,1,0,\cdots 0)}$, up to some polynomial expression in the $\varphi_a$. 

An important difference with SQCD with fundamental matter is that because 
\begin{equation}
(\varphi_a - \varphi_b)^2 u^a_+ u^b_+=  \left[(\varphi_a - \varphi_b)^2 - m^2 \right]u_+^{a,b}  
\end{equation}
we cannot generally obtain all dressed monopoles of charge $(1,1,0,\cdots,0)$ in a simple way from the $U^{(k)}_\pm(z)$. 
Thus we expect the ring of functions to have extra generators besides the coefficients of $Q(z)$ and $U_\pm(z)$. 

\subsection{Quantization}
\label{sec:quant-SQCD}

Applying the general prescription of Section \ref{sec:quant-nonabel} to SQCD with $N_f$ fundamental flavors produces an abelianized algebra $\CA_C^{\rm abel}$ with generators $\hat u_a^\pm$, $\hat\varphi_a$, and the inverses of W-boson masses $(\hat\varphi_a-\hat\varphi_b)^{-1}$. To simplify notation, we remove the `hats' from these operators.  They obey quantized relations
\be \label{qSQCD-1}
[\HAT\varphi_a,\HAT\varphi_b]=0\,,\qquad[\HAT\varphi_a,\HAT u_b^\pm] = \pm\epsilon \delta_{a,b}\HAT u_b^\pm\,,\qquad \HAT u_a^\pm\HAT u_a^\mp=\frac{-P(\HAT\varphi_a\mp\tfrac\epsilon2)}{\prod_{b\neq a}(\HAT\varphi_a-\HAT\varphi_b)(\HAT\varphi_a-\HAT\varphi_b\mp\epsilon)}\,,
\ee
where $P(z)=\prod_{\alpha=1}^{N_f}(z-m_\alpha)$ as usual. In addition, for $a\neq b$, the products $\HAT u_a^\pm\HAT u_b^\pm = \HAT u_{ab}^\pm/[(\HAT \varphi_b-\HAT\varphi_a)(\HAT\varphi_a-\HAT\varphi_b\mp\epsilon)]$ and $\HAT u_a^\pm \HAT u_b^\mp =\HAT u_{ab}^{\pm\mp}$ determine the commutators
\be  \label{qSQCD-2}  [\HAT u_a^\pm,\HAT u_b^\mp]=0\,,\qquad (\HAT\varphi_a-\HAT\varphi_b)[\HAT u_a^\pm,\HAT u_b^\pm] =
 \pm\epsilon[\HAT u_a^\pm,\HAT u_b^\pm]_+\,,  
\ee
where $[x,y]_+ := xy+yx$. If we choose only the $\HAT u_a^\pm$ as our generators (ignoring $\HAT u_{ab}^\pm$, etc.), there are also Serre-like relations of the form $[\HAT u_a^\pm,[\HAT u_b^+,\HAT u_c^+]]=0$.

We expect the nonabelian quantized algebra $\CA_C \subset \CA_C^{\rm abel}$ to be generated by quantized versions of the classical generators \ie\ quantized versions of the coefficients of $Q(z)$ and $U^\pm(z)$ from \eqref{QUflavor}, namely
\be \HAT Q(z) = \prod_{a=1}^{N_c} (z-\HAT\varphi_a)\,,\qquad \HAT U^\pm(z) = \sum_{a=1}^{N_c} \HAT u_a^\pm \prod_{b\neq a}(z-\HAT\varphi_b)\,, \ee
Noting that the dressing factors $\prod_{b\neq a}(z-\varphi_a)$ commute with the abelian operators $u_a$, so there are no ordering ambiguities. By virtue of \eqref{qSQCD-1} and \eqref{qSQCD-2}, these polynomials obey
\begin{subequations} \label{quantumQQ}
\be
 \begin{array}{c} \HAT Q(z+\tfrac\epsilon2) \HAT{\wt Q}(z-\tfrac\epsilon2) - \HAT U^+(z+\tfrac\epsilon2)\HAT U^-(z-\tfrac\epsilon2) = \HAT P(z-\tfrac\epsilon 2)\,, \\[.2cm]
\HAT Q(z-\tfrac\epsilon2) \HAT{\wt Q}(z+\tfrac\epsilon2) - \HAT U^-(z-\tfrac\epsilon2)\HAT U^+(z+\tfrac\epsilon2) = \HAT P(z+\tfrac\epsilon 2)\,, \end{array}
\ee
for a suitable auxiliary polynomial $\HAT {\wt Q}(z)$. Similarly,
\be \begin{array}{c}  \HAT{\wt Q}(z-\tfrac\epsilon2) \HAT Q(z+\tfrac\epsilon2) - \HAT U^+(z-\tfrac\epsilon2)\HAT U^-(z+\tfrac\epsilon2) = \HAT P(z-\tfrac\epsilon 2)\,, \\[.2cm]
 \HAT{\wt Q}(z+\tfrac\epsilon2) \HAT Q(z-\tfrac\epsilon2) - \HAT U^-(z+\tfrac\epsilon2)\HAT U^+(z-\tfrac\epsilon2) = \HAT P(z+\tfrac\epsilon 2)\,. \end{array}
\ee
\end{subequations}
The auxiliary polynomial contains dressed monopoles in the adjoint representation, with all necessary quantum corrections included
\be \HAT{\wt Q}(z+\tfrac\epsilon2) = \sum_{a\neq b}\HAT u_+^a\HAT u_-^b \prod_{c\neq a,\,c\neq b}(z-\HAT\varphi_c)+ \cdots\,. \ee

Other polynomial relations are deformed in a similar nice manner. 
For example, when $N_f=0$, if we define
\begin{align}
\HAT U_+^{(k)}(z) &= \sum_a \HAT \varphi_a^k \HAT v^+_a \prod_{b \neq a} (z - \HAT\varphi_b)
\end{align}
then 
\begin{align}
\HAT U_+^{(2)}(z+ \tfrac\epsilon2) \HAT U_+(z- \tfrac\epsilon2)-& \HAT U_+^{(1)}(z+ \tfrac\epsilon2) \HAT U_+^{(1)}(z- \tfrac\epsilon2)  + \epsilon \,\HAT U_+^{(1)}(z+ \tfrac\epsilon2) \HAT U_+(z- \tfrac\epsilon2)   \cr &
 =\HAT Q(z + \tfrac\epsilon2) \sum_{a\neq b} \HAT \varphi_a(\HAT \varphi_a- \HAT \varphi_b -\epsilon) \HAT u_a^+ \HAT u_b^+ \prod_{c \neq a, c\neq b} (z - \HAT \varphi_c- \tfrac\epsilon2)  \cr
 &  =\HAT Q(z + \tfrac\epsilon2) \sum_{a\neq b} \frac{\HAT \varphi_a}{\HAT \varphi_a- \HAT \varphi_b} \HAT u_{a,b}^+  \prod_{c \neq a, c\neq b} (z - \HAT \varphi_c- \tfrac\epsilon2)   \cr
  &  =\HAT Q(z + \tfrac\epsilon2) \sum_{a\neq b}\HAT u_{a,b}^+  \prod_{c \neq a, c\neq b} (z - \HAT \varphi_c- \tfrac\epsilon2)
\end{align}
gives us the quantized generating function of monopoles in the second minuscule representation.

\subsubsection{Quantizing Atiyah-Hitchin and $U(2)$ theory}

As a quick example, we quantize the Coulomb branch of the pure $PSU(2)$ and $U(2)$ theories. For $PSU(2)$, the ring  $\C[\CM_C^{\rm abel}]\subset \C[\CM_C]$ was described in Section \ref{sec:AH}. The quantization $\CA_C^{\rm abel}$ is generated by operators $\HAT v^\pm$ and $\HAT \varphi^{\pm 1}$, subject to
\be [\HAT\varphi,\HAT v^\pm] = \pm\epsilon \HAT v^\pm\,,\qquad \HAT v^+\HAT v^- = \frac{-1}{\HAT\varphi(\HAT\varphi-\epsilon)}\,, \ee
The nonabelian subalgebra $\CA_C\subset \CA_C^{\rm abel}$ is generated by $\HAT \Phi = \HAT\varphi^2$, $\HAT Y = \HAT v^++\HAT v^-$, and $\HAT Z = \HAT \varphi(\HAT v^+-\HAT v^-)$, which obey
\be [\HAT\Phi,\HAT Y] = 2\epsilon \HAT Z-\epsilon^2\HAT Y\,,\qquad [\HAT \Phi,\HAT Z]=2\epsilon\HAT\Phi\HAT Y-\epsilon^2\HAT Z\,,\qquad [\HAT Y,\HAT Z]=-\epsilon \HAT Y^2\,,\ee
together with the constraint
\be \HAT Z^2+\epsilon \HAT Z\HAT Y -\Phi\HAT Y^2 = 4\,, \ee
which quantizes the Atiyah-Hitchin manifold.

For pure $U(2)$ theory, the abelian algebra $\CA_C^{\rm abel}$ is generated by $\HAT\varphi_1,\HAT\varphi_2,\HAT u_1^\pm,\HAT u_2^\pm$, subject to the general relations \eqref{qSQCD-1}. In particular,
\be \HAT u_1^+\HAT u_1^- = \HAT u_2^-\HAT u_2^+ = \frac{-1}{(\HAT \varphi_1-\HAT \varphi_2)(\HAT \varphi_1-\HAT \varphi_2-\epsilon)}\,,\qquad  \HAT u_1^-\HAT u_1^+ = \HAT u_2^+\HAT u_2^- = \frac{-1}{(\HAT \varphi_1-\HAT \varphi_2)(\HAT \varphi_1-\HAT \varphi_2+\epsilon)}\,, \notag \ee
and $(\HAT\varphi_1-\HAT\varphi_2)[\HAT v_1^+,\HAT v_1^-]=\epsilon[\HAT v_1^+,\HAT v_1^-]_+$, $(\HAT\varphi_1-\HAT\varphi_2)[\HAT v_1^\pm,\HAT v_2^\pm]=\pm\epsilon[\HAT v_1^\pm,\HAT v_2^\pm]_+$. The nonabelian subalgebra $\CA_C$ is generated by coefficients of the scattering matrix
\be \HAT S(z) = \begin{pmatrix} \HAT Q(z) & \HAT U^+(z) \\ \HAT U^-(z) & \HAT{\wt Q}(z) \end{pmatrix}
 = \begin{pmatrix} (z-\HAT\varphi_1)(z-\HAT\varphi_2) & \HAT u_1^+(z-\HAT\varphi_2)+\HAT u_2^+(z-\HAT\varphi_1) \\
\HAT u_1^-(z-\HAT\varphi_2)+\HAT u_2^-(z-\HAT\varphi_1) & (\HAT u_1^++\HAT u_2^+)(\HAT u_1^++\HAT u_2^-)\end{pmatrix}\,,\ee
which obey relations \eqref{quantumQQ} with $P=1$. In particular, we can take as generators $\HAT\Phi_1=\HAT\varphi_1+\HAT\varphi_2$, $\HAT\Phi_2=\HAT\varphi_1\HAT\varphi_2$, $\HAT V^\pm = \HAT u_1^\pm+\HAT u_2^\pm$, and $\HAT W^\pm = \HAT \varphi_2\HAT u_1^\pm + \HAT\varphi_1\HAT u_2^\pm$, subject to
\be \HAT W^+(\HAT W^-+\epsilon\HAT V^-)= \HAT\Phi_2\HAT V^+\HAT V^--1\,,\qquad \HAT V^+\HAT W^-+\HAT W^+\HAT V^- = (\HAT\Phi_1-\epsilon)\HAT V^+\HAT V^-\,.\ee
The commutation relations are the straightforward generalization of \eqref{PoissonU2}, with the RHS rescaled by $\epsilon$.

\subsection{Twistor space}
\label{sec:twistor-SQCD}

We can specialize our general prescription for the twistor space to, say, $U(N_c)$ SQCD with $N_f$ fundamental hypermultiplets.
First, the coefficients of $z^k$ in the polynomial $Q_\zeta(z)$ will have to live in $O(2N_c-2k)$
bundles over the twistor sphere. If we replace $z$ with a variable $z_\zeta$ living in 
an $O(2)$ bundle, we can simply state that by requiring $Q_\zeta(z_\zeta)$ 
to live in $O(2N_c)$. 

As for $U^\zeta_\pm(z_\zeta)$, we can use the following gluing condition:
\begin{equation} \label{Uzeta}
\tilde z_{\wt \zeta}  =  \zeta^{-2} z_\zeta\,, \qquad \qquad \wt U_{\wt \zeta}^\pm(\tilde z_{\wt \zeta}) =  \zeta^{-N_f} e^{\pm \frac{1}{g^2} \frac{z_\zeta}{\zeta}}U_{\zeta}^\pm(z_{\zeta}) \quad \mathrm{mod} \quad Q_\zeta(z_\zeta)\,. \end{equation}
The notation means that we should expand $e^{\pm \frac{1}{g^2} \frac{z_\zeta}{\zeta}}U_{\zeta}^\pm(z_{\zeta})$ as a Taylor series in $z$, and reduce it to back to a polynomial in $z$ modulo $Q(z)$. The coefficients of this polynomial will be infinite series in the coefficients of the original $U_\zeta^\pm$ and $Q_\zeta$.

This prescription follows from the general abelianization proposal of Section \ref{sec:twist-gen}. To see the relation, write $U^\pm(z) = \sum_a u_a^\pm \prod_{b\neq a}(z-\varphi_b)$, and observe that working modulo $Q(z) = \prod_a(z-\varphi_a)$ we have
\begin{align}  e^{\frac{1}{g^2}\frac{z}{\zeta}} U(z)  \overset{\text{mod $Q$}}{=} & \sum_a e^{\frac{1}{g^2}\frac{\varphi_a}{\zeta}}u_a^\pm \prod_{b\neq a}(z-\varphi_b)
 \notag \\ 
 = \;\;\,&\sum_a e^{\frac{1}{2g^2\zeta}\{(\varphi,\varphi),-\}}u_a^\pm \prod_{b\neq a}(z-\varphi_b) \notag \\
  = \;\;\, & e^{\frac{1}{2g^2\zeta}\{(\varphi,\varphi),-\}} U(z)\,.
\end{align}
In the form \eqref{Uzeta}, the gluing condition also agrees with the standard construction of the twistor space of $N_c$ smooth $PSU(2)$ monopoles (when $N_f=0$) \cite{AH}.

The transformation of the remaining component $\wt Q$ of the scattering matrix follows from the transformation of $Q$ and $U_\pm$. 
In particular,  $\wt Q$ is multiplied by $\zeta^{2 N_c - 2N_f}$ and shifted by an intricate polynomial function of $U_\pm$ and $Q$.
 
\subsubsection{Example: $U(2)$ theory}

Consider pure $U(2)$ theory, whose chiral ring was described in Section \ref{sec:U2PSU2}. The transition functions for the operators $\Phi_{1,\zeta},\Phi_{2,\zeta},V_\zeta^\pm,W_\zeta^\pm$ are
\be
\begin{array}{c} \wt z_{\wt \zeta} = \zeta^{-2}z_\zeta\,,\qquad
 \wt \Phi_{1,\wt\zeta} = \zeta^{-2}\Phi_{1,\zeta}\,,\qquad \wt \Phi_{2,\wt\zeta} = \zeta^{-4}\Phi_{2,\zeta}\,, \\[.2cm]
 \wt V_{\wt \zeta}^\pm \wt z_{\wt \zeta} - \wt W_{\wt \zeta}^\pm = e^{\frac 1{g^2}\frac{z_\zeta}{\zeta}}
 (V_\zeta^\pm z_\zeta - W_\zeta^\pm)\quad {\rm mod}\;\; z_\zeta^2 - \Phi_{1,\zeta}z + \Phi_{2,\zeta}\,.
\end{array}
\ee
Explicitly (omitting $\pm$ superscripts),
\begin{align} \wt V_{\wt \zeta}^\pm &= e^{\frac{1}{g^2}\frac{\varphi_{1,\zeta}}{\zeta}}u_{1,\zeta} + e^{\frac{1}{g^2}\frac{\varphi_{2,\zeta}}{\zeta}}u_{2,\zeta} = \sum_{n=0}^\infty \frac{1}{g^{2n}\zeta^nn!}(\varphi_{1,\zeta}^nu_{1,\zeta} + \varphi_{2,\zeta}^n u_{2,\zeta}) \notag \\
&= V_\zeta + \frac{1}{g^2\zeta} (\Phi_{1,\zeta} V_\zeta-W_\zeta) + \frac1{2g^4\zeta^2}(\Phi_{1,\zeta}^2V_\zeta-\Phi_{2,\zeta}V_\zeta-\Phi_{1,\zeta}W_\zeta) + \ldots,
\label{Vexp}
\end{align} 
and similarly
\be \wt W_{\wt \zeta}^\pm = \sum_{n=0}^\infty \frac{1}{g^{2n}\zeta^nn!}(\varphi_{1,\zeta}^n\varphi_{2,\zeta}u_{1,\zeta} + \varphi_{2,\zeta}^n \varphi_{1,\zeta} u_{2,\zeta}) =  W_\zeta + \frac{1}{g^2\zeta}\Phi_{2,\zeta}V_\zeta + \ldots.
\label{Wexp}\ee
The operators appearing in the expansions in equations \eqref{Vexp} and \eqref{Wexp} are monopole operators in complex structure $\zeta$ of dressed by polynomials of the scalar $\varphi_\zeta$ labelled by $(n,0)$ and $(n,1)$ respectively.

\section{Linear quivers of unitary groups}
\label{sec:quiver}

\begin{figure}[htb]
\centering
\includegraphics[width=2.2in]{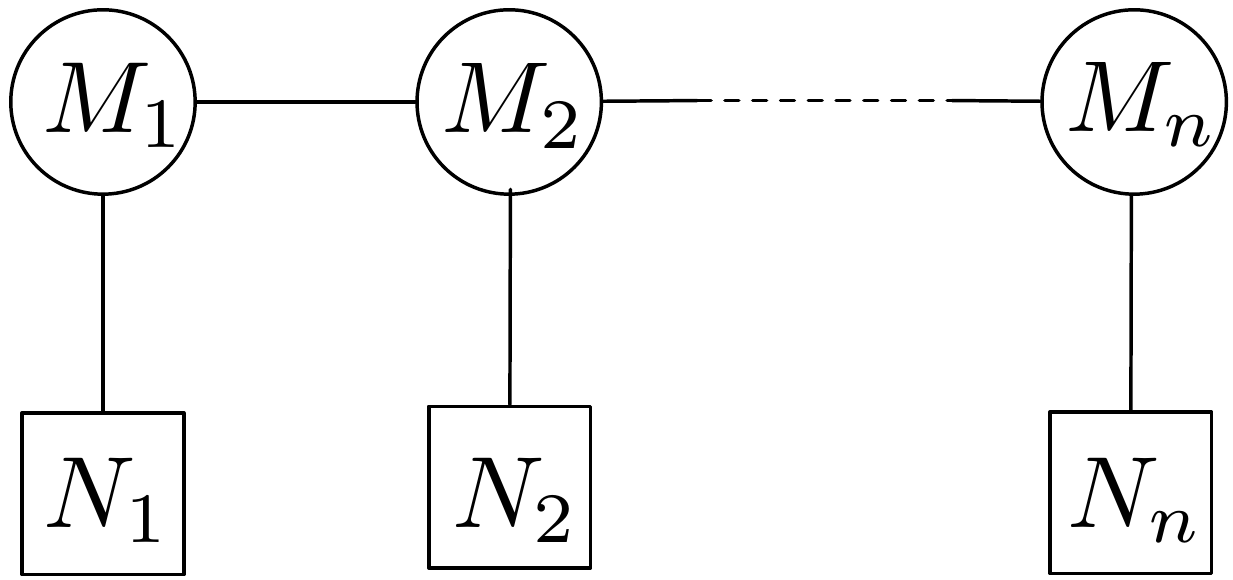}
\caption{The quiver gauge theory $T^\nu_\mu$.}
\label{fig:quiver}
\end{figure}

We now study 3d $\CN=4$ linear quivers with unitary gauge groups. The general $A_n$-type quiver, shown in Figure \ref{fig:quiver}, defines a theory whose gauge group is $G=\prod_{i=1}^n U(M_i)$, with hypermultiplets transforming in a representation $\CR$ that is a sum of $N_i$ copies of the fundamental representation of each $U(M_i)$ plus a sum of bifundamental representations for all adjacent nodes,
\be \label{q-rep}
\CR = R\oplus R^* 
\,,\qquad R := \bigoplus_{i=1}^n \text{Hom}(\C^{M_i},\C^{N_i})\;\oplus\; \bigoplus_{i=1}^{n-1} \text{Hom}(\C^{M_i},\C^{M_{i+1}})\,.
\ee

It is often convenient to regroup the data $(N_i,M_i)$ into a pair of weights $(\nu,\mu)$ of $\mathfrak{sl}(n+1)$. Let $\{\omega_i\}_{i=1}^n$ denote the fundamental weights of $\mathfrak{sl}_{n+1}$, such that $\omega_i$ is the highest weight of the $i$-th antisymmetric power of the fundamental representation; and let $\{\alpha_i\}_{i=1}^n$ denote the simple roots. Then $\nu = \sum_i N_i \, \omega_i$ is a dominant weight and $\mu = \nu - \sum_i M_i \, \alpha_i$ labels a weight space in the irreducible representation of highest weight $\nu$. Inverting these relations, we can recover the original data from the formulae $N_i = (\nu , \alpha_i)$ and $M_i = (\nu-\mu,\omega_i)$, where $(\;,\;)$ is the Cartan-Killing form.%
\footnote{The Cartan-Killing form is normalized so that $(\omega_i , \alpha_j) = \delta_{ij}$ and $( \alpha_i , \alpha_j ) = \kappa_{ij}$ where $\kappa_{ij}$ is the Cartan matrix.}. %
We henceforth denote the quiver gauge theory as $T^\nu_\mu$.

We can of course also read the quiver from right to left rather than left to right. This has the effect of replacing the weights by their conjugates: $(\mu,\nu)\to(-w_0\mu,-w_0\nu)$, where $w_0$ is the longest element of the Weyl group. The theories $T^\nu_\mu$ and $T^{-w_0\nu}_{-w_0\mu}$ and their moduli spaces are equivalent.

The infrared physics of $T^\nu_\mu$ depends on the R-charges $\Delta_i$ (under a subgroup $U(1)_C\subset SU(2)_C$)
of undressed monopole operators labelled by the fundamental cocharacter of each $U(M_i)$ factor of the gauge group. These are \cite{GW-Sduality}
\be \Delta_i = N_i - 2M_i + M_{i-1}+ M_{i+1} = (\mu,\alpha_i)\,.\ee
The theory is called good if $\Delta_i\geq 0$ for all nodes $i$, or equivalently if $\mu$ is also a dominant integral weight. In this case, $T^\nu_{\mu}$ is expected to flow to a CFT in the infrared, whose operators all have non-negative dimension. Mirror symmetry applies naturally to good A-type quivers, mapping them to dual A-type quivers of length $\tilde n+1 = \sum_i N_i$.%
\footnote{In mathematical works \cite{Maffei-A, MirkovicVybornov} it was argued that the \emph{Higgs} branches of certain linear quivers are isomorphic to transverse slices and to monopole moduli spaces (in the guise of affine Grassmannians). Together with mirror symmetry, this should imply the equivalence of the Coulomb-branch descriptions mentioned below.} %
The $i$-th node of a quiver is called balanced if $\Delta_i=0$ and the quiver is called fully balanced if all nodes are balanced, which is equivalent to $\mu=0$. Balanced nodes signal a low-energy nonabelian enhancement of Coulomb-branch isometry group $G_C$.

The Higgs branch $\CM_H$ of $T^\nu_\mu$ is the hyperk\"ahler quotient $\CR/\!/\!/G$, which is known as a Nakajima quiver variety \cite{Nakajima-quiver}. The Higgs branch chiral ring $\C[\CM_H]$ can be computed by enumerating all gauge invariant combinations of the hypermultiplet scalars and imposing the complex moment map relations.

The Coulomb branch $\CM_C$ of $T^\nu_\mu$ has several descriptions. Some descriptions are valid for general quivers, while others apply only to good quivers, in the IR limit. 
It follows from a type IIB brane construction and S-duality that $\CM_C$ is the moduli space of BPS monopoles for the gauge group $PSU(n+1)$ in the presence of Dirac monopole singularities \cite{ChalmersHanany, HananyWitten}.  
(As we show in Section \ref{sec:Grass}, in the case of good quivers, this monopole moduli space also appears as a slice in the affine Grassmannian for $PSL(n+1,\C)$.)

Via a Nahm transform, the same construction also offers a description of $\CM_C$ as a moduli space of solutions to Nahm equations on a chain of segments, with appropriate boundary conditions. If the quiver is good, the description can be simplified considerably in the IR, to the moduli space of solutions of 
$SL(p)$ Nahm equations on a single segment, where $p = \sum_{i=1}^n i N_i$ \cite{GW-boundary, GW-Sduality}. 
This moduli space is the intersection of a nilpotent orbit and a transverse slice inside the Lie algebra $\mathfrak{sl}(p,\C)$. 
The equivalence of these descriptions is quite nontrivial, and only expected to hold for A-type (\ie\ linear) quivers.%
\footnote{More generally, moduli spaces of ADE monopoles occur as Coulomb branches of unitary quivers in the shape of ADE Dynkin diagrams \cite{Tong-ADE}. In contrast, the intersections of transverse slices and nilpotent orbits in simple Lie algebras of non-A type do not all have a construction as moduli spaces of (Lagrangian) 3d gauge theories. In type D, they are realized using quivers with $SO/USp$ gauge groups \cite{GW-Sduality}.}

Our goal in this section is to give a direct construction of the chiral ring $\C[\CM_C]$, using the methods of Section \ref{sec:NA}. We will propose in Section \ref{sec:gen-quiv} that $\C[\CM_C]$ is generated as a Poisson algebra by dressed monopole operators labelled by the fundamental cocharacter at each node; and that $\C[M_C]$ is generated as a ring by dressed monopole operators labelled by a sum of fundamental cocharacters at subsets of multiple nodes.
We then proceed to check this proposal in a variety of examples, using the known descriptions of $\CM_C$ mentioned above.  Finally, we will discuss the Poisson structure on $\C[\CM_C]$ and its quantization, which for a general good quiver produces a central quotient of a shifted Yangian, and in special cases a finite W-algebra. This lets us connect to the mathematical and mathematical-physics works \cite{GanGinzburg, GKLO-Yangian, KWWY}.

\subsection{Flavor symmetries}

The Higgs-branch flavor group of $T^\nu_\mu$ is $G_H=PS\big[\prod_{i=1}^n U(N_i)\big]=\big(\prod_{i=1}^n U(N_i)\big)/U(1)$ with $U(N_i)$ acting on the $N_i$ fundamental hypermultiplets at the $i$-th node. The overall quotient by $U(1)$ is due to mixing with the dynamical gauge symmetry. Correspondingly, the theory admits triples of mass parameters valued in the Cartan $\mathfrak t_H$, split into real and complex parts. We label the complex masses at the $i$-th flavor node as $m_{i,\alpha}$ for $1\leq \alpha \leq N_i$. These are defined up to an overall shift.

The Coulomb-branch flavor group $G_C$ of the theory $T^\nu_\mu$ has rank $n$, and its maximal torus $\mathbb T_C\simeq U(1)^n$ contains the topological symmetries for each gauge node.
The twisted masses for these topological symmetries are $n$ triples of FI parameters. In the infrared of a good theory $T^\nu_\mu$, the symmetry group is enhanced to the Levi subgroup of $SU(n+1)$ that stabilizes $\mu\in \mathfrak{su}(n+1)^*$.
Explicitly, if the \emph{un}balanced nodes separate the $n$ nodes of the quiver into partitions of size $k_j\geq 0$, the symmetry group is $S\big[\prod_j U(k_j+1)\big]$. For example, a quiver with $\vec \Delta=(*,0,0,*,0,*)$, where $*$ denotes unbalanced nodes, has $\vec k = (0,2,1,0)$, and the enhanced symmetry is $S[U(1)\times U(3)\times U(2)\times U(1)]\subset SU(7)$.
Since the chiral ring $\C[\CM_C]$ is independent of gauge couplings, it has an action of the enhanced, infrared symmetry group --- in fact, of a complexification thereof. Below, we will simply use `$G_C$' to denote the enhanced group.

\subsection{Abelian coordinates}
\label{sec:abel-quiv}

Following Section \ref{sec:NA}, we want to embed the full Coulomb branch chiral ring $\C[\CM_C]$ into a larger chiral ring $\C[\CM_C^{\rm abel}]$ of an abelianized theory. The abelian chiral ring is generated by eigenvalues $\varphi_{i,a}$ of the adjoint scalars at each $i$-th gauge node, with $1\leq a\leq M_i$; by the inverses of W-boson masses $(\varphi_{i,a}-\varphi_{i,b})^{-1}$, $a\neq b$; and by monopole operators $u^\pm_A$ labelled by cocharacters $A$ of $G = \prod_{i=1}^n U(M_i)$. The generators satisfy relations of the form \eqref{rel-nonab}. 

In order to generate $\C[\CM_C^{\rm abel}]$ as a \emph{Poisson} algebra, it is sufficient to take a much smaller set of generators, consisting of the $\varphi_{i,a}$ together with abelian monopole operators $u_{i,a}^\pm$ labelled by the fundamental cocharacters $1\leq a\leq M_i$ of each $U(M_i)$ gauge group. These abelian monopole operators satisfy%
\footnote{In the convention of Section \ref{sec:abelianize}, the RHS of \eqref{quiverrelations} would include an extra sign $(-1)^{M_{i-1}+M_i}$. We absorb this sign into the $u_{i,a}^-$ operators (\cf\ \eqref{sign-SQCD}), obtaining the more uniform expression above.}
\be
u^+_{i,a}u^-_{i,a} = \frac{P^{i,a}_{\rm hypers}(\varphi,m)}{P^i_{\rm W}(\varphi)} =  - \frac{P_i(\varphi_{i,a};m) Q_{i-1}(\varphi_{i,a};\varphi) Q_{i+1}(\varphi_{i,a};\varphi)}{\prod_{b \neq a}(\varphi_{i,a}-\varphi_{i,b})^2} \,,
\label{quiverrelations}
\ee
where $P_i(z;m) := \prod_{\alpha=1}^{N_i}(z-m_{i,\alpha})$ and $Q_i(z;\varphi) := \prod_{a=1}^{M_i}(z-\varphi_{i,a})$ are the matter and gauge polynomials at the $i$-th node. (We commonly denote these $P_i(z)$ and $Q_i(z)$.)
Among these generators, the only nonzero Poisson brackets are
\be
\begin{aligned}
\{  \varphi_{i,a} , u^\pm_{i,a} \} & = \pm  u_{i,a}^{\pm} \\
\{ u^+_{i,a} , u^-_{i,a} \} & =  - \frac{\partial}{\partial\varphi_{i,a}} \bigg[\frac{P^{i,a}_{\rm hypers}(\varphi,m)}{P^i_{\rm W}(\varphi)}\bigg] \\
\{ u^\pm_{i,a} , u^\pm_{j,b} \} & = \pm \kappa_{ij} \frac{u^\pm_{i,a} u^\pm_{j,b} }{\varphi_{i,a} - \varphi_{j,b} }\,.
\end{aligned}
\label{quiverpoisson}
\ee
Strictly speaking, the RHS of the $\{ u^\pm_{i,a} , u^\pm_{j,b} \}$ brackets contain operators $u_A^\pm$ labelled by cocharacters that are sums of the fundamental cocharacters $(i,a)$ and $(j,b)$; this is precisely how the Poisson brackets are able to produce missing ring generators. In the abelian chiral ring $\C[\CM_C^{\rm abel}]$, these operators satisfy relations of the form $(\varphi_{i,a}-\varphi_{j,b})u_A^\pm \sim u_{i,a}^\pm u_{j,b}^\pm$. Throughout this section, we represent such additional generators by rational functions $u_A^\pm \to \frac{u^\pm_{i,a} u^\pm_{j,b} }{\varphi_{i,a} - \varphi_{j,b} }$, in order to simplify notation and to make the relations among them manifest.

\subsection{Generating the chiral ring}
\label{sec:gen-quiv}

As in Section \ref{sec:SQCD}, we can further introduce the polynomials
\be
U_i^\pm(z)= \sum_{a=1}^{M_i} u^{\pm}_{i,a} \prod_{b \neq a}(z - \varphi_{i,b} )\,,
\ee
which are generating functions for non-abelian monopole operators at the $i$-th node of fundamental charge $A=(\pm1,0,\ldots,0)$, and dressing factors $n=(0,1,\ldots,1,0,\ldots,0)$. The relations~\eqref{quiverrelations} can then be reformulated as the statement that the polynomials $P_i(z)  Q_{i-1}(z)Q_{i+1}(z)+U_i^+(z) U_i^-(z)$ are divisible by $Q_i(z)$ for all $i$; or equivalently that there exist auxiliary polynomials $\wt Q_i(z)$ of degree at most $\mathrm{max}(M_i-2,M_i+\Delta_i)$ such that
\be
Q_i(z) \wt Q_i(z) + U_i^+(z) U_i^-(z) =  P_i(z)  Q_{i-1}(z)Q_{i+1}(z) \, .
\label{quiverrelations2}
\ee
The auxiliary polynomial $\wt Q_i(z)$ is a generating function for monopole operators of adjoint cocharacter $A=(1,0,\ldots,0,-1)$ for $U(M_i)$, dressed by electric factors labelled by $n=(0,1,\ldots,1,0,\ldots,0,0)$,
up to some gauge-invariant polynomials in $\varphi$.

For a single node, we argued in Section \ref{sec:SQCD} that the Coulomb branch chiral ring is generated by the coefficients of polynomials $Q(z)$ and $U^\pm(z)$. Here we propose that the coefficients of $Q_i(z)$ and $U_i^\pm(z)$ for $i=1,...,n$ are sufficient to generate the chiral ring $\C[\CM_C]$ of a theory $T^\nu_\mu$ as a \emph{Poisson algebra}. In other words, the ring generators of $\C[\CM_C]$ all arise by taking successive Poisson brackets of the  dressed fundamental monopole operators and invariant polynomials in $\varphi_i$ at single nodes.

Of course, we do not generally expect that the ring generators of $\C[\CM_C]$ can all be expressed as products of the operators charged at single nodes. Indeed, the Poisson brackets of operators charged at single nodes produce operators charged at multiple nodes, which cannot be constructed as simple products. In our examples below, we will test the proposal that $\C[\CM_C]$ can be generated as a ring by monopole operators with sums of fundamental charges at subsets of the nodes.

\subsection{Predictions from string theory: Monopole scattering}
\label{sec:monopoles}

The type IIB brane constructions of $\mathcal{N}=4$ linear quiver gauge theories suggest that their Coulomb branches are moduli spaces of BPS monopoles. We briefly review the brane constructions following Hanany and Witten \cite{HananyWitten} (see also \cite{GK-3d} for an applicable recent review), describe the monopole scattering matrix, and show how the coefficients of the scattering matrix generate the chiral ring $\C[\CM_C]$.

\begin{figure}[htb]
\centering
\includegraphics[width=6in]{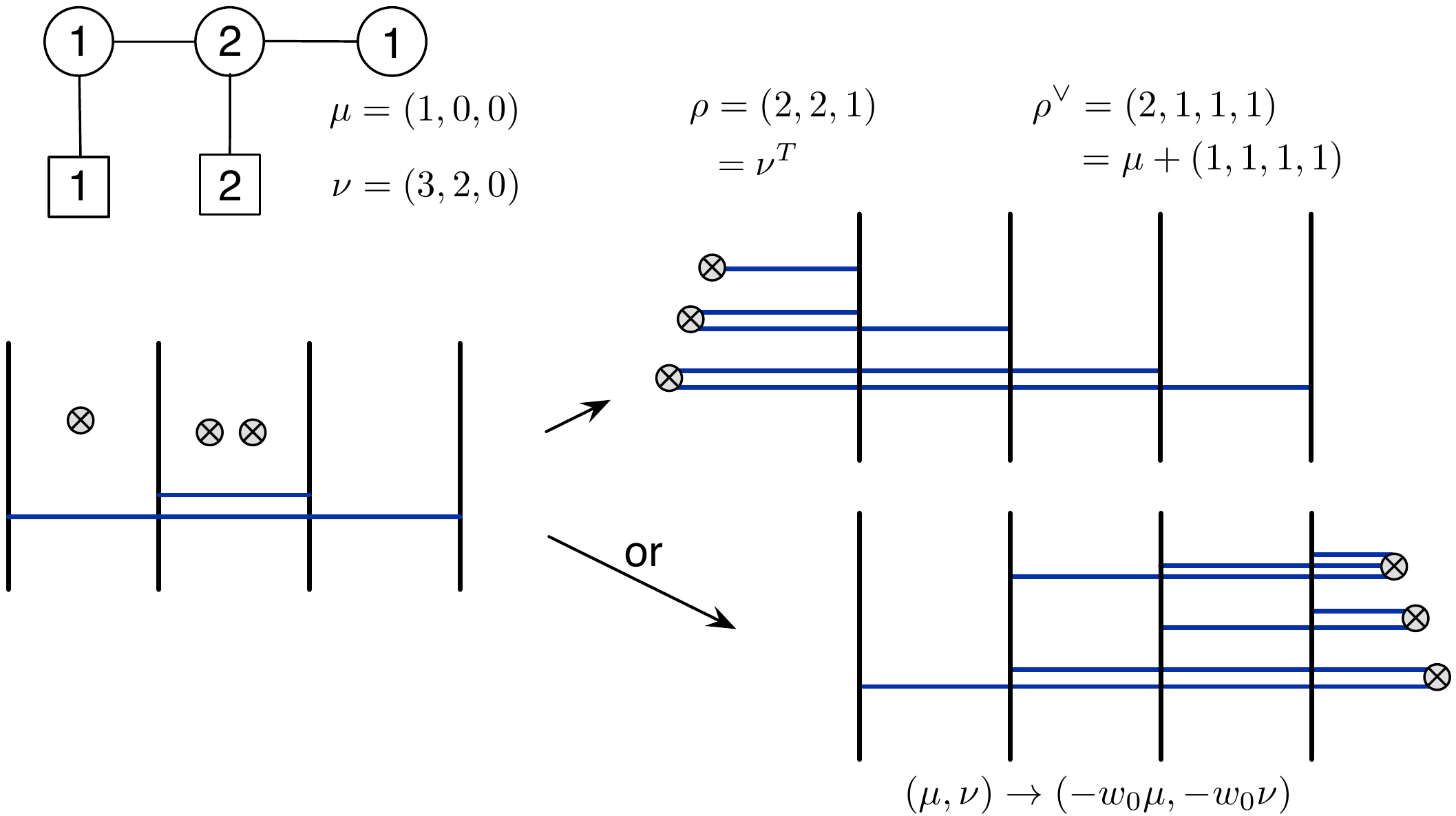}
\caption{A type IIB brane construction for a 3d quiver gauge theory, and the transition that leads to monopoles and to Nahm equations (Section \ref{sec:slices}). The $x^3-x^7$ plane is shown, with D3 branes has horizontal line segments, NS5 branes as vertical lines, and D5 branes as crosses. Here $\mu$ and $\nu$ are written as partitions (Young diagrams).}
\label{fig:branes-q}
\end{figure}

The linear quiver $T^\nu_\mu$ can be engineered by a brane construction in type IIB string theory, as shown in an example on the left of Figure \ref{fig:branes-q}. In the general case, there are $(n+1)$ NS5 branes spanning the directions $x^0,x^1,x^2,x^4,x^5,x^6$ and separated in the $x^3$ direction. The $U(M_i)$ gauge symmetry at the $i$-th node of the quiver is realized on the worldvolume of $M_i$ D3 branes suspended between the $i$-th and $(i+1)$-th NS5 branes and spanning the $x^0,x^1,x^2$ directions. In addition there are $N_i$ $D5$ branes located on the segment of the $x^3$ direction between the $i$-th and $(i+1)$-th NS5 brane and spanning the $x^0,x^1,x^2,x^7,x^8,x^9$ directions.

 Perfoming an S-duality transformation turns the NS5 branes into a stack of $n+1$ D5 branes, which at low energies supports a maximally supersymmetric 6d super-Yang-Mills theory at low energies with gauge group $G'=PSU(n+1)$.%
\footnote{A priori, the group is $U(n)$, but the center $U(1)$ plays no role, since no fields are charged under it. (Alternatively, there are no dynamical monopoles for $U(1)$.) We will be interested in configurations where a single semi-infinite D3 brane can intersect the stack of D5 branes. Such configurations will necessarily correspond to singular BPS monopole solutions for $PSU(n+1)$ (rather than, say, $SU(n+1)$) once the center is factored out.} %
The D3 branes intersecting the stack of D5 branes in codimension 3 then behave like BPS monopoles in the 6d SYM theory with magnetic charge $q_\infty = -\mu$. We refer the reader to Appendix \ref{app:monopoles} for further details and a summary of our notation.
This suggests that the Higgs branch of the worldvolume theory on the D3 branes, which is equivalent to the Coulomb branch of our original theory, should be the corresponding moduli space of BPS monopoles.

If there are additional D5 branes providing fundamental hypermultiplets, they can be moved to infinity $x^3 \to - \infty$ as on the right of  Figure \ref{fig:branes-q}. Alternatively, we could pull the D5 branes to $x^3\to +\infty$, which conjugates the weights $(\mu,\nu)\to(-w_0\mu,-w_0\nu)$.
Performing an S-duality transformation, this leads to semi-infinite D3 branes intersecting the stack of D5 branes, which behave as singular BPS monopole solutions in the 6d SYM theory~\cite{CherkisKapustin-mon}. For a general quiver, there are $N_i = (\nu,\alpha_i)$ singular monopole solutions labelled by each fundamental cocharacter $\omega_{n+1-i}$ of $PSU(n+1)$. In the language of Appendix \ref{app:monopoles}, we have $\sum_a q_a =  \sum_i N_i \omega_{n+1-i}$ and $\sum_a q^-_a =  - \sum_i N_i \omega_i = - \nu$. The positions of the singular monopoles in the $x^4,x^5,x^6$ directions are the triplets of hypermultiplet mass parameters. 

A precise definition of the moduli space of singular monopoles for any compact simple $G$ can be found in \cite{MRB-monopole, MRB-monopole2}, where it is shown that the complex dimension of this moduli space is $\langle \rho, q_\infty - \sum_aq_a^- \rangle  = \langle \rho, \nu - \mu \rangle = \sum_i M_i$. The agrees with the complex dimension of the Coulomb branch of $T^\nu_\mu$.

\subsubsection{The scattering matrix}
\label{sec:scat-quiver}

As a complex manifold, the moduli space of singular monopoles is parametrized by a meromorphic scattering matrix $S(z) \in G_\C' = PSL(n+1,\C)$.
In terms of the brane setup, $z = x^4+ix^5$ (say) parametrizes a chosen complex plane within the $x^4,x^5,x^6$ directions.
Generalizing the structure in Section \ref{sec:pureUN}, the scattering matrix is defined modulo holomorphic upper-triangular matrices acting on the right and holomorphic lower-triangular matrices acting on the left.
It must satisfy two boundary conditions, described in detail in Appendix \ref{app:monopoles}.

First, near a singular monopole of charge $\omega_j$ at $z_0$, the scattering matrix should behave as 
\be S(z) \sim g(z) (z-z_0)^{-w_0\omega_j } g'(z) \, , 
\ee
where $g,g'\in G_\C'$ are meromorphic and regular at  $z=z_0$. Here we follow a standard notation: for any cocharacter $\lambda \in \text{Hom}(\C^*,\mathbb T_{G_\C'}) \subset \text{Hom}(\C^*,G_\C')$, we write $x^\lambda := \lambda(x)$ for the image%
\footnote{As in Section \ref{sec:SQCDmatter}, if we view the gauge group as $G_\C' = SL(n+1,\C)/\Z_{n+1}$, the elements $x^\lambda$ may appear to involve $(n+1)$-th roots. Such roots do not appear in the faithful adjoint representation of the group, and they can also be removed by viewing $G_\C'\simeq GL(n+1,\C)/\C^*$ and rescaling.} %
of $x\in \C^*$; also, $w_0 \, \omega_j = - \omega_{n+1-j}$ is reflection by the longest element of the Weyl group.
Altogether, the Dirac singularities require the scattering matrix to be of the form
\be \textstyle  S(z) = g(z) \Big[ \prod_{i=1}^n P_i(z)^{ - w_0 \omega_j} \Big] g'(z)\,, \ee
where $g,g'\in G_\C'$ are holomorphic and $P_i(z) = \prod_{\alpha=1}^{N_i}(z-m_{i\alpha})$ is the characteristic polynomial at the $i$-th flavor node. When all masses are set to zero, we simply have $\prod_iP_i(z)^{ - w_0 \omega_i } = z^{-w_0\nu}$. As in Section \ref{sec:SQCDmatter}, the nonsingular matrices $g(z)$ and $g'(z)$ encode smooth monopoles surrounding the Dirac singularities.

Second, near infinity, the scattering matrix must behave as
\be S(z) \sim h(z) \, z^{-\mu} \, h'(z)\,, \ee
where $h$ and $h'$ are now respectively lower and upper diagonal gauge transformations that are regular in the vicinity of $z \to \infty$. 

To describe the boundary condition explicitly, we first use the scaling freedom in $PSL(n+1,\C) \simeq PGL(n+1,\C) = GL(n+1,\C)/\C^*$ to normalize $S(z)\in GL(n+1,\C)$ so that it has determinant
\be \det\,S(z) = \prod_{i=1}^n P_i(z)^{n+1-i}\,. \ee
Then all entries of $S(z)$ become polynomials (instead of rational functions).
The leading principal minors of $S(z)$, which are invariant under unipotent triangular gauge transformations, must have the form
\be \textstyle  S^{1,\ldots,i}_{1,\ldots,i}(z) = Q_i(z)\, \prod_{j=1}^i P_j(z)^{i-j}\,, \label{Qfixed} \ee
where the $Q_i(z)$ are polynomials of degree $M_i$. We can use (constant) diagonal gauge transformations to fix the $Q_i(z)$ to be monic. As the notation suggests, we propose that $Q_i(z)$ be identified with the characteristic polynomial $Q_i(z) = \prod_{a=1}^{M_i}(z-\varphi_{i,a})$ at the $i$-th node of the quiver. The $\varphi_{i,a}$ are now interpreted as the positions of the D3 branes suspended between the $i$-th and $(i+1)$-th NS5 brane in the complex $z$-plane. 

After fixing the leading principal minors as above, we still have the freedom to multiply $S(z)$ by unipotent holomorphic matrices. This redundancy can be used to systematically reduce the elements of the scattering matrix $S(z)$ to polynomials in $z$.

To begin, the $i\times i$ minors $S^{1,\ldots,i}_{1,\ldots,i-1,i+1}(z)$ and $S^{1,\ldots,i-1,i+1}_{1,\ldots,i}(z)$ transform respectively under left and right multiplication by addition of a holomorphic function multiplied by $S^{1,\ldots,i}_{1,\ldots,i}(z)$. We may therefore transform these minors to be of the form
\be
\textstyle
S^{1,\ldots,i}_{1,\ldots,i-1,i+1}(z) = {U^+_i(z)} \prod_{j=1}^i P_j(z)^{i-j}\,,\qquad
S^{1,\ldots,i-1,i+1}_{1,\ldots,i}(z) = {U^-_i(z)} \prod_{j=1}^i P_j(z)^{i-j}\,,
\label{Ufixed}
\ee
where $U^\pm_j(z)$ are polynomials of degree $M_i-1$. Assuming that the coordinates $\varphi_{i,a}$ are distinct, these polynomials are uniquely determined by their values at the $\varphi_{i,a}$. We can therefore introduce coordinates $u^\pm_{i,a}$ such that $U^\pm_i(\varphi_{i,a}) = u_{i,a} \prod_{j\neq i}(\varphi_{i,a}-\varphi_{i,b})$. The coordinates $u^\pm_{i,a}$ are identified with the abelian monopole coordinates at the $i$-th node of the quiver.

The coordinates $\varphi_{i,a}$, $u^\pm_{i,a}$ are not independent: from the identity
\be
S_{1,\ldots,i}^{1,\ldots,i}(z)S_{1,\ldots,i-1,i+1}^{1,\ldots,i-1,i+1}(z) - S^{1,\ldots,i}_{1,\ldots,i-1,i+1}(z) S^{1,\ldots,i-1,i+1}_{1,\ldots,i}(z) = S_{1,\ldots,i-1}^{1,\ldots,i-1}(z)S_{1,\ldots,i+1}^{1,\ldots,i+1}(z)
\ee
we find
\be
Q_i(z) \tilde{Q}_i(z) - U^+(z)U^-(z) = Q_{i-1}(z) Q_{i+1}(z) P_i(z) 
\label{QQtilde}
\ee
where
\be
\textstyle
S^{1,\ldots,i-1,i+1}_{1,\ldots,i-1,i+1}(z) = \wt Q_i(z)  \prod_{j=1}^i P_j(z)^{i-j}
\ee
and $\wt Q_i(z)$ is a monic polynomial of degree of degree $\Delta_i+M_i$ if $\Delta_i \geq -1$ and $M_i-2$ otherwise. Evaluating the polynomial equation \eqref{QQtilde} at the positions $z = \varphi_{i,a}$ we recover the relation \eqref{quiverrelations} obeyed by the coordinates $\varphi_{i,a}$, $u^+_{i,a}$ and $u^-_{i,a}$. 

For a quiver of rank $n=1$, as discussed in Section \ref{sec:SQCD}, this completely fixes the redundancy in $S(z)$ and the Coulomb branch chiral ring is generated by the polynomials $Q(z)$, $U^+(z)$, $U^-(z)$ and $\wt Q(z)$ at the single node subject to the relation~\eqref{QQtilde}. However, for quivers of higher rank, there are additional generators that cannot be obtained from these polynomials. There is also further gauge redundancy. The redundancy can be systematically fixed by specifying the degrees of a set of remaining independent minors of $S(z)$. We illustrate the procedure in the case of rank 2 quivers.

\subsubsection{Rank-2 quivers}

For $n=2$, we normalize the scattering matrix $S(z)\in GL(3,\C)/C^*$ so that $\det S(z) = P_1(z)^2P_2(z)$. We can fully fix the gauge redundancy by parametrizing the scattering matrix as 
\be
S(z) = 
\begin{bmatrix}
\, Q_1(z) \, & \, U^+_1(z) \, & \, U^+_{12}(z) \,  \\
\, U^-_1(z) \, & \, \wt Q_1(z) \,& \, \tilde U^+_2(z) \,\\
\, U^-_{12}(z) \, &  \, \tilde U^-_2(z) \, &\,  \wt Q_{12}(z) \, 
\end{bmatrix} \, ,
\ee
with 
\begin{align}
Q_1(z) \wt Q_1(z) - U^+_1(z) U^-_1(z) &= P_1(z) Q_2(z) \cr
Q_1(z) \tilde U^+_2(z) - U^+_{12}(z) U^-_1(z) &= P_1(z) U^+_{2}(z) \cr
Q_1(z) \tilde U^-_2(z) - U^-_{12}(z) U^+_1(z) &= P_1(z) U^-_{2}(z) \cr
Q_1(z)  \wt Q_{12}(z) - U^+_{12}(z) U^-_{12}(z) &= P_1(z) \wt Q_{2}(z)\,,
\end{align}
where, in addition to above conditions on the degrees of $Q_i,U_i^\pm$ and $\wt Q_i$, the polynomials  $U_{12}^+(z)$ are required to have degree at most $M_1-1$.

The position of a polynomial in the scattering matrix is related to the charge of the corresponding chiral operators 
under the Coulomb branch isometries. The relations above suggest that $U^\pm_{12}(z)$ should be a generating function 
for dressed monopole operators of minimal charges $(\pm 1,0,\cdots,0)$ at both nodes,
which cannot be obtained from products of monopole operators at individual nodes. 

Explicitly, plugging the roots $\varphi_{1,a}$ of $Q_1(z)$ in the second line, we find
\be
U^+_{12}(\varphi_{1,a}) u^-_{1,a} \prod_{b\neq a} (\varphi_{1,a}-\varphi_{1,b}) = -P_1(\varphi_{1,a}) U^+_{2}(\varphi_{1,a})
\ee
\ie 
\be
\frac{U^+_{12}(\varphi_{1,a})}{\prod_{b\neq a} (\varphi_{1,a}-\varphi_{1,b})} = \frac{U^+_{2}(\varphi_{1,a})}{Q_2(\varphi_{1,a})}u^+_{1,a}
\ee
which allows us to write 
\be
U^+_{12}(z) = \sum_{a,a'}  \frac{u^+_{1,a} u^+_{2,a'}}{\varphi_{1,a}-\varphi_{2,a'}}\prod_{b\neq a} (z-\varphi_{1,b})\,.
\ee
This is the desired generating function.
The presence of $(\varphi_{1,a} - \varphi_{2,b})^{-1}$ is a remainder 
of the fact that the coefficients of $U^+_{12}(z)$ must be included as new generators of the chiral ring.

Armed with an expression for $U^+_{12}(z)$ we can derive $\tilde U^+_2(z)$ and identify it with the generating function of 
monopoles of charges $(1,-1,0,\cdots,0)$ at the first node and $(1,0,\cdots,0)$ at the second node, dressed by the characteristic polynomial of the $U(N_1-2)$ scalars at the first node. The identification will hold up to possible terms with monopole charge $(1,0,\cdots,0)$ at the second 
node only. 

We can do a similar analysis for  $U^-_{12}(z)$ and $\tilde U^-_2(z)$. Finally, $\wt Q_{12}(z)$ captures monopole operators of 
charges $(1,-1,0,\cdots,0)$ at the both nodes, dressed by the characteristic polynomial of the $U(N_1-2)$ scalars at the first node.
The identification holds up to possible terms with monopole charge $(1,-1,0,\cdots,0)$ at the first or second 
node only, and operators of charge $0$. 

This analysis completely fixes the redundancy in the scattering matrix and provides an explicit description of $\C[\CM_C]$ in terms of generators and relations.

We will briefly specialize further to two simple examples that are useful for comparing the Coulomb branch to Nahm equations in Section~\ref{sec:slices}.

\subsubsection{Abelian quiver}

Consider the balanced abelian quiver of rank $n=2$, with gauge group $U(1) \times U(1)$ and one fundamental hypermultiplet at each node. In other words, $\vec N=\vec M=(1,1)$ and $\nu=\omega_1+\omega_2$ while $\mu=0$. We normalize the scattering matrix so that $\det S(z) = P_1(z)^2P_2(z)= (z-m_1)^2(z-m_2)$, where $m_i$ are the complex mass parameters.
We shift the masses to obey $m_1+m_2=0$.
Gauge-fixing the scattering matrix as above, we find
\be
S(z) = \gamma_1(z)  \begin{bmatrix}
\, z - \varphi_1 \, & \, u^+_1 \, & \, \frac{u_1^+u_2^+}{\varphi_1 - \varphi_2} \,  \\
\, u^-_1 \, & \, z - m_1 +\varphi_1 - \varphi_2 \,& \, u^+_2 \,\\
\, \frac{u_1^-u_2^-}{\varphi_1 - \varphi_2} \, &  \, u_2^- \, &\,  z +\varphi_2 \, 
\end{bmatrix} \, .
\ee
The Coulomb branch is generated by the components of the scattering matrix, subject to the familiar relations
\be
u^+_1u^-_1 = - (\varphi_1-\varphi_2)(\varphi_1 - m_1) \qquad u^+_2u^-_2 = - (\varphi_2-\varphi_1)(\varphi_2 - m_2) \, .
\ee

\subsubsection{$T[U(3)]$}

Now consider the balanced quiver with $\vec N=(1,2)$ and $\vec M=(0,3)$, \ie\ $\nu=3\omega_2$ and $\mu=0$. The theory $T^\nu_\mu$ is commonly called $T[SU(3)]$ or $T[U(3)]$. In this case we have $\det S(z) = P_1(z)^2P_2(z) = (z-m_1)(z-m_2)(z-m_3)$, and we shift the complex masses to obey $\sum_\alpha m_\alpha=0$.
We now find
\be
S(z) =  \gamma_1(z)  \begin{bmatrix}
\, z-\varphi_1 \, & \, u^+_1 \, & \,   \sum\limits_{a=1}^2 \frac{u_1^+ u_{2,a}^+}{\varphi_1 - \varphi_{2,a}} \,  \\
\,  u^-_1 \, & \, z + \varphi_1-\varphi_{2,1}-\varphi_{2,1} \, & \, \sum\limits_{a=1}^2 u^+_{2,a} \\
\, \sum\limits_{a=1}^2 \frac{u_1^- u_{2,a}^-}{\varphi_1 - \varphi_{2,a}} \, & \, \sum\limits_{a=1}^2 u^-_{2,a} &  z+\varphi_{2,1}+\varphi_{2,2}
\end{bmatrix}\, .
\ee
The Coulomb branch is generated by the coefficients of this scattering matrix subject to the relations \eqref{quiverrelations} for the abelian coordinates.

\subsection{Slices in the affine Grassmannian}
\label{sec:Grass}

When a quiver is good (\ie\ $\mu$ is a dominant weight), the moduli space of scattering matrices obeying the boundary conditions of Section \ref{sec:scat-quiver} coincides with a slice in the affine Grassmannian for $G_\C'=PSL(n+1,\C)$. Roughly speaking, the affine Grassmannian is a certain quotient of the loop group $LG_\C'$. It is related to monopole moduli space because a given monopole configuration in $\R^3$ defines a $G_\C'$ bundle on the $S^2\simeq \cp^1$ surrounding the monopoles, and the transition function for this bundle on the equator of $\cp^1$ is a map $S^1\to G_\C'$, \ie\ an element of $LG_\C'$. This element of $LG_\C'$ is essentially our scattering matrix $S(z)$.

To see the correspondence in detail, let us formalize (and algebrize) the notion of a scattering matrix. For any algebraic group $K$, let $K[z]$, $K[[z]]$, and $K(\!(z)\!)$ denote (respectively) the groups defined over the polynomial ring $\C[z]$, the ring of formal Taylor series $\C[[z]]$, and ring of formal Laurent series $\C(\!(z)\!)$. Let $K_{(1)}[[z^{-1}]]$ denote the first congruence subgroup of $K[[z^{-1}]]$, \ie\ the subgroup of elements that become the identity when $z^{-1}\to 0$. Also, for our particular group $G_\C'$ with fixed maximal torus $\mathbb T'$, let $B^\pm$ and $N^\pm$ be the upper/lower Borel and unipotent subgroups.

The scattering matrix $S(z)$, defined modulo triangular gauge transformations, is an element of the double quotient
\be  \mathbb S :=   B^-[z]\bs G_\C'(\!(z)\!) / B^+[z]\,. \label{BGB} \ee
With all masses set to zero, the boundary condition at singularities requires $S(z)$ to lie in (the closure of) the subset
\be  \mathbb S^{-w_0\nu} :=  G_\C'[z] z^{-w_0\nu}\,G_\C'[z]\,, \ee
while, as long as $\mu$ is dominant, the boundary condition at infinity requires $S(z)$ to lie in the subset
\be \mathbb S_{-\mu} := (G_\C')_{(1)}[[z^{-1}]] z^{-\mu}\,.\ee
Here we implicitly consider both $\mathbb S^{-w_0\nu}$ and $\mathbb S_{-\nu}$ modulo triangular gauge transformations \eqref{BGB}, so that they are subsets of $\mathbb S$. Then the moduli space of scattering matrices is the intersection
\be \CM_C \simeq \overline{\mathbb S^{-w_0\nu}} \cap \mathbb S_{-\mu}\, \subset \mathbb S\,. \ee
As a consistency check, notice that if $\mu=\nu$ (so all gauge nodes in the quiver have rank $M_i=0$), the intersection is trivial.

It is also useful to note that a scattering matrix written explicitly in the form $S(z) = g_1(z) z^{-\mu}$ with $g_1(z)\in (G_\C')_{(1)}[[z^{-1}]]$ is already partially gauge fixed. Acting on the left, only the identity element of $B^-[z]$ preserves this form. Acting on the right, as long as $\mu$ is dominant, the unipotent elements $N^+[z]\subset B^+[z]$ preserve this form.

We want to compare the above with the so-called ``thick'' affine Grassmannian
\be {\rm Gr} = G_\C'(\!(z)\!) / G_\C'[z]\,.\ee
Following \cite{KWWY}, we consider the intersection ${\rm Gr}^{-w_0\nu}_{-\mu} = \overline{{\rm Gr}^{-w_0\nu}}\cap {\rm Gr}_{-\mu}$, where
\be 
  {\rm Gr}^{-w_0\nu} := G_\C'[z] z^{-w_0\nu}\,,\qquad {\rm Gr}_{-\mu} = (G_\C')_{(1)}[[z^{-1}]]z^{-\mu}\ee
are subsets of ${\rm Gr}$ (with an implicit quotient by $G_\C'[z]$ on the right).
Such slices played an important role in the geometric Satake correspondence \cite{Lusztig-slice, Ginzburg-loop, MirkovicVilonen} (which, physically, describes S-duality of loop operators in 4d super-Yang-Mills theory \cite{Kapustin-Witten}), and were quantized in \cite{KWWY}. We claim that
\be  \CM_C \simeq \overline{\mathbb S^{-w_0\nu}} \cap \mathbb S_{-\mu} \simeq {\rm Gr}^{-w_0\nu}_{-\mu} \,.\ee
To see the equivalence, simply observe that an element $g(z)$ of ${\rm Gr}_{-\mu}$ that is written explicitly in the form $g(z) = g_1(z) z^{-\mu}$ for $g_1(z)\in (G_\C')_{(1)}[[z^{-1}]]$ is partially gauge fixed: since $\mu$ is dominant, the subgroup of $G_\C'[z]$ that preserves this form when acting on the right is $N^+[z]$. This is exactly the gauge-fixed form of the scattering matrix $S(z)$ that we found above.

\subsection{Predictions from string theory: Nahm equations}
\label{sec:slices}

The scattering analysis of Section \ref{sec:monopoles} showed us that dressed monopole operators charged at a single node of a quiver are not sufficient to generate the chiral ring $\C[\CM_C]$. It did suggest, as proposed in Section \ref{sec:gen-quiv}, that these operators could generate $\C[\CM_C]$ as a Poisson algebra, and that the ring generators are in fact dressed monopole operators with fundamental charge at sequences of nodes.

We now consider a rather different description of the Coulomb branch of $T^\nu_\mu$ as a moduli space of solutions to $SL(p)$ Nahm equations on an interval, where $p=\sum_{i=1}^n iN_i$ \cite{GW-Sduality}. The description (using a single interval) is valid as long as the quiver is good, \ie\ $\mu$ is a dominant weight. It identifies the Coulomb branch as an intersection of a nilpotent orbit and a transverse slice in the complexified Lie algebra $\mathfrak{sl}_{n+1}^\C$, and provides further evidence in support of our proposal for generators of $\C[\CM_C]$.

To see how this description comes about from the brane construction, we must reformulate the data $(\nu,\mu)$ of a good or balanced quiver as a pair of partitions $(\RHO,\RHO^\vee)$. 
It is convenient to assign a linking number $r_i$ to each $D5$ brane and a linking number $r_j^\vee$ to each $NS5$ brane, as described (\eg) in \cite{GK-3d}:
\be \begin{array}{ll} r_i &= \#(\text{D3 attached to the right})-\#\text{(D3 attached to the left)} + \#\text{(NS5 on the left}) \\[.1cm]
r_j^\vee &= \#(\text{D3 attached to the left})-\#\text{(D3 attached to the right)} + \#\text{(D5 on the right})\,.
\end{array}
\notag
\ee
The dominant weight $\nu$ is recovered as follows: $(\nu,\alpha_i)=N_i$ is the number of times $i$ appears in the list of D5 brane linking numbers $(r_1,r_2,\ldots,r_{|\nu|})$ where $|\nu| = \sum_i N_i$ is the total number of D5 branes. The second weight $\mu$ is recovered from the NS5 brane linking numbers by the formula $(\mu,\alpha_j) = r_j^\vee-r_{j+1}^\vee$. Both linking numbers obey $ \sum_i r_i = \sum_j r_j^\vee = \sum_i i N_i $. 

The infrared physics is independent of the positions of the 5-branes in the $x^3$ direction, provided additional D3 branes are inserted whenever a D5 and NS5 brane cross, in such a way that the linking numbers $(r_i,r_j^\vee)$ are preserved. In particular, the D5 branes can be moved all the way to the left in the $x^3$ direction, as shown in Figure \ref{fig:branes-q} on page \pageref{fig:branes-q}. In this configuration, we find $p = \sum_i i N_i$ D3 branes attached to D5 branes on the left and NS5 branes on the right. The pattern in which the D3 branes end on the two sides can be encoded in two partitions of $p$, commonly called $\RHO = (r_1,\ldots,r_{|\nu|})$ (for the D5 branes) and $\RHO^\vee = (r_1^\vee,\ldots,r_{n+1}^\vee)$ (for the NS5 branes).

The partitions $\RHO$ and $\RHO^\vee$ can also be related directly to the weights $\mu$ and $\nu$ as follows: the Young diagram for $\RHO^T$ labels the irreducible representation of $SU(n)$ with highest weight $\nu$; while the Young diagram for $\RHO^\vee$ is obtained by setting the Young diagram for $\mu$ atop a block of width $n+1$ and height $M_n$ (the rank of the last gauge node),
\be \RHO^T = \nu\,,\qquad \RHO^\vee = \mu+\big[\text{$(n+1)\times M_n$ block}\big]\,. \ee

In the stretched configuration, the D3 branes support a 4d $\CN=4$ super-Yang-Mills theory with gauge group $U(p)$. The 4d theory lives on an interval, with a boundary condition on the left corresponding to a Nahm pole of type $\RHO^T$ and a boundary condition on the right corresponding to a modified Neumann boundary condition of type $\RHO^\vee$ \cite{GW-boundary, GW-Sduality}. The moduli space of our 3d theory $T^\nu_\mu$ can be identified as the space of supersymmetric field configurations of the 4d SYM theory that are compatible with the two boundary conditions. The result of that calculation predicts that, in the infrared, the Coulomb branch $\CM_C$ is the intersection of the closure of a nilpotent orbit $\ol{\CN_{\RHO^T}}$ and a transverse (\eg\ Slodowy) slice $\CS^{\RHO^\vee}$ inside the Lie algebra $\mathfrak{sl}_p^\C$,
\be \CM_C \simeq \CS^{\RHO^\vee}\cap \ol{\CN_{\RHO^T}}\,. \label{orbits} \ee

In detail, recall that any partition $\RHO$ of $p$ labels an embedding of $\mathfrak{sl}_2^\C$ into $\mathfrak{sl}_p^\C$. (If $\RHO = (\rho_1,...,\rho_d)$, then $\mathfrak{sl}_2$ is embedded using the direct sum of its representations of dimension $\rho_i$ for each $i$.) Let $e,f,h$ denote the standard generators of $\mathfrak{sl}_2^\C$, and $\RHO(e),\RHO(e),\RHO(e)$ their images in $\mathfrak{sl}_n^\C$. The nilpotent orbit $\CN_{\RHO}$ is the orbit of the element $\RHO(e)\subset \mathfrak{sl}_p^\C$ under the adjoint action of $SL(p,\C)$. A slice $\CS^{\RHO}$ transverse to this orbit can be obtained in many different ways, all symplectomorphic. One may, for example, take $\CS^{\RHO}$ to be the affine subspace $\RHO(e)+ \text{ker}\,\text{ad}(\RHO(f))$ inside $\mathfrak{sl}_p^\C$; this is the standard Slodowy slice, with a manifest Poisson structure \cite{GanGinzburg}. In the moduli space \eqref{orbits}, we are taking the intersection of the nilpotent orbit labelled by $\RHO^T$ with the slice transverse to a (necessarily smaller) nilpotent orbit labelled by $\RHO^\vee$. For example, the setup of Figure \ref{fig:branes-q} leads to
\be \CS^{\RHO^\vee}= \begin{pmatrix}\alpha &0&0&0&1\\
*&*&*&*&0 \\
*&*&*&*&0 \\
*&*&*&*&0 \\
*&*&*&*& \alpha \end{pmatrix}\,,\qquad
\CN_{\RHO^T} = \CN_{\nu} = g \begin{pmatrix} 0&1&0&0&0\\0&0&1&0&0\\0&0&0&0&0\\
0&0&0&0&1\\0&0&0&0&0 \end{pmatrix}g^{-1}\quad (g\in SL(5,\C))\,.
\ee

Strictly speaking, the description  \eqref{orbits} of $\CM_C$ holds only when mass parameters of $T^\nu_\mu$ are turned off. Indeed, the nilpotent orbits $\CN_{\RHO^\vee}$ are singular. In the presence of complex masses, the nilpotent orbits are deformed to orbits of semisimple elements, with eigenvalues specified by the masses. In the presence of real masses, the orbits are resolved via the Springer resolution.
 
The 3d mirrors of type-A quiver gauge theories can quickly be read off from a brane construction \cite{dBHOY, dBHOO}. Mirror symmetry descends from S-duality of type IIB string theory, which exchanges D5 and NS5 branes while preserving D3 branes. Therefore, mirror symmetry simply exchanges the partitions $\RHO\leftrightarrow\RHO^\vee$.

We now study in detail several families of theories for which the partition $\rho^\vee=(1,1,...,1)$ is trivial, along with several more general examples.

\subsubsection{General structure for trivial $\rho^\vee$: theories $T^\RHO$}
\label{sec:Trho}

The balanced quivers with $M_n=1$ (meaning that $\mu=0$ and $p=\sum_i i N_i = n+1$, or equivalently that $\nu = \sum_i M_i\alpha_i$ has a single copy of $\alpha_n$) form an especially nice class of examples. In this case, in the notation of Section \ref{sec:slices}, the partition $\RHO^\vee$ just equals $(1,1,...,1)$, and the transverse slice $\CS^{\RHO^\vee}$ is all of $\mathfrak{sl}_p^\C=\mathfrak{sl}_{n+1}^\C$.  The Coulomb becomes the entire nilpotent orbit
\be \CM_C \simeq \ol{\CN_\nu}\,, \label{MCnilp} \ee
or its resolution in the presence of mass parameters. Theories of this type were called $T^{\RHO}[U(n+1)]$ in \cite{GW-Sduality}. Here we will argue that for these theories the proposal of Section \ref{sec:gen-quiv} is correct: namely, the chiral ring $\C[\CM_C]$ is generated as a Poisson algebra by monopoles operators with fundamental charge at single nodes, and is generated as a ring by monopole operators with fundamental charge at sequences of neighboring nodes.

Since the $\mu=0$, the symmetry acting on the Coulomb branch of $T^{\RHO}$ via hyperk\"ahler isometries is the the fully enhanced group $G_C = SU(n+1)$. The action has a complex moment map $\bm \mu \in(\mathfrak{sl}_{n+1}^\C)^* \simeq \mathfrak{sl}_{n+1}^\C$, where we use the Cartan-Killing form to identify the Lie algebra with its dual. The moment map is a traceless $(n+1)\times(n+1)$ matrix, whose entries are elements of the chiral ring $\C[\CM_C]$. Moreover, in the absence of masses, the moment map is nilpotent, and parameterizes the entire orbit \eqref{MCnilp} \cite{GW-Sduality}. (With masses turned on, the moment map instead parametrizes the semisimple deformation of \eqref{MCnilp}.) Thus, the entries of the moment map actually \emph{generate} the entire ring $\C[\CM_C]$. We just need to understand what these entries are.

Let $E_{ij}\in \mathfrak{gl}_{n+1}$ be the matrix with entry $+1$ in the $i$-th row and $k$-th column, and zero elsewhere. Let $H_i:=E_{i,i}-E_{i+1,i+1}$, $E_i = E_{i,i+1}$, and $F_i:=E_{i+1,i}$ be the standard Chevalley-Serre generators of $\mathfrak {sl}_{n+1}$. By definition, the functions
\be  h_i = \Tr(\bm \mu H_i)\,,\qquad e_i = \Tr(\bm \mu E_i)\,,\qquad f_i = \Tr(\bm \mu F_i) \qquad (1\leq i\leq N)\ee
must generate the (complexified) infinitesimal action of the symmetry group $SU(n+1)$ on the chiral ring $\C[\CM_C]$, acting via the Poisson bracket.  In particular, they themselves must obey
\be
\begin{aligned}
\{h_i,e_j\}&=\kappa_{ij}e_j \\
\{h_i,f_j\}&=-\kappa_{ij}f_j \\
\{e_i,f_j\}&=\delta_{ij}h_i \\
\mathrm{ad}(e_i)^{1-\kappa_{ij}} e_j &=0 \quad i \neq j\\
\mathrm{ad}(f_i)^{1-\kappa_{ij}} f_j &=0 \quad i \neq j
\end{aligned}
\label{eq:classicalsl}
\ee
where $\kappa$ is the Cartan matrix and $ad(x)y := \{x,y\}$.

We propose, in the notation of Section \ref{sec:abel-quiv}, that
\be
h_i = \sum_{j=1}^n \kappa_{ij} \sum_{a=1}^{M_i} \varphi_{j,a} - \sum_{\alpha=1}^{N_i}m_{i,\alpha}  \,,\qquad e_i = \sum_{a=1}^{M_i} u^+_{i,a}\,, \qquad f_i = \sum_{a=1}^{M_i} u^-_{i,a}\, .
\label{hefdefinition}
\ee
In other words, the $e_i$ and $f_i$ are nonabelian monopole operators labelled by the fundamental cocharacters $A=(\pm 1,0,...,0)$ at the $i$-th node of the quiver.
It follows from \eqref{quiverpoisson} that the relations \eqref{eq:classicalsl} hold.

The $h_i$ determine the diagonal components of the moment map $\bm\mu$, while the $e_i,f_i$ determine the components directly above and below the diagonal. However, since $\bm\mu$ itself transforms in the adjoint representation of the symmetry algebra $\mathfrak{sl}_{n+1}^\C$, the remaining entries of $\bm\mu$ can all be expressed as successive Poisson brackets of the $e_i$ and~$f_i$.
This demonstrates that monopole operators with fundamental charge at single nodes are sufficient to generate $\C[\CM_C]$ as a Poisson algebra. Moreover, the successive Poisson brackets are precisely the monopole operators with a sum of charges at sequences of neighboring nodes, which altogether generate $\C[\CM_C]$ as a ring. These are all the Coulomb branch operators of R-charge $2$. 

To describe $\bm\mu$ more explicitly, let us define
\be \varphi_i := \sum_{a=1}^{M_i}\varphi_{i,a}\,,\qquad m_i := \sum_{\alpha=1}^{N_i} m_{i,\alpha}\,,\qquad (\kappa^{-1}m)_i = \sum_{j=1}^n (\kappa^{-1})_{ij}m_j\,,\ee
and denote the (undecorated) nonabelian monopole operators with a sum of charges $(\pm 1,0,...,0)$ at a set of consecutive nodes $\{i,i+1,...,i+k\}$ as
\be V^\pm_{[i:i+k]} := \sum_{a_0,a_1,...,a_k} \frac{u^\pm_{i,a_0} u^\pm_{i+1,a_1}\cdots u^\pm_{i+k,a_k}}{(\varphi_{i,a_0}-\varphi_{i+1,a_1})(\varphi_{i+1,a_1}-\varphi_{i+2,a_2})\cdots (\varphi_{i+k-1,a_{k-1}}-\varphi_{i+k,a_k})}
\ee
Thus, for example, $V^+_{[i:i]} = \sum_a u^+_{i,a}=e_i$ and $V^-_{[i:i]}=f_i$. Observe that
\be \{V^\pm_{[i,i+k']},V^\pm_{[i+k',i+k]}\} = \mp V^\pm_{[i:i+k]}\,, \label{V-quiver} \ee
and in particular that all the $V^\pm_{[i:i+k]}$ are generated by taking successive Poisson brackets of the $e_i$ or $f_i$. Then the components of the moment map are
\be \label{mu-nonabelian}
 \bm\mu^i{}_j = \begin{cases} \varphi_i-\varphi_{i-1} + (\kappa^{-1}m)_i- (\kappa^{-1}m)_{i-1}& i=j \\
 (-1)^{i-j-1} V^-_{[i:j-1]} & i < j \\
 (-1)^{i-j-1} V^+_{[j:i-1]} & i > j\,.
 \end{cases} \ee
For example, when $n=3$, the $\mathfrak{sl}_3^\C$--valued moment map is
\be \bm\mu =  \begin{pmatrix} \varphi_1 +\frac13(2m_1+m_2)  & V^-_{[1:1]} & - V^-_{[1:2]} \\
   V^+_{[1:1]} & \varphi_2-\varphi_1 + \frac13(m_2-m_1) & V^-_{[2:2]} \\
   -V^+_{[1:2]} & V^+_{[2:2]} & -\varphi_2 - \frac13(m_1+2m_2)
\end{pmatrix}\,.  \ee

\subsubsection{Trivial $\RHO^\vee$: Abelian Quiver}
\label{sec:abel-quiver}

\begin{figure}[htb]
\centering
\includegraphics[width=2.6in]{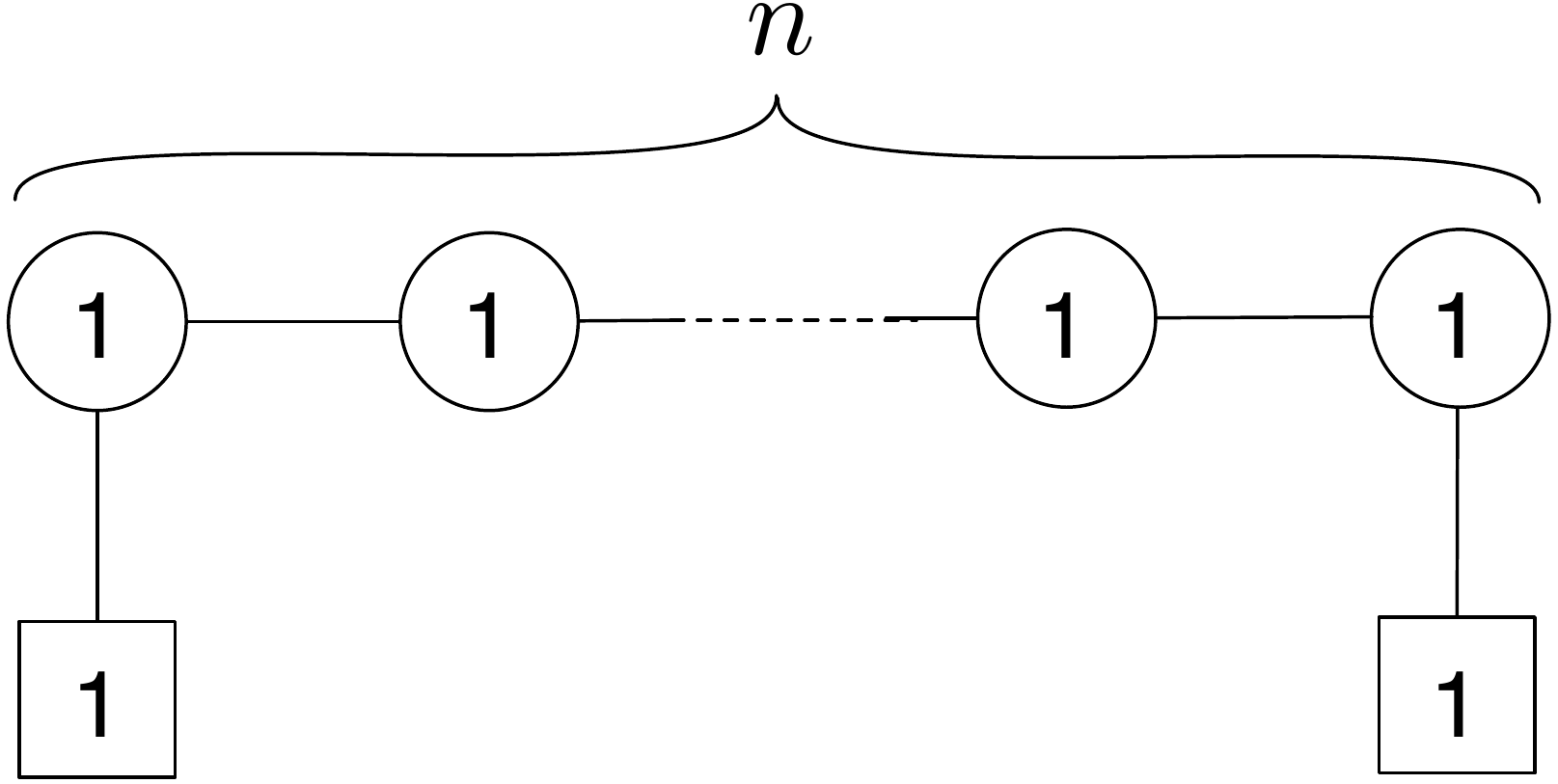}
\caption{The balanced abelian quiver, with $\nu=(2,1,...,1)$ and $\mu=0$}
\label{fig:abel-quiv}
\end{figure}

An extreme case of a quiver with trivial $\rho^\vee$ is the abelian quiver of rank $n$, with gauge group $G=U(1)^n$ and, necessarily, a single hypermultiplet at the initial and final node (Figure \ref{fig:abel-quiv}). Although this quiver falls under the analysis of Section \ref{sec:abelian}, we revisit it here with a focus on the IR symmetry enhancement. The quiver is balanced and the Coulomb branch symmetry is enhanced to $G_C=SU(n+1)$. This is a theory of type $T^{\RHO}[U(n+1)]$ with partition $\RHO^T = \nu = (2,1,...,1)$ (of length $n$).

The abelian coordinates are $\varphi_i$ and $u^\pm_i$ for $i=1,...,n$, with Poisson brackets
\be
\begin{aligned}
\{  \varphi_{i} , u^\pm_{i} \} & = \pm u_i^{\pm} \\
\{ u^+_{i} , u^-_{i} \} & =  2\varphi_i-\varphi_{i-1}-\varphi_{i+1} \\
\{ u^\pm_{i} , u^\pm_{j} \} & = \mp \frac{u^\pm_{i} u^\pm_{j} }{\varphi_{i} - \varphi_{j} } \qquad |i - j| = 1\,,
\end{aligned}
\label{abelianquiverpoisson}
\ee
and relations
\be
u^+_i u^-_i = -(\varphi_i-\varphi_{i-1})(\varphi_i - \varphi_{i+1}) \, ,
\label{abelianrelation}
\ee
where we have included the masses of the hypermultiplets at the first and last node as $\varphi_0:=m_1$ and $\varphi_{i+1}=m_n$. The generators of the $\mathfrak{sl}_{n+1}^\C$ symmetry algebra are $h_i = 2\varphi_i - \varphi_{i-1} - \varphi_{i+1}$, $e_i = u^+_i$, and $f_j := u^-_i$, and more generally the  $\mathfrak{sl}_{n+1}^\C$-values moment map has entries
\be
{\bm\mu^i}_j  = \begin{cases} \varphi_1- \frac{1}{n+1}(nm_1+m_n) & i=j=1\\
 \varphi_i-\varphi_{i-1} +\frac{1}{n+1}(m_1-m_n) & 1<i=j<n+1 \\
 -\varphi_i +\frac{1}{n+1}(m_1+nm_n) & i=j=n+1 \\
  \displaystyle (-1)^{i-j+1} \frac{u^-_i u^-_{i+1} \cdots u^-_{j-1}}{(\varphi_i - \varphi_{i+1}) \cdots(\varphi_{j-2}-\varphi_{j-1})}  &  i < j \\
\displaystyle (-1)^{i-j+1} \frac{u^+_j u^+_{j+1} \cdots u^+_{i-1}}{(\varphi_j - \varphi_{j+1}) \cdots(\varphi_{i-2}-\varphi_{i-1})}  & j < i \, .
\end{cases}
\label{momentabelian2}
\ee
Subject to \eqref{abelianrelation}, this moment map parametrizes a nilpotent orbit (or its semisimple deformation) of type $\nu=(2,1,...,1)$. It is the minimal nontrivial nilpotent orbit. Direct calculation shows that the minimal and characteristic polynomials of $\bm \mu$ are
\be \big(\bm \mu-\tfrac1{n+1}(m_1-m_n)\big)\big(\bm\mu+\tfrac{n}{n+1}(m_1-m_n)\big) = 0\,,\ee
\be \det(\bm\mu-x)= \tfrac1{(n+1)^2}\big(x-\tfrac1{n+1}(m_1-m_n)\big)^n\big(x+\tfrac{n}{n+1}(m_1-m_n)\big)\,.\ee
as expected.

The mirror of this theory is SQED: $U(1)$ gauge theory with $n+1$ hypermultiplets. The Higgs branch moment map of the mirror theory has components $X_i Y_j - \frac{1}{N} \delta_{ij} \sum_n X_n Y_n$. Equating this with the Coulomb branch moment map~\eqref{momentabelian2} and identifying the mirror FI parameter with $m_1-m_n$, we have constructed the mirror map explicitly for this mirror pair.

\subsubsection{Trivial $\RHO^\vee$: $T[U(n+1)]$}

\begin{figure}[htb]
\centering
\includegraphics[width=3in]{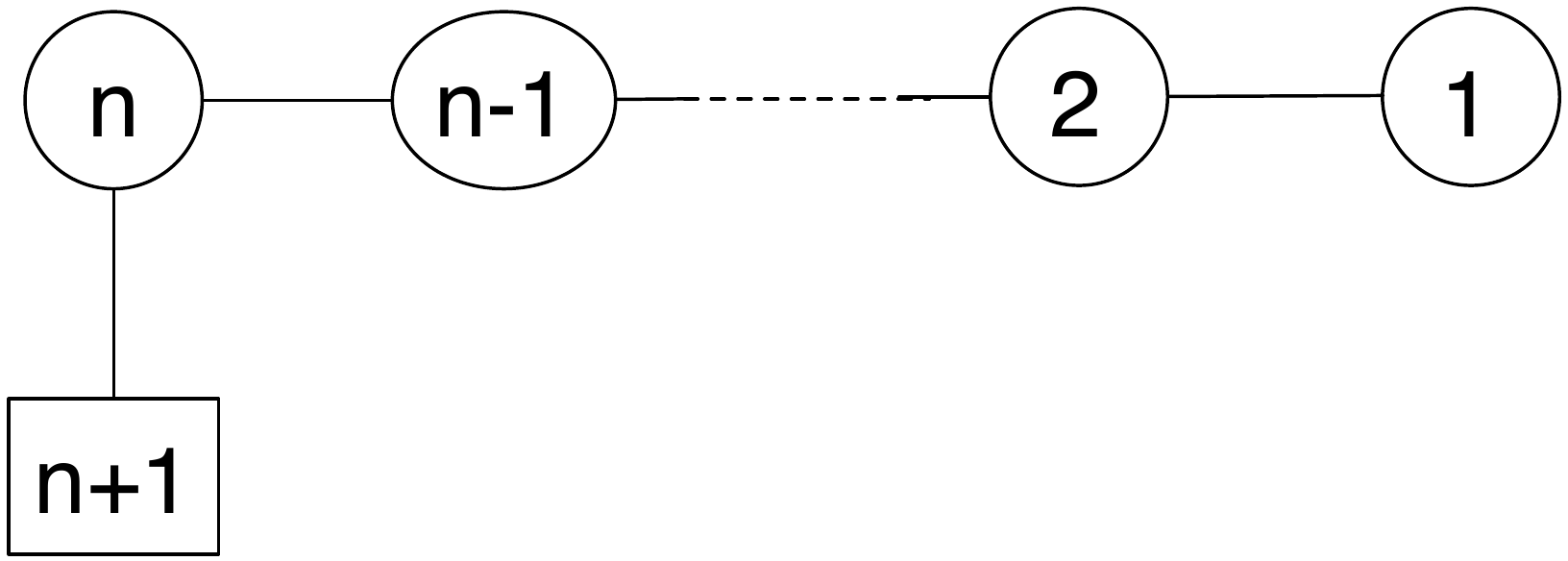}
\caption{The $T[U(n+1)]$ quiver, with $\nu=(n+1)$ and $\mu=0$}
\label{fig:TSUn}
\end{figure}

At the other extreme is the theory $T[U(n+1)]$, which is $T^{\RHO}[U(n+1)]$ with $\RHO^T = \nu = (n+1)$, coming from the triangular quiver of Figure \ref{fig:TSUn}. Its Coulomb branch is a deformation of the maximal nilpotent orbit in $\mathfrak{sl}_{n+1}^\C$.  There are $n+1$ mass parameters $m_\alpha:=m_{1,\alpha}$ for the hypermultiplets at the first node of the quiver, defined up to an overall shift, which parameterize the eigenvalues of the deformed orbit.

The moment map $\bm\mu$ takes the general form \eqref{mu-nonabelian}. Its diagonal components are
\be \textstyle \bm\mu^i{}_i = \varphi_i-\varphi_{i-1} +\tfrac{1}{n+1}(1-(n+1)\delta_{i,1})\sum_\alpha m_\alpha\,,\ee
and its off-diagonal components are nonabelian monopole operators charged at consecutive nodes, as in \eqref{V-quiver}. The characteristic and minimal polynomials are identical. Letting $M=\sum_{\alpha=1}^{n+1} m_\alpha$, we have
\be \prod_{\alpha=1}^{n+1} \big(\bm\mu+m_\alpha-\tfrac{1}{n+1}M\big) = 0\,. \label{TUnpoly} \ee
(We have verified numerically that the abelianized chiral ring relations imply \eqref{TUnpoly} for $n\leq 8$.)

The statement that the full chiral ring $\C[\CM_C]$ is generated by the entries of the moment map is a rather nontrivial statement for nonabelian quivers. Indeed, even for a single node, it is not obvious how the all the coefficients of the polynomials $Q_i(z),U_i^\pm(z)$ are obtained from $\bm\mu$. We can illustrate how this works in the simple example of $T[U(3)]$ theory. With masses turned off, the moment map takes the form
\be \bm\mu = \begin{pmatrix} \varphi_{1,1}+\varphi_{1,2}
& u^-_{1,1}+u^-_{1,2} & -\frac{u^-_{1,1}u^-_2}{\varphi_{1,1}-\varphi_2} - \frac{u^-_{1,2}u^-_2}{\varphi_{1,2}-\varphi_2} \\
u^+_{1,1}+u^+_{1,2} & -\varphi_{1,1}-\varphi_{1,2}+\varphi_2 
 & u^-_2 \\
-\frac{u^+_{1,1}u^+_2}{\varphi_{1,1}-\varphi_2}- \frac{u^+_{1,2}u^+_2}{\varphi_{1,2}-\varphi_2} & u^+_2 & -\varphi_2
\end{pmatrix}\,,
\ee
and contains the leading coefficients of $Q_1(z)$ and $U_1^\pm(z)$. The subleading coefficient of $Q_1(z)$ is
\be -\bm\mu^2{}_3\bm\mu^3{}_2 -(\bm\mu^3{}_3)^2-\bm\mu^3{}_3\bm\mu^1{}_1 = \varphi_{1,1}\varphi_{1,2}\,. \ee
The subleading coefficient of $U_1^+(z)$, a dressed monopole operator, is
\be \bm\mu^2{}_3\bm\mu^3{}_1 -\bm\mu^3{}_3\bm\mu^2{}_1 = u^+_{1,1}\varphi_{1,2}+u^+_{1,2}\varphi_{1,1}\,.\ee

\subsubsection{Intermediate orbits}

\begin{figure}[htb]
\centering
\includegraphics[width=4.7in]{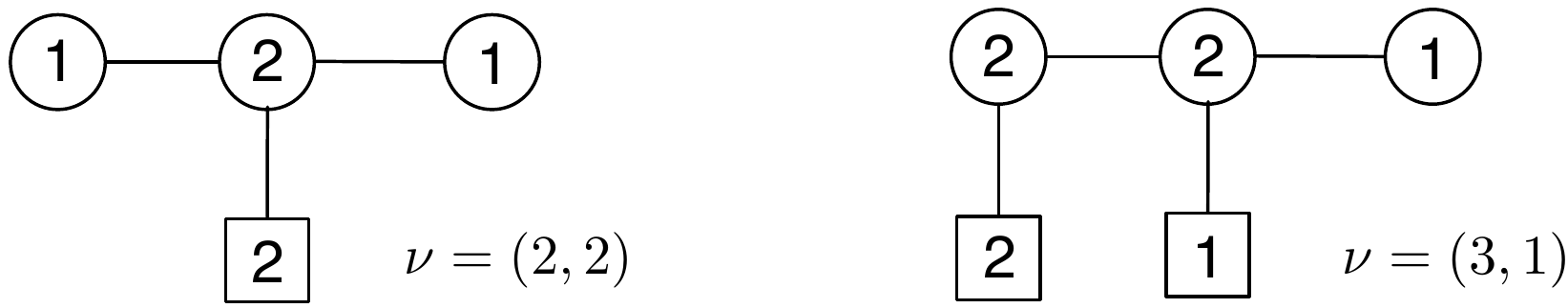}
\caption{Quivers whose Coulomb branches are intermediate nilpotent orbits.}
\label{fig:quiv-intermed}
\end{figure}

The first examples of orbits that are neither minimal nor maximal occur for quivers of rank $n=3$, \ie\ orbits in  $\mathfrak sl_4^\C$. The two relevant quivers are shown in Figure \ref{fig:quiv-intermed}.

On the left we expect an orbit of type $\nu=(2,2)$. Letting $m_{2,1}$ and $m_{2,2}$ denote the mass parameters at the middle node, and using the shift symmetry to set $m_{2,1}+m_{2,2}=0$,
 we indeed find that the moment map $\bm\mu$ constructed as in \eqref{mu-nonabelian} satisfies
\be
\begin{array}{c}
\det(\bm \mu-x) = \big(x+m_{2,1}\big)^2\big(x+m_{2,2}\big)^2\,, \\[.2cm]
\big(\bm \mu+m_{2,1}\big)\big(\bm \mu+m_{2,2}\big) =0\,.
\end{array}
\ee
On the right, we expect a slightly larger orbit of type $\nu=(3,1)$. Again, we use the shift symmetry to set the sum of masses $m_{1,1}+m_{1,2}+m_2=0$. We find that the moment map satisfies
\be
\begin{array}{c}
\det(\bm \mu-x) = \big(x+\tfrac34m_2\big)^2\big(x+m_{1,1}-\tfrac14m_2\big)\big(x+m_{1,2}-\tfrac14m_2\big)\,, \\[.2cm]
\big(\bm\mu+\tfrac34m_2\big)\big(\bm\mu+m_{1,1}-\tfrac14m_2\big)\big(\bm\mu+m_{1,2}-\tfrac14m_2\big) =0\,.
\end{array}
\ee

\subsubsection{General slices and nilpotent orbits}

Finally, we consider the simplest examples of quivers (balanced and unbalanced) for which $\RHO^\vee$ is nontrivial, and whose Coulomb branches are described as the intersection of a nontrivial slice (\ie\ not all of $\mathfrak{sl}_p^\C$) with some nilpotent orbit. One point that we wish to illustrate is that the entries of the moment map for the Coulomb-branch flavor symmetry $G_C$ are \emph{not} usually enough to generate the chiral ring. This is particularly obvious for unbalanced quivers with $G_C\simeq U(1)^n$, since the moment map in this case is much too small; but even for balanced quivers with $\RHO^\vee\neq (1,1,...,1)$ additional operators are needed. We do verify in these examples that the additional operators are still of the form proposed in Section \ref{sec:gen-quiv}.

\begin{figure}[htb]
\centering
\includegraphics[width=5.2in]{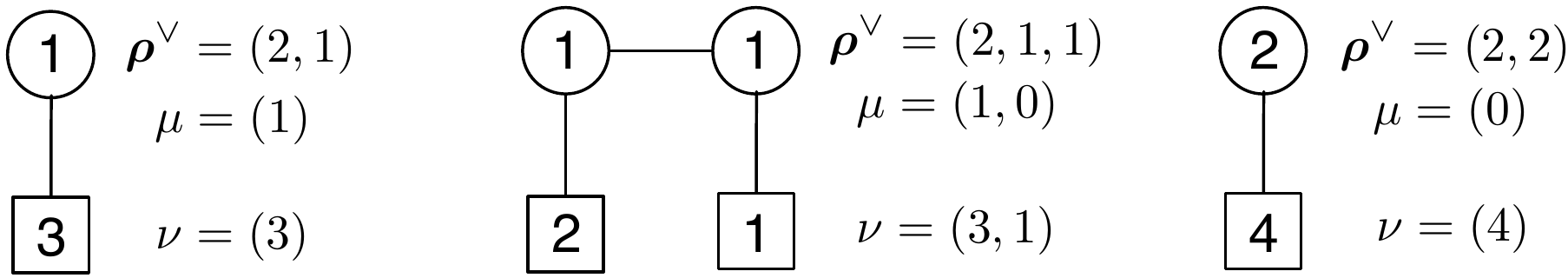}
\caption{Three quivers for which $\CM_C$ involves a nontrivial slice $\CS^{\RHO^\vee}$.}
\label{fig:slice}
\end{figure}

The simplest example is the quiver on the left of Figure \ref{fig:slice}, \ie\ SQED with three hypermultiplets. It is a good but unbalanced quiver, with $\RHO^T=\nu = (3)$, $\mu = (1)$, and $\RHO^\vee=(2,1)$. Thus, $\CM_C$ should be the intersection of a nontrivial slice of type $(2,1)$ and the maximal nilpotent orbit in $\mathfrak{sl}_3^\C$. We find that the slice is parametrized as
\be \bm\sigma  =  \begin{pmatrix} \frac\varphi 2 & 0 & 1 \\ u^+ & \varphi & 0 \\
 -\frac34\varphi^2 + m_1^2+m_2^2+m_3^2 \;& -u^- & \frac\varphi2 \end{pmatrix}\,, \ee
where $u^\pm$ and $\varphi$ are the known generators of the abelian chiral ring.
Subject to the standard relations $u^+u^- =- (\varphi-m_1)(\varphi-m_2)(\varphi-m_3)$ and $m_1+m_2+m_3=0$, the matrix $\bm\sigma$ obeys
\be (\bm\sigma+m_1)(\bm\sigma+m_2)(\bm\sigma+m_3)=0\,. \ee
When the masses are zero, it lies in the maximal nilpotent orbit.

The quiver in the center of Figure \ref{fig:slice} is another abelian theory, with gauge group $U(1)\times U(1)$. This time, we expect that $\CM_C$ is the intersection of a slice labelled by $\RHO^\vee=(2,1,1)$ with the sub-maximal nilpotent orbit $\RHO^T=\nu = (3,1)$.
The chiral ring $\C[\CM_C]$ is generated as a Poisson algebra by $\varphi_1,\varphi_2$ and $u_1^\pm,u_2^\pm$ subject to $u_1^+u_1^- = - (\varphi-m_{1,1})(\varphi-m_{1,2})(\varphi_1-\varphi_2)$ and $u_2^+u_2^-=-(\varphi_2-m_2)(\varphi_2-\varphi_1)$. To generate $\C[\CM_C]$ as a ring, we need an additional pair of monopole operators $u_3^\pm:=\pm\{u_1^\pm,u_2^\pm\} = (u_1^\pm u_2^\pm)/(\varphi_1-\varphi_2)$. All these operators fit together into the slice
\be \bm\sigma = \begin{pmatrix} \frac{\varphi_1}2 +\frac{m_2}{4}& 0 & 0 & 1 \\
 - u_3^+ & - \varphi_2 +\frac{m_2}{4} & u^+_2 & 0 \\
 u_1^+ & u_2^- & -\varphi_1+\varphi_2 -\frac{3m_2}{4}& 0 \\
 -\frac34\varphi_1^2+\frac12(m_{1,1}^2+m_{1,2}^2+m_2^2) & u_3^- & - u_1^- & \frac{\varphi_1}{2} +\frac{m_2}{4}\end{pmatrix}\,. \ee
Having shifted the masses so that $m_{1,1}+m_{1,2}+m_2=0$, this matrix obeys
\be \begin{array}{c}
 \det(\bm\sigma-x) = \tfrac14(x+\tfrac34 m_2)^2(x+m_{1,1}-\tfrac14m_2)(x+m_{1,2}-\tfrac14m_2)\,, \\[.2cm]
 (\bm\sigma+\tfrac34 m_2)(\bm\sigma+m_{1,1}-\tfrac14m_2)(\bm\sigma+m_{1,2}-\tfrac14m_2)  = 0\,,
 \end{array}
\ee
as appropriate for a deformation of the nilpotent orbit of type $(3,1)$. Note that, since this quiver is partially balanced, the gauge symmetry is partially enhanced to $G_C=U(1)\times SU(2)$. The moment map for the $SU(2)$ part is sitting in the middle $2\times2$ block of $\bm\sigma$.

Finally, the nonabelian quiver on the right of Figure \ref{fig:slice} is fully balanced, so $G_C=SU(2)$; however, since $\dim_\C\CM_C=4$, the moment map is not sufficient to parametrize the Coulomb branch. Indeed, $\CM_C$ is the intersection of a nontrivial slice of type $\RHO^\vee=(2,2)$ with the maximal nilpotent orbit ($\nu=(4)$) inside $\mathfrak{sl}_4^\C$.%
\footnote{The Coulomb branch of this theory was studied in great detail using mirror symmetry in \cite{HM-intersections}.}

Since the theory in question is just $U(2)$ SQCD with two fundamental hypermultiplets, we already know from the scattering theory of Section \ref{sec:SQCD} how to generate its Coulomb branch: the generators are the operators
\be V^\pm = u_1^++u_2^+\,,\qquad W^\pm = u_1^+\varphi_2+u_2^+\varphi_1\,,\qquad \Phi_1 = \varphi_1+\varphi_2\,,\qquad \Phi_2 = \varphi_1\varphi_2\,, \ee
which appear as coefficients of the polynomials $Q(z) = z^2-\Phi_1 z+\Phi_2$ and $U^\pm(z) = V^\pm z - W^\pm$, and satisfy $Q(z)\wt Q(z)-U^+(z)U^-(z) = z^4$, or alternatively the abelianized relations
$u_a^+u_a^- = -{\varphi_a^4}/{(\varphi_1-\varphi_2)^2}.$ (Here we will set all masses to zero for simplicity.)
The auxiliary polynomial $\wt Q(z)=z^2+\Phi_1 z+V_{(1,1)}$, contains the monopole operator with adjoint cocharacter $V_{(1,-1)} = V^+V^-+\Phi_1^2-\Phi_2$. All these operators fit together in the slice (in slightly nonstandard form)
\be \bm\sigma = \left(\begin{array}{cc@{\;\;\;}c@{\;\;\;}c} \Phi_1 & V^+ & 1 & 0 \\
 U^- & -\Phi_1 & 0 & 1 \\[.1cm]
 -V_{(1,-1)} & W^+ & 0 & 0 \\[.1cm]
 W^- & -\Phi_2 & 0 & 0 \end{array}\right)\,, \ee
which is nilpotent as desired: the minimal polynomial is $\bm\sigma^4=0$. Note that the $G_C=SU(2)$ moment map sits in the upper $2\times 2$ block.

\subsection{General Nahm transform}
\label{sec:nahm}

In the previous section, we matched elements of transverse slices and Coulomb-branch operators by hand.
There is actually a general strategy one can employ to do so, building on the identification 
of the Coulomb branch as a moduli space of monopoles: we can simply do a Nahm transform 
of the holomorphic data 
\footnote{This is a holomorphic version of the Nahm transform that the one of the authors learnt from a beautiful paper he can no longer find. We would be delighted if the authors of this reference could make themselves known to us.}.
We first illustrate this for a single-node quiver ($M_1=N_c,\,N_1=N_f)$, and then explain the general process.

As the determinant of the rescaled scattering matrix $S(z)$ equals $P(z)$, 
$S(m_a)$ must have a null vector. This idea can be promoted to a statement about polynomial matrices: 
there must be a $2 \times N_f$ matrix $M(z)$ of polynomials of degree up to $N_f-1$, 
such that 
\begin{equation}
S(z) M(z) = 0 \;\;\mathrm{mod} \,P(z)\,.
\end{equation}
If the masses $m_a$ are distinct, we can determine $M(z)$ by, say, the constraint that 
\begin{equation}
S(m_a) M(m_a) = 0 \,.
\end{equation}
Thus $M(m_a)$ decomposes as the outer product of the null vector of $S(m_a)$ and 
an arbitrary vector with $N_f$ components. Given any choice of $N_f$ linearly independent such vectors, 
we can reconstruct $M(z)$ from the $N_f$ values $M(m_a)$. 

For example, we could set 
\be
M(z) = \sum_a  {- U^+(m_a)  \choose Q(m_a) } \otimes w_a \prod_{b \neq a} (z - m_b)
\ee
for a basis of vectors $w_a$ in $\C^{N_f}$. 
Without loss of generality we can take $w_a = (1,m_a,m_a^2,\cdots)$ 
so that 
\be
M(z) = \begin{pmatrix} - U^+(z) & - z U^+(z) & - z^2 U^+(z) & \cdots \cr Q(z) & z Q(z) & z^2 Q(z) & \cdots \end{pmatrix} \,\mathrm{mod} \,P(z)\,.
\ee
This expression remains valid for general values of the masses and we can multiply it by some $GL(N_f)$ matrix to get a general solution.  

We can write the relation between $S(z)$ and $M(z)$ more explicitly as
\begin{equation}
S(z) M(z) = P(z) \wt M(z)
\end{equation}
in order to define a second $2 \times N_f$ matrix $\wt M(z)$. 

As $\wt M(z)$ has precisely $N_f^2$ coefficients, it seems reasonable to use our $GL(N_f)$ gauge freedom to fix it to a given form. 
For example, 
\be
\wt M(z) = \begin{pmatrix} 1 & z & \cdots & z^{N_c-1} & 0 & \cdots &0 \cr 0 & 0 & \cdots & 0 & 1 & z & \cdots & z^{N_f - N_c-1} \end{pmatrix} 
\ee
We can do so, say, by the choice 
\be
M(z) = \begin{pmatrix} \wt Q(z) & z \wt Q(z) & \cdots & z^{N_c-1} \wt Q(z) & - U^+(z) & - z U^+(z) 
& \cdots & z^{N_f - N_c-1} U^+(z) \cr - U^-(z) & - z U^-(z) & \cdots & - z^{N_c-1} U^-(z) & Q(z) & z Q(z) 
& \cdots & z^{N_f - N_c-1} Q(z)\end{pmatrix} \,.
\ee

The matrix $z M(z) \,\mathrm{mod} \,P(z)$ also satisfies the same linear equations as $M(z)$. 
Thus it should be linearly related to $M(z)$:
\be
z M(z) = M(z) \bm\sigma \,\mathrm{mod} \,P(z)
\ee
for some $N_f \times N_f$ constant matrix $\bm\sigma$. 
If we write 
\be
z M(z) = M(z) \bm\sigma + P(z) \bm \tau
\ee
then we also find
\begin{equation}
 z \wt M(z) =\wt M(z) \bm\sigma + S(z) \bm \tau\,.
\end{equation}
With the gauge fixing above for $\wt M(z)$ we find 
\be
\bm \tau = \begin{pmatrix} 0 & 0 & \cdots & 1 & 0 & \cdots &0 \cr 0 & 0 & \cdots & 0 & 0 & 0 & \cdots & 1 \end{pmatrix}\,.
\ee

Expanding in powers of $z$ we have
\be
\bm \sigma_i^j = \begin{cases} \delta_{i+1,j} + Q^{(i)} \delta_{N,j} + U_+^{(i)} \delta_{N_f,j} & i<N \\
 \delta_{i+1,j} + U_-^{(i-N)} \delta_{N,j}  + \wt Q^{(i-N)} \delta_{N_f,j} & i \geq N\,, \end{cases}
\ee
where the upper indices in parenthesis indicate the coefficients of the polynomials in the scattering matrix. 
This is a slice transverse to the nilpotent orbit with Jordan blocks of size $N$ and $N_f - N$, 
though not presented quite in the standard form. 

For distinct $m_a$, it is obvious that $m_a - \bm\sigma$ has a left null vector $w_a$, and thus 
$\bm\sigma$ has eigenvalues $m_a$, as desired. If $P(z)$ has a root of degree $k$ at some $z=m$, 
we can expand $M(z)$ in powers of $(z-m)$ to find the generators of a size $k$ Jordan block for 
$\bm\sigma$. Thus $\bm\sigma$ belongs to the correct orbit. 
 
The extension to a general good linear quiver is rather straightforward. 
We can first rescale the scattering matrix $S(z) \in PGL(n+1,\C)$ by an appropriate power of the $P_i(z)$ to make it a polynomial matrix with%
\footnote{For the most direct comparison with Section \ref{sec:slices}, one should either take the inverse of the scattering matrix $S(z)^{-1}$, normalized such that $\det S(z)^{-1} = \prod_i P_i(z)^i$; or equivalently act everywhere with the Weyl reflection $-w_0$. Here we follow notation from Section \ref{sec:monopoles}.}
\be 
\textstyle \det S(z) = \prod_{i=1}^{n} P_i(z)^{n+1-i}  
\ee
and denote as $d_i$ the degrees of the diagonal elements of $S(z)$. 

Then we can write an equation 
\begin{equation}
S(z) M(z) = \det S(z) \wt M(z)
\end{equation}
where the degree of $M(z)$ is $p-1$ while the degree of the $i$-th row of $\wt M(z)$ is $d_i-1$.

We can gauge fix the $GL(p)$ ambiguity by fixing $\wt M(z)$ completely, 
with the $i$-th row non-zero only between the $(1+\sum_{j=1}^{i-1} d_j)$-th and the $(\sum_{j=1}^i d_j)$-th locations,
with increasing powers of $z$ from $1$ to $z^{d_i-1}$. 
Then $M(z)$ can be found simply by multiplying $\wt M(z)$ by the matrix of minors of $S(z)$, i.e. the rescaled scattering matrix $\gamma_1(z) S(z)$. 
 
We can then define $\bm \sigma$ by 
\be
z M(z) = M(z) \bm\sigma + \det S(z) \bm \tau
\ee
and thus
\begin{equation}
z \wt M(z) =\wt M(z) \bm\sigma + S(z) \bm \tau
\end{equation}
After gauge-fixing $\wt M(z)$ we find a simple $\bm \tau$, with a single element $1$ at the $(\sum_{j=1}^i d_j)$-th location of the $i$-th row. 
Thus $\bm\sigma$ takes again the form of a raising operator with Jordan blocks of size $d_i$, 
plus a transverse contribution controlled by $\tau$ and the coefficients of $S(z)$. 
In this manner, we can reproduce our previous examples.

\subsection{Quantization and Yangians}
\label{sec:quiv-Poisson}

We have seen that in many cases it is much easier to generate the chiral ring $\C[\CM_C]$ as a Poisson algebra rather than a ring. For example, for balanced quivers with trivial $\rho^\vee$ (Section \ref{sec:Trho}) the Poisson algebra is simply generated by $h_i$, $e_i$ and $f_i$ for $1 \leq i \leq n$.
In this section we present generators of the Poisson algebra for a general good quiver, and show how its quantization produces a central quotient of a (shifted) Yangian for $\mathfrak{sl}_{n+1}$. We follow the mathematical work \cite{KWWY}, which studied the Poisson structure and quantization of slices in the affine Grassmannian. (We identified such slices with $\CM_C$ for a good quiver in Section \ref{sec:Grass}).

\subsubsection{Poisson algebra revisited}

We revisit the Poisson algebra of a generic good quiver, describing its generators and relations in a completely uniform way  --- at the cost of working with infinitely many generators. 

Let us first focus on balanced quivers. We can introduce the generating functions
\be
H_i(z) = P_i(z) \prod_{j} Q_j(z)^{-\kappa_{ij}} \qquad E(z) = \frac{U^+_i(z)}{Q_i(z)} \qquad F(z) = \frac{U^-_i(z)}{Q_i(z)} \, ,
\label{eq:balancedgenerators}
\ee
which are to be thought of as formal power series in $z^{-1}$. At leading order they reproduce the generators $h_i$, $e_i$ and $f_i$ defined in~\eqref{hefdefinition}. The subleading terms in $H_i(z)$ contain polynomials $p_n(\varphi_i)$ with $n=(1,1,..,1,0,...,0)$. The subleading terms in $E_i(z)$ and $F_i(z)$ contain monopoles operators of fundamental magnetic weight at the $i$-th node, dressed with $\Tr(\varphi^k)$.

As formal power series in $z^{-1}$ and $w^{-1}$, the Poisson brackets of these generating functions take the  form
\be
\begin{aligned}
\{ H_i(z) , E_j(w) \} & = - \kappa_{ij} H_i(z) \frac{E_j(z)-E_j(w)}{z-w} \\
\{ H_i(z) , F_j(w) \} & =  \kappa_{ij} H_i(z) \frac{F_j(z)-F_j(w)}{z-w} \\
\{ E_i(z) , F_j(w) \} & = - \delta_{ij} \frac{H_i(z)-H_i(w)}{z-w} \,,
\end{aligned}
\label{eq:classicalyangian}
\ee
where $\kappa_{ij}$ is the Cartan matrix,
together with a multitude of Serre-like relations discussed in \cite{KWWY}.

A similar presentation of the Poisson algebra for a generic good quiver can be made. We can introduce the generating functions
\be
H_i(z) = z^{-\Delta_i}P_i(z) \prod_{j=1}^r Q_j(z)^{-\kappa_{ij}} \qquad E(z) = \frac{U^-_i(z)}{Q_i(z)} \qquad F(z) = \frac{U^+_i(z)}{Q_i(z)}
\ee
where the additional factor of $z^{-\Delta_i}$ compared to equation~\eqref{eq:balancedgenerators} ensures that the first generating function has an expansion of the form $H_i(z) = 1 + \mathcal{O}(z^{-1})$.

The Poisson brackets appearing in the first two lines of equations~\eqref{eq:classicalyangian} are unchanged for a generic good quiver. However, in order to express the Poisson bracket $\{ E_i(z),F_i(w)\}$ cleanly, one should introduce a shifted generator $J_{i}(z)$ defined as follows. If the expansion of $H_{i}(z)$ takes the form
\be
H_i(z) = \sum_{n=0}^\infty H_i^{(n)} z^{-n}
\label{Hdef}
\ee
then we define
\be
J_{i}(z) := \sum_{n=0}^\infty H_i^{(n+\Delta_i)} z^{-n}\, .
\label{Jdef}
\ee
The Poisson brackets can then be expressed
\be
\begin{aligned}
\{ H_i(z) , E_j(w) \} & = - \kappa_{ij} H_i(z) \frac{E_j(z)-E_j(w)}{z-w} \\
\{ H_i(z) , F_j(w) \} & =  \kappa_{ij} H_i(z) \frac{F_j(z)-F_j(w)}{z-w} \\
\{ E_i(z) , F_j(w) \} & = - \delta_{ij} \frac{J_{i}(z)-J_{i}(w)}{z-w} \, .
\end{aligned}
\label{eq:shiftedyangianpoisson}
\ee

\subsubsection{Quantization}
\label{sec:quiv-quant}

In the presence of an Omega background, the generators of the abelianized chiral ring obey (we omit `hats' for clarity)
\begin{align}
[\HAT\varphi_{i,a},\HAT\varphi_{i,b}] & =0 \cr
[\HAT\varphi_{i,b},\HAT u^\pm_{i,a}] & = \pm \, \epsilon \, \delta_{ab} \, \HAT u^\pm_{i,a} \cr
\HAT u^+_{i,a} \HAT u^-_{i,a} & = - \frac{P_i(\varphi_{i,a}  - \frac{\epsilon}{2} ) \HAT Q_{i-1}(\varphi_{i,a} - \frac{\epsilon}{2} ) \HAT Q_{i+1}(\varphi_{i,a} - \frac{\epsilon}{2} )}{\prod_{b \neq a}(\varphi_a - \varphi_b)(\varphi_a - \varphi_b - \epsilon )} \cr
\HAT u^-_{i,a} \HAT u^+_{i,a} & = - \frac{P_i(\varphi_{i,a} +\frac{\epsilon}{2} ) \HAT Q_{i-1}(\varphi_{i,a} +\frac{\epsilon}{2} ) \HAT Q_{i+1}(\varphi_{i,a} +\frac{\epsilon}{2} )}{\prod_{b \neq a}(\varphi_{i,a} - \varphi_{i,b})(\varphi_{i,a} - \varphi_{i,b} + \epsilon )} \, .
\label{eq:quiverquant}
\end{align}
 
Consider again a balanced quiver of rank $n$. We again define generators $h_i$, $e_i$ and $f_i$ (with $i =1,\ldots,n$) as in equation~\eqref{hefdefinition}. It is then straightforward to check as a consequence of the relations~\eqref{eq:quiverquant} that
\be
\begin{aligned}
\left[ h_i,e_j \right]&= \epsilon \kappa_{ij}e_j \\
\left[ h_i,f_j \right] &=-\epsilon \kappa_{ij}f_j \\
\left[ e_i,f_j \right] &=\epsilon \delta_{ij}h_i \\
\mathrm{ad}(e_i)^{1-\kappa_{ij}} e_j &=0 \quad i \neq j\\
\mathrm{ad}(f_i)^{1-\kappa_{ij}} f_j &=0 \quad i \neq j
\end{aligned}
\ee
where now $ad(x)y = \left[ x,y \right]$. These are the defining equations of $\mathfrak{sl}_{n+1}$ in the Chevalley-Serre basis. The algebra of Coulomb branch operators in the Omega background provides a representation of the universal enveloping algebra $U(\mathfrak{sl}_{n+1})$. 

To include dressed monopole operators, we will introduce the generating functions
\be
\begin{aligned}
H_i(z) & = P_i(z)\frac{\HAT Q_{i-1}(z-\frac{\epsilon}{2}) \HAT Q_{i-1}(z+\frac{\epsilon}{2}) }{\HAT Q_i(z) \HAT Q_i(z+\epsilon)} \\
E(z) & = \frac{1}{\HAT Q_i(z)}\HAT U^-_i(z) = \sum_{a=1}^{M_i} \frac{1}{z- \HAT \varphi_{i,a} } \HAT u^-_{i,a} \\
F(z) & = \HAT U^+_i(z) \frac{1}{\HAT Q_i(z)} = \sum_{a=1}^{M_i} \frac{1}{z- \HAT \varphi _{i,a} + \epsilon} \HAT u^+_{i,a}
\end{aligned}
\label{gfbalanced}
\ee
deforming those defined in~\eqref{eq:balancedgenerators}. In particular, we must now pay careful attention to the ordering of the operators in the definitions of the generating functions $E(z)$ and $F(z)$. We have checked by expanding in $z^{-1}$ and $w^{-1}$ that the commutation relations of these generators are given by
\be
\begin{aligned}
\left[ H_i(z) , E_j(w) \right] & = - \frac{\epsilon}{2} \kappa_{ij} \frac{ \left[ H_i(z) , E_j(z)-E_j(w) \right]_+}{z-w} \\
\left[ H_i(z) , F_j(w) \right] & =  - \frac{\epsilon}{2} \kappa_{ij} \frac{ \left[ H_i(z) , F_j(z)-F_j(w) \right]_+ }{z-w} \\
\left[ E_i(z) , F_j(w) \right] & = - \epsilon \delta_{ij} \frac{H_i(z)-H_i(w)}{z-w} \, .
\end{aligned}
\label{eq:yangian}
\ee
together with Serre-like relations, where $[x,y]_+ := x y + y x$ is defined as the anti-commutator. 

The commutation relations in \eqref{eq:yangian} are those of the Yangian $Y(\mathfrak{sl}_{n+1})$. The representation of these commutation relations found by quantizing a moduli space of monopoles was introduced in~\cite{GKLO-Yangian}, where a proof of~\eqref{eq:yangian} can be found. As pointed out in~\cite{KWWY}, this is in fact a representation of a central quotient of the Yangian labelled by the weight $\nu$ of a balanced quiver.

For a generic good quivers, we define
\bea
H_i(z) = z^{-\Delta_i} P(z)\frac{\HAT Q_{i-1}(z-\frac{\epsilon}{2}) \HAT Q_{i-1}(z+\frac{\epsilon}{2}) }{\HAT Q_i(z) \HAT Q_i(z+\epsilon)}
\eea
with $E_i(z)$ and $F_i(u)$ defined as in equation~\eqref{gfbalanced}. To express the commutators cleanly, one must again define a shifted generator $J_{i}(z)$ as defined in equations~\eqref{Hdef} and~\eqref{Jdef}. Expanding in $z^{-1}$ and $w^{-1}$ one checks order by order that
\be
\begin{aligned}
\left[ H_i(z) , E_j(w) \right] & = - \frac{\epsilon}{2} \kappa_{ij} \frac{ \left[ H_i(z) , E_j(z)-E_j(w) \right]_+}{z-w} \\
\left[ H_i(z) , F_j(w) \right] & =  - \frac{\epsilon}{2} \kappa_{ij} \frac{ \left[ H_i(z) , F_j(z)-F_j(w) \right]_+ }{z-w} \\
\left[ E_i(z) , F_j(w) \right] & = - \epsilon \delta_{ij} \frac{J_{i}(z)-J_{i}(w)}{z-w} \, .
\end{aligned}
\label{eq:shiftedyangian}
\ee
A proof of these commutation relations can be found in \cite{KWWY}, where it is also explained that they coincide with a central quotient of a `shifted' Yangian of $\mathfrak{sl}_{n+1}$ (where the shifting is labelled by the non-zero weight $\mu$), extending the constructions of~\cite{GKLO-Yangian}.%
\footnote{Shifted Yangians were first introduced and studied in~\cite{BrundanKleshchev-YangW,BrundanKleshchev-reps}. The shifted Yangians constructed there arise in our construction from the theories $T^\rho[U(n+1)]$ of Section \ref{sec:Trho}, \ie\ are balanced theories with trivial $\rho^\vee$, whose Coulomb branches are isomorphic to nilpotent orbits.
It is known that such shifted Yangians are isomorphic to finite W-algebras $W(\rho)$ \cite{dBT-W}.
The more general shifted Yangians that arise in the case of generic good quivers were introduced in~\cite{KWWY}.}

\subsection{Twistor space}
\label{sec:twistor-quiver}

The twistor space for the Coulomb branch of a generic rank $n$ linear quiver is given by a straightforward extension of the construction in the case of SQCD. The transition functions for the polynomials $Q_j(z)$, $U^\pm(z)$ are given by
\be
\begin{aligned}
\wt Q_{j,\tilde\zeta}( \tilde z_{\zeta} ) & = \zeta^{-2M_j} Q_{j,\zeta}(z_\zeta) \\
\tilde U^\pm_{j,\tilde \zeta} (\tilde z_{\tilde \zeta}) & = \zeta^{-2M_j - \Delta_j} e^{\pm \frac{z_\zeta}{g^2\zeta}} \, U^\pm_{j,\zeta}(z_\zeta) \quad \mathrm{mod} \quad Q_{j,\zeta}(z_\zeta) \, .
\label{polytransfquiver}
\end{aligned}
\ee
for $j=1,\ldots,n$, together with $\wt z_{\wt \zeta} = \zeta^{-2}z_\zeta$. These transition functions define a pair of rank $M_j$ vector bundles $V^\pm(2M_j+\Delta_j) \to Y_j $ at the $j$-th node of the quiver, where $Y_j = \oplus_{l=1}^j \mathcal{O}(2l)$. The constraints are promoted to
\be
U^+_\zeta(z_\zeta) U^-_\zeta(z_\zeta) = Q_{j-1,\zeta}(z_\zeta) Q_{j+1,\zeta}(z_\zeta) P_\zeta(z_\zeta) \quad \mathrm{mod} \quad Q_\zeta(z_\zeta) \, .
\label{quiverconstraint}
\ee
To find a complete description of the twistor space, one can also easily write down the transition functions for all components and minors of the scattering matrix.

\acknowledgments{We are grateful to C. Beem, K. Costello, J. Hilburn, G. Moore, H. Nakajima, T. Nevins, N. Seiberg, B. Webster, and E. Witten for many insightful discussions during the preparation of this paper.
Part of this work was developed during visits of TD to the Perimeter Institute.
Research at Perimeter Institute is supported by the Government of Canada through Industry Canada and by the Province of Ontario through the Ministry of Economic Development \& Innovation.
TD is supported by DOE grant DE-SC0009988, and also in part by ERC Starting Grant no. 335739 ``Quantum fields and knot homologies,'' funded by the European Research Council under the European Union's Seventh Framework Programme.
MB gratefully acknowledges support from IAS Princeton through the Martin A. and Helen Choolijan Membership.}

\appendix

\section{Singular monopoles}
\label{app:monopoles}

In this appendix we review some aspects of moduli spaces of singular monopoles, and explain how to describe them as complex symplectic manifolds. Namely, we construct the scattering matrix that encodes their complex structure, and derive the Poisson bracket in terms of the scattering matrix.

\subsection{Data}

We consider moduli spaces of solutions to the Bogomolnyi equations $F = * D \, \Phi$ on $\mathbb{R}^3$ for a connection $D$ in a $G$-bundle and an adjoint valued scalar field $\Phi$. Under favorable conditions, these moduli spaces are hyperk\"ahler manifolds. We are mainly interested in solutions of these equations in the presence of singular monopoles. The relevant moduli spaces have been defined in reference \cite{MRB-monopole} (see also \cite{MRB-monopole2}). Let us briefly review the data that define such a moduli space:

\begin{enumerate}
\item A number of singular monopoles each labelled by a position $\vec x_a \in \mathbb{R}^3$ and an element $q_a$ of the cocharacter lattice $\Lambda_G := \mathrm{Hom}(U(1),\mathbb T_G) \subset \mathfrak t_G$, defined up to Weyl transformations. The charge $q_a$ defines an embedding of an abelian Dirac monopole into the gauge group $G$, specifying the singular boundary conditions
\be
\Phi = - \frac{q_a}{2 r_a} + \CO(r_a^{-1/2})\qquad \vec x\to \vec x_a\,,
\label{bc1}
\ee
where $r_a = | \vec x - \vec x_a|$. The coordinates $\vec x_a$ are a hyperk\"ahler triplet of deformation parameters for the moduli space.
\item A magnetic charge $q_\infty$ and asymptotic value $\Phi_\infty$ valued in the Cartan subalgebra and defined up to Weyl transformations. This data determines the asymptotic behavior of solutions to be
\be
\Phi = \Phi_\infty - \frac{q_\infty}{2 r} + \CO(r^{-1-\delta}) \qquad r \to \infty
\label{bc2}
\ee
for any $\delta >0$ where $r=|\vec x|$. We will always assume that the centralizer of $\Phi_\infty$ is the maximal torus $T \subset G$. Global consistency requires that the combination $q_\infty - \sum_a q_a$ is an element of the coroot lattice.
\end{enumerate} 

The dimension of the moduli space has been computed in \cite{MRB-monopole}. The asymptotic value $\Phi_\infty$ defines a system of positive roots such that $\alpha \in \Delta^+$ iff $\langle \alpha , \Phi_\infty \rangle > 0$. We can then define $q^-_a$ to be the unique element of the Weyl orbit of $q_a$ such that $\langle \alpha ,q^-_a \rangle < 0$ for all $\alpha\in \Delta^+$. The complex dimension of the moduli space is $2 \langle \rho , \tilde q_\infty \rangle$ where $\rho$ is the Weyl vector and $\tilde q_\infty := q_\infty - \sum_a q_a^-$ is called the relative magnetic charge.

We will now focus on the gauge group $G = PSU(N)$. To simplify notation, we use the Cartan-Killing form $(-,-)$ to identify the Lie algebra and its dual. The fundamental weights are denoted $\{ \omega_j \}_{j=1}^{N-1}$ and simple positive roots are $\{ \alpha_j \}_{j=1}^{N-1}$. We then have the inner products $(\omega_i , \alpha_j)=\delta_{ij}$ and $(\alpha_i,\alpha_j) = \kappa_{ij}$, where $\kappa_{ij}$ is the Cartan matrix. When required, we will choose a representation by $N\times N$ anti-hermitian traceless matrices where $(a,b)=-\mathrm{Tr}(a b)$. Without loss of generality, we can take the asymptotic scalar to have the form $\Phi_\infty = \mathrm{diag}(i\phi_1,\ldots,i\phi_N)$ where $\sum_i \phi_i = 0$ and $\phi_i > \phi_{i+1}$. The positive simple roots with respect to $\Phi_\infty$ are then the matrices $\alpha_j= i(E_{i,i} - E_{i+1,i+1})$.

For $G = PSU(N)$, the cocharacter lattice is equal to the coweight lattice.  We will consider solutions with $N_j$ Dirac singularities labelled by each fundamental weight $\omega_{N-j}$. We can then identify $\sum_a q_a =  \sum_j N_j \omega_{N-j}$ and hence 
\be
\sum_a q^-_a =  - \sum_j N_j \omega_j
\ee
where we used the formula $w_0 \, \omega_j = - \omega_{N-j}$ to reflect the magnetic weight of each singular monopole into the negative Weyl chamber, with $w_0$ being the longest element of the Weyl group. The overall relative magnetic weight is an element of the root lattice,
\be
\tilde q_{\infty} =  \sum_j M_j \alpha_j\, .
\ee
It is expected that the moduli space is non-empty provided that all $M_j\geq0$. In relating the moduli space of singular monopoles with the Coulomb branch of a three-dimensional linear quiver gauge theory, $M_j$ is the rank of the $j$-th node and $N_j$ the number of hypermultiplets at this node.  In the notation of Section~\ref{sec:quiver}, we have $\nu = -\sum_a q_a^-$ and $\mu=-q_\infty$ which agrees with the expected dimension $2\langle \lambda - \mu,\rho\rangle$ of the Coulomb branch.

\subsection{Hyperk\"ahler quotient}

The moduli space can be described as an infinite-dimensional hyperk\"ahler quotient as follows. We first introduce a flat hyperk\"ahler structure on the configuration space of fields $(A,\Phi)$ obeying the boundary conditions \eqref{bc1} and \eqref{bc2} with metric
\be
g = - \Tr \int_{\mathbb{R}^3 - \{x_a\}} d^3x \left( \delta A_i \otimes \delta A_i + \delta \Phi \otimes \delta \Phi \right)
\ee
and K\"ahler forms
\be
\omega_i = \Tr \int_{\mathbb{R}^3 - \{x_a\}} d^3x \left( \delta\Phi \wedge \delta A_i  + \frac{1}{2}\epsilon_{ijk} \delta A_j \wedge \delta A_k  \right) \, .
\label{kahlerforms1}
\ee

This is acted upon by the group $\CG$ of local gauge transformations that are trivial on the sphere at infinity $|\vec x|\to\infty$ and leave $q_a$ invariant at each singular point $x_a$. Although the moduli space depends only on the Weyl orbit of $q_a$ it is convenient to fix a representative and consider gauge transformations that leave it invariant. Local gauge transformations act as $\delta A_i = - D_i \lambda$, $\delta \Phi = [\lambda,\Phi]$. Contracting the vector field generating this local gauge transformation with the k\"ahler forms \eqref{kahlerforms1} and integrating by parts we find the corresponding moment maps
\be
\mu_i[\lambda] =\mathrm{Tr} \int d^3 x \, \lambda \left(  \frac{1}{2} \epsilon_{ijk} F_{jk} -  D_i \Phi \right) \, .
\ee
In fact, the integration by parts produces a boundary term $ \int d^2 S_i \, \mathrm{Tr}(\lambda \Phi)$ which could receive contributions from the singular points $\vec x_a$ and the sphere at infinity $|\vec x| \to \infty$. These contributions vanish since $\Phi \to 0$ at each $\vec x_a$ and we restrict to local gauge transformations $\lambda$ acting trivially at infinity. Thus the vanishing of the moment maps indeed imposes the Bogomolnyi equations and the moduli space is an infinite-dimensional hyperk\"ahler quotient.

\subsection{Scattering data}

To describe the moduli space of singular monopoles as a complex symplectic manifold we can instead consider a holomorphic symplectic quotient. We choose a particular complex structure on the configuration space of fields $(A,\Phi)$ such that the holomorphic coordinates are $\CA_{\bar z} = A_1+iA_2$ and $\CA_3 = A_3 - i \Phi$ (we denote the corresponding covariant derivatives by $\CD_{\bar z}$ and $\CD_3$). Then
\be
\Omega := \omega_1+i\omega_2 = i \, \mathrm{Tr} \int d^3 x \, \delta\CA_3  \wedge \delta\CA_{\bar z}
\ee
is a holomorphic symplectic form in this complex structure. We now consider complexified local gauge transformations acting on these complex coordinates by $\delta \CA_{\bar z} = - \CD_{\bar z} \lambda$ and $\delta \CA_{t} = - \CD_3 \lambda$ with moment map
\be
\mu=\mu_1+i\mu_2=i \, \mathrm{Tr} \int d^3 x \, \left[ \CD_3 , \CD_{\bar z} \right] \, .
\ee
We construct the moduli space of singular monopoles as a complex symplectic manifold by imposing the complex moment map condition $[ \CD_{t} , \CD_{\bar z} ]=0$ and taking the quotient by the group of local complexified gauge transformations.

We \emph{assume} here that Donaldson's theorem \cite{Donaldson-scattering} generalizes to the case of singular monopoles: namely, that the points of the hyperk\"ahler quotient are in 1-1 correspondence with points of the holomorphic symplectic quotient (subject to an appropriate stability condition), and hence both describe the same moduli space. It would be interesting to prove this.

We want to provide an explicit description of the moduli space by reformulating it in terms of holomorphic `scattering data'. We first review the case of smooth monopoles without singularities. The complex moment map can be solved locally by finding group-valued functions $g(\vec x)$ such that
\be
\mathcal{D}_{\bar z} = g \, \partial_{\bar z} \, g^{-1}\,, \qquad \mathcal{D}_{t} = g \, \partial_{t} \, g^{-1} \, .
\label{triv}
\ee
The function $g(\vec x)$ is determined locally up to right multiplication by a group-valued function holomorphic in $z$. We consider two sets of distinguished solutions  $g_\pm(\vec x) $ that tend towards `abelian' solutions $g^\infty_\pm$ as $t \to \pm \infty$, which are determined by the asymptotic behavior of $(\Phi,A)$. The holomorphic scattering data $S(z)$ is then the transition function relating these distinguished solutions.

We explain this in more detail for $G=PSU(N)$. We can construct a group-valued function $g(\vec x)$ solving the moment map condition locally from $N$ independent solutions $\ell_1,\ldots,\ell_N$ of the associated linear problem $\mathcal{D}_{\bar z}\,\ell = \mathcal{D}_t \,\ell = 0$, normalized such that $\langle \ell_1, \ldots,\ell_N\rangle=1$. Note that here we are identifying the complexified gauge group as $G_{\mathbb{C}}=SL(N,\mathbb{C})/\mathbb{Z}_N$. We are free to consider holomorphic linear transformations of $\ell_1,\ldots,s_\ell$ preserving the normalization condition, which corresponds to right multiplication of $g(\vec x)$ by holomorphic group-valued functions. Therefore we may equivalently work with normalized sections of the associated linear problem.

We now want to consider distinguished sets of solutions $\ell^\pm_1,\ldots,\ell_N^\pm$ with asymptotic behavior at $x^3\to \pm\,\infty$ specified by the boundary condition \eqref{bc2}. Let us focus on $x^3 \to \infty$. We can choose an asymptotic gauge where $\CA_3\to -i (\Phi_\infty - q_\infty/2x^3)$. Then we have asymptotic solutions
\be
\ell^{+,\infty}_a  = e^{-\phi_a t} t^{q_a/2} e_a
\ee
where $e_a$ is the standard unit basis of $\mathbb{C}^N$ and we write $\Phi_\infty=\mathrm{diag}(i\phi_1,\ldots,ia_N)$ and $q_\infty = (i q_1,\ldots,iq_N)$. We can now find $N$ solutions to the linear problem $\ell_1^+,\ldots,\ell^+_N$ with this asymptotic behavior $\ell^+_a \to \ell^{+,\infty}_a$ as $x^3 \to \infty$. Since $\phi_1 > \phi_2 >\cdots > \phi_N$, $s^+_1$ is unambiguously determined as the section with the fastest exponential deacy at $x^3\to\infty$. However, $s^+_2$ is determined only up to adding $s^+_1$ multiplied by an arbitrary function of $z$. More generally, the sections $\ell_a^+$ are determined only up to $\ell^+_a \to {h_a}^b \, \ell_b^+$ where $h$ is a lower triangular matrix with unit diagonal and holomorphic dependence on $z$. Similarly we have sections $\ell^-_a$ defined up to $\ell^-_a \to {h_a}^b \, \ell_b^-$ with $h$ is now an upper triangular matrix with unit diagonal.

The two sets of solutions are related by $\ell_a^+ = {S_a}^b(z) \ell^-_a$ where the matrix $S(z)$ has unit determinant due to the normalization $\langle \ell_1^\pm, \ldots, \ell_N^\pm\rangle = 1$ and depends holomorphically on $z$. From the indeterminacy of the sections $\ell_a^\pm$, $S(z)$ is defined only up to multiplication on the left by holomorphic lower-triangular matrices with unit diagonal, and on the right by holomorphic upper-triangular matrices with unit diagonal. The leading principal minors $S^{1,\ldots,i}_{1,\ldots,i}(z)$ are invariant under these transformations and encode the magnetic charge $q_\infty$ as follows: $S^{1,\ldots,i}_{1,\ldots,i}(z)=Q_i(z)$ is a monic polynomial degree $M_i$. In particular,
\be
S^{1,\ldots,i}_{1,\ldots,i}(z) \to z^{(q_\infty, \omega_i)} \qquad |z| \to \infty\, .
\ee
where since there are no singular monopoles $q_\infty = \sum_j M_j \alpha_j$.

As described in the main text (see Sections \ref{sec:pureUN} and \ref{sec:monopoles}), the redundancy by left and right multiplication can be fixed by specifying the degrees of further non-principal minors. The moduli space is then generated as a complex variety by the coefficients of the components of the matrix $S(z)$. The holomorphic Poisson bracket is described below in Section \ref{sec:monopole-poisson}.

In the presence of singular monopoles the above construction is modified slightly. As described above, we consider for the case $G=PSU(N)$ that we have $N_j$ singularities of fundamental weight $\omega_{N-j}$ so that $\sum_a q_a^{-} = -\sum_j N_j \omega_j$ and $q_\infty = \sum_jM_j\alpha_j-\sum_j N_j \omega_j$. We denote the positions of the singularities by $m_{i,\alpha}$ with $a=1,\ldots,N-1$ and $\alpha=1,\ldots,N_j$ and introduce monic polynomials $P_j(z) = \prod_{\alpha=1}^{N_j}(z-m_{j,\alpha})$. Then we conjecture that the leading principle minors have the form
\be
S^{1,\ldots,i}_{1,\ldots,i}(z) = \frac{Q_i(z)}{\prod_j P_j(z)^{\kappa^{-1}_{ij}}} \, .
\label{pminor-fixed}
\ee
where $Q_i(z)$ is again a monic polynomial of degree $M_i$. In particular, note that
\be
\begin{aligned}
S^{1,\ldots,i}_{1,\ldots,i}(z) & \to z^{(q_\infty, \omega_i)} \qquad && z \to \infty \\
S^{1,\ldots,i}_{1,\ldots,i}(z) & \to (z-z_a)^{(q_a^{-}, \omega_i)}\qquad && z \to z_a \, .
\end{aligned}
\ee
The redundancy of $S(z)$ by left and right multiplication can again be fixed by specifying the form of further non-principle minors, as described in Sections \ref{sec:SQCDmatter} and \ref{sec:monopoles}.

Above, we implicitly viewed the complexified gauge group as $G_{\mathbb{C}} = SL(N,\mathbb{C}) / \mathbb{Z}_N$. In particular, the fractional powers appearing in the formula \eqref{pminor-fixed} are only well defined in the quotient by $\mathbb{Z}_N$. It is often more convenient to work with the equivalent $G_{\mathbb{C}}=PGL(N,\mathbb{C})/\mathbb{C^*}$ in order to remove the algebraic functions in the denominator of \eqref{pminor-fixed}. Indeed, using the $\mathbb{C}^*$ freedom, we can multiply the scattering matrix $S(z)$ by the factor $\prod_j P_j(z)^{\kappa^{-1}_{1,j}}$. Then principal minors then have the form
\be
S^{1,\ldots,i}_{1,\ldots,i}(z) = Q_i(z) \prod_{j=1}^iP_j(z)^{i-j}
\ee
and in particular $\det S(z) = \prod_{j=1}^{N-1} P_j(z)^{N-j}$. After fixing the redundancy by left and right multiplication, the components of the scattering matrix are then polynomial functions of $z$.

\subsection{Poisson structure}
\label{sec:monopole-poisson}

The complex Poisson bracket on the monopole moduli space can be computed by lifting the holomorphic functions on the monopole moduli space to 
functionals of $\CD_3$, $\CD_{\bar z}$ (unconstrained by $[\CD_3, \CD_{\bar z}]=0$) and computing the Poisson bracket with the flat complex symplectic form above. 

Here we are concerned with explaining why the Poisson bracket for the scattering data $S(\sigma)$ takes the form 
\begin{equation}
\{ S^a_b(z), S^c_d(w) \} = \frac{S^c_b(z) S^a_d(w)  - S^a_d(z) S^c_b(w) }{z - w}
\label{smatrixyangian}
\end{equation}
which is reminiscent of the Yangian of $\mathfrak{sl}(N,\mathbb{C})$. This formula is true only if properly understood: $S(z)$ is not 
well-defined by itself, but only up to multiplication by lower and upper triangular matrices with unit diagonal from the left and from the right. The Poisson bracket only works properly for gauge invariant functionals of $S(z)$.

At first sight, \eqref{smatrixyangian} seems paradoxical: the scattering data for $\CD_3$ is computed from the parallel transport 
along the $x^3$ axis, \ie\ using only the holomorphic coordinate~$\CA_3$. Naively, this gives a lift of $S(z)$ to a 
functional of $\CA_3$ only, but then the Poisson bracket with itself would be zero. 

The crucial subtlety is that the ``bare'' scattering data for $\CD_3$ only takes the form of a holomorphic 
function of $z$ with a precise choice of holomorphic framing. In other words, the scattering data itself 
is ambiguous by multiplication by triangular matrices with a generic dependence on $z$ and it becomes holomorphic only if 
such ambiguity is fixed appropriately. That introduces a dependence on $\CA_{\bar z}$ which allows a non-trivial 
Poisson bracket. 

One can consider quantities which are invariant under multiplication by triangular matrices, such as the leading principal minors:
determinants $Q_i(z)$ of the $i \times i$ minors built from the first $i$ rows and columns of $S(z)$, 
which can be normalized to be monic polynomials in $z$. The coefficients of the polynomials $Q_i(z)$ 
lift to functionals of $\CA_3$ only and should therefore Poisson commute. This is indeed consistent with the Poisson bracket \eqref{smatrixyangian}: labelling minors by multi-indices we have
\begin{multline}
\{ S^{a_1, \cdots a_i}_{b_1,\cdots b_i}(z), S^{c_1, \cdots c_j}_{d_1, \cdots, d_j}(w) \} = \\\sum_{m,n} \frac{S^{a_1, \cdots, c_n, \cdots, a_i}_{b_1,\cdots b_i}(z) S^{c_1, \cdots, a_m, \cdots c_j}_{d_1, \cdots, d_j}(w)  - S^{a_1, \cdots a_i}_{b_1,\cdots,d_n,\cdots b_i}(z) S^{c_1, \cdots c_j}_{d_1, \cdots,b_m, \cdots, d_j}(w) }{z - w} \, ,
\end{multline}
and in particular
\begin{equation}
\{ Q_i(z), Q_j(w) \} =0 \, .
\end{equation}
It can be shown further that the coefficients of the $D_i(\sigma)$ give a maximal set of Poisson-commuting functions on the moduli space. 

The notion of holomorphic framing can be formalized with the help of holomorphic Wilson lines in the $\sigma$ plane for $\CD_{\bar z}$. Let us first recall that the standard Wilson loop for $\CD_3$ along the $x^3$ direction at fixed $z$, $\bar z$ is
\begin{equation}
W_{\R}(x^3_f,z, x^3_i) = \mathrm{P\,exp}\int_{x^3_i}^{x^3_f} \CA_3(x^3,z, \bar z) dx^3 \equiv \sum_{n=0}^{\infty} \prod_{m=1}^n \int_{x^3_{m-1}}^{x^3_{m+1}} dx^3_m \overleftarrow{\prod_{m=1}^n}  \CA_3(x^3_m,z, \bar z)
\end{equation}
where in the $n$-th term of the sum we have $x^3_0 = x^3_i$ and $x^3_{n+1} = x^3_f$.

The Wilson loop is built in such a way to satisfy $\CD_{x^3_i} W_{\R}(x^3_f,z, x^3_i)=0$ and 
 $\CD_{x^3_f} W_{\R}(x^3_f,z, x^3_i)=0$ where the covariant derivatives act on the right and on the left respectively.
It is also gauge-covariant: gauge transformations of $\CA_3$ act on $W_{\R}(x^3_f,z, x^3_i)$
by gauge transformations at the endpoints.  Furthermore, it obeys the concatenation property $W_{\R}(x^3_f,z, x^3)W_{\R}(x^3,z, x^3_i)=W_{\R}(x^3_f,z, x^3_i)$. 

Another extremely useful property is that under small deformations $\delta \CA_3$ we have
\be
\delta W_{\R}(x^3_f,z, x^3_i) = \int^{x^3_f}_{x_i^3} dx \, W_{\R}(x^3_f,z, x) \, \delta\CA_3(x,z,\bar z) \, W_{\R}(x ,z, x^3_i) \, .
\label{var1}
\ee
Using this property and integrating by parts once it is straightforward to show that
\be
\begin{aligned}
\CD_{\bar z} W_{\R}(x^3_f,z, x^3_i) & := \partial_{\bar z}W_{\R}(x^3_i,z, x^3_f) + \CA_{\bar z}(x^3_f,z,\bar z) W_{\R}(x^3_i,z, x^3_f) - W_{\R}(x^3_i,z, x^3_f) \CA(x^3_i,z,\bar z) \\
& = \int^{x^3_f}_{x_i^3} W_{\R}(x^3_f,z, x ) \left[ \CD_{\bar z} , \CD_3 \right](x,z,\bar z) W_{\R}(x,z, x^3_i)
\end{aligned}
\ee
which vanishes due to the complex moment map condition. Thus the Wilson line $W_{\R}(x^3_i,z, x^3_f)$ 
is locally covariantly holomorphic in $z$. 

The holomorphic Wilson line operator for $\CD_{\bar z}$ at fixed $x^3$ is 
\begin{equation}
W_{\C}(z_f, x^3, z_i) = \sum_{n=0}^{\infty}  \int \prod_{m=1}^n d z_m d \bar z_m \frac{(z_f - z_i)}{\prod_{m=0}^n (z_{m+1} - z_m)}\overleftarrow{\prod_{m=1}^n}  \CA_{\bar z}(x^3,z_m, \bar z_m)
\end{equation}
where in the $n$-th term of the sum we have $z_0=z_i$ and $z_{n+1}=z_f$.

The holomorphic Wilson line operator is built to satisfy $\CD_{\bar z_i} W_{\C}(z_f, x^3, z_i)=0$ and $\CD_{\bar z_f} W_{\C}(z_f, x^3, z_i)=0$ where the covariant derivatives act on the right and on the left respectively. 
Furthermore, we have gauge covariance and the concatenation property $W_{\C}(z_f, x^3, z)W_{\C}(z, x^3, z_i)=W_{\C}(z_f, x^3, z_i)$.
The holomorphic Wilson loop is essentially a way to introduce a global holomorphic frame on the $z$ plane.

Under small variations $\delta\CA_{\bar z}$ we have
\be
\delta W_{\C}(z_f, x^3, z_i) = \int dz \, d\bar z\, \frac{(z_f-z_i)}{(z_f-z)(z-z_i)} W_{\C}(z_f, x^3, z) \delta \CA_{\bar z}(x^3,z,\bar z)  W_{\C}(z, x^3, z_i) \, .
\label{var2}
\ee
Using this formula and integrating by parts we find
\be
\begin{aligned}
\CD_3 W_{\C}(z_f, x^3, z_i) & := \partial_3 W_{\C}(z_f, x^3, z_i)+  \CA_3(x^3;z_f, \bar z_f) W_{\C}(z_f, x^3, z_i)  - W_{\C}(z_f, x^3, z_i) \CA_3(x^3;z_i, \bar z_i) \\
& = \int dz \, d\bar z\, \frac{(z_f-z_i)}{(z_f-z)(z-z_i)} W_{\C}(z_f, x^3, z ) \left[ \CD_3 , \CD_{\bar z} \right](x,z,\bar z) W_{\C}(z, x^3, z_i)
\end{aligned}
\ee
which again vanishes on imposing the complex moment map. Thus the holomorphic Wilson line $W_{\C}(z_f, x^3, z_i)$ is locally covariantly constant in the $x^3$ direction.

When defined on a $\cp^1$, the holomorphic Wilson loop covariant under gauge transformations which are well-defined
on $\cp^1$. When defined on the complex plane $\sigma$, the holomorphic Wilson loop will not be 
gauge covariant under gauge transformations which are non-trivial at infinity. 
Such gauge transformations are crucial for the definition of the monopole moduli space: 
the monopole solutions have boundary conditions at infinity given by 
a constant abelian vev for $\CA_3$ plus an abelian monopole charge. As usual, to avoid Dirac strings the abelian monopole connection is 
defined on the two half-spaces, glued by a gauge transformation which is non-trivial at infinity. 

Although the holomorphic Wilson loop is locally covariantly constant as a function of $x^3$, 
the non-trivial gauge transformations required in a monopole background will introduce extra jumps 
as one parallel-transports it from $x^3\ll 0$ to $x^3\gg 0$. 

Now we are ready to introduce the object which computes the parallel transport for the operator $\CD_3$ 
in a globally holomorphic frame and thus can be used to compute the scattering data $S(\sigma)$: 
\begin{equation}
W_{z_f, x^3_f;z_i, x^3_i}(z)= W_{\C}(z_f, x_f^3, z) W_{\R}(x^3_f,z, x^3_i)W_{\C}(z, x_i^3, z_i)
\end{equation}

If $[\CD_3, \CD_{\bar z}]=0$, in the absence of monopole charge, one could just 
parallel transport the holomorphic Wilson loops to a common position, 
deforming the above expression to a $z$-independent quantity 
$W_{\R}(x^3_f,z_f, x^3)W_{\C}(z_f, x^3, z_i) W_{\R}(x^3,z_i, x^3_i)$.

In the presence of monopole charge (and/or Dirac singularities), the deformation is not possible, and 
$W_{z_f, x^3_f;z_i, x^3_i}(z)$ is an holomorphic function of $z$. 
If we take $x^3_f \gg 0$ and $x^3_i \ll 0$, send $z_{i,f}\to \infty$ and regularize $W_{z_f, x^3_f;z_i, x^3_i}(z)$ in a standard way
in the limit, we obtain the scattering data $S(z)$.
 
The crucial observation to compute the Poisson brackets postulated above is that Poisson brackets between 
Wilson lines only get a contribution from intersection points of a real and a holomorphic Wilson lines. Indeed, using formulae \eqref{var1} and \eqref{var2} it is straightforward to check that
\begin{equation}
\begin{aligned}
\{ W_{\R}(x^3_f,z, x^3_i), W_{\C}(z_f, x^3,z_i) \} =  \frac{(z_f - z_i)}{(z_f - z)(z - z_i)}W_{\R}(x^3_f,z, x^3) W_{\C}(z, x^3,z_i ) \\
\otimes W_{\C}(z_f, x^3,z) W_{\R}(x^3,z, x^3_i)
\end{aligned}
\end{equation}
if and only if $x^3$ is included in the interval $(x^3_i, x^3_f)$ and zero otherwise. Note that we have suppressed the gauge indices here to streamline notation. For simplicity, we write the Poisson bracket relevant for $GL(N,\mathbb{C})$ as the determinant factor will decouple from the final formula \eqref{spoissonfinal}.

Thus we can compute the Poisson bracket, assuming, say, the order $x^3_i, \tilde x^3_i, x^3_f, \tilde x^3_f$ and $x^3_f \gg 0$ and $x^3_i \ll 0$, 
$\tilde x^3_f \gg 0$ and $\tilde x^3_i \ll 0$:
i.e. 
\begin{align}
&\{ W_{z_f, x^3_f;z_i, x^3_i}(z), W_{\tilde z_f, \tilde x^3_f;\tilde z_i, \tilde x^3_i}(\tilde z) \} = \cr
& \frac{z_f - z}{(z_f - \tilde z)(\tilde z - z)} W_{z_f, x^3_f;\tilde z_i, \tilde x^3_i}(\tilde z) \otimes W_{\tilde z_f, \tilde x^3_f;z_i, x^3_i}(z) - \cr 
 &\frac{\tilde z - \tilde z_i}{(\tilde z- z)(z - \tilde z_i)} W_{z_f, x^3_f;\tilde z_i, \tilde x^3_i}(z) \otimes W_{\tilde z_f, \tilde x^3_f;z_i, x^3_i}(\tilde z) 
\end{align}
and finally the desired
\begin{align}
&\{ W_{\infty, x^3_f;\infty, x^3_i}(z), W_{\infty, \tilde x^3_f;\infty, \tilde x^3_i}(\tilde z) \} = \cr
& \frac{1}{(\tilde z - z)} W_{\infty, x^3_f;\infty, \tilde x^3_i}(\tilde z) \otimes W_{\infty, \tilde x^3_f;\infty, x^3_i}(z) - \cr 
 &\frac{1}{(\tilde z- z)} W_{\infty, x^3_f;\infty, \tilde x^3_i}(z) \otimes W_{\infty, \tilde x^3_f;\infty, x^3_i}(\tilde z) 
 \label{spoissonfinal}
\end{align}

Before sending the $z_{i,f}$ and $\tilde z_{i,f}$ to infinity, there is some dependence on the order of 
the holomorphic Wilson loops along $x^3$. Alternatively, we can attach some extra real Wilson loops 
at $\sigma_{i,f}$ and $\tilde \sigma_{i,f}$ from some reference locations $\pm L$ to $x^3_{i,f}$, $\tilde x^3_{i,f}$
to improve our observable further. The extra stubs correct the rational functions in the Poisson bracket to 
remove the dependence on the order of 
the holomorphic Wilson loops along~$x^3$

\section{Equivariant integrals and monopole operators of higher charges}
\label{app:equivariant}

In this appendix, we compare the quantized abelianization map from the main text to preliminary equivariant localization calculations.

\subsection{Pure $PSU(2)$ theory}

The relations in the abelianized algebra are
\begin{align}
u_+ u_- &= -\frac{1}{\varphi(\varphi-\epsilon)} \cr
u_- u_+ &= -\frac{1}{\varphi(\varphi+\epsilon)} \cr
[\varphi,u_\pm] &= \pm \epsilon u_\pm\,.
\end{align}
The basic non-abelian operators, which generate the quantization $\CA_H$ of $\C[\CM_C]$, are
\begin{equation}
\Phi = \varphi^2  \qquad M_{1,0}=Y =  u_+ + u_-  \qquad M_{1,1} =Z= (u_+ - u_-)\varphi  = \varphi(u_+ - u_-) - \epsilon M_{1,0}\,.
\end{equation}
Observe
\be
M_{1,0}^2 = u_+^2 - \frac{1}{\varphi(\varphi-\epsilon)}- \frac{1}{\varphi(\varphi+\epsilon)}+ u_-^2= u_+^2 - \frac{2}{\varphi^2 - \epsilon^2} + u_-^2\,.
\ee
We can identify this with $M_{2,0}$. The charge $0$ part should be computable from the moduli space of one smooth 
monopole in the presence of a charge $2$ Dirac singularity, which is $\CM_2^0 = \C^2/\Z_2$. This space should be resolved, and the resolution $T^* \cp^1$ 
precisely corresponds to separating the charge 2 singularity into two charge $1$ singularities. The two terms above $\frac{1}{(\pm \varphi)(\mp \varphi - \epsilon)}$
account for the tangent bundle at the two equivariant fixed points of $T^* \cp^1$.

In general, if we expand out $M_{1,0}^A$ we find ${A \choose k}$ terms of abelian charge $A - 2 k$. These 
terms can be associated to the ${A \choose k}$ equivariant fixed points in the resolution of the moduli space of $k$ smooth monopoles in th presence of a charge $A$ 
Dirac singularity. The resolution, of course, precisely corresponds to splitting the charge $A$ Dirac singularity into $A$ charge $1$ Dirac singularities. 
Thus a specific term $u_+ u_+ u_- \cdots$ (etc.) corresponds a fixed point where the smooth monopoles screen the singularities 
contributing, say, $u_+$ to the product and not the singularities contributing, say, $u_-$ to the product. 
 
We can compute some products of monopoles with higher charges. 
\be
M_{1,0}M_{1,1} = u_+^2 \varphi + \frac{\varphi}{\varphi(\varphi-\epsilon)}- \frac{\varphi}{\varphi(\varphi+\epsilon)}- u_-^2 \varphi= u_+^2 \varphi - \frac{2 \epsilon}{\varphi^2 - \epsilon^2} - u_-^2 \varphi
\ee
and 
\be
M_{1,1}M_{1,0} = (\varphi-\epsilon)  u_+^2 - \frac{\varphi- \epsilon}{\varphi(\varphi-\epsilon)}+ \frac{\varphi+\epsilon}{\varphi(\varphi+\epsilon)}- (\varphi+\epsilon) u_-^2 = (\varphi-\epsilon)  u_+^2 - (\varphi+\epsilon) u_-^2
\ee
Thus $M_{1,0}M_{1,1} - M_{1,1}M_{1,0} = -\epsilon M_{2,0}$. Either of the two products could be a candidate for $M_{2,1}$, and they coincide in the classical ring. 
Also recall that 
\be
M_{1,1}^2 = u_+^2 \varphi(\varphi+\epsilon) + \frac{\varphi(\varphi-\epsilon)}{\varphi(\varphi-\epsilon)}+ \frac{\varphi(\varphi+\epsilon)}{\varphi(\varphi+\epsilon)}+ u_-^2 \varphi(\varphi-\epsilon)=M_{1,0}^2 (\Phi- \epsilon^2)+\epsilon M_{1,1} M_{1,0}  + 4
\ee

In general, as we resolve the monopole moduli space $\CM_A^B$ by splitting the Dirac singularity of charge $A$ into $A$ Dirac singularities of charge $1$, we encounter an ambiguity: the resolved space has $A$ natural line bundles $E_{1,i}$, associated to individual Dirac singularities. Each line bundle can be thought as a resolution of the natural line bundle $E_A$ on $\CM_A^B$. Thus we have 
a certain degree of ambiguity in defining $M_{A,\varphi^n}$: the canonical line bundle $E_A^{\otimes n}$ can be regularized to a generic $\otimes_i E_{1,i}^{n_i}$, corresponding to 
defining $M_{A,\varphi^n} = \prod_i M_{1,n_i}$. The resulting operators differ, even for $\epsilon =0$, by multiples of monopole operators of lower magnetic charge. 

\subsubsection{Pure $PSU(2)$ in four dimensions}

We can lift the above analysis a four-dimensional $PSU(2)$ theory on $(\R^2\times S^1)_\epsilon\times \R$.
The abelian relations are
\begin{align}
u_+ u_- &= -\frac{1}{(1-e^{-\varphi})(1-e^{\varphi-\epsilon})} \cr
u_- u_+ &= -\frac{1}{(1-e^{-\varphi- \epsilon})(1-e^{\varphi})} \cr
[\varphi,u_\pm] &= \pm \epsilon u_\pm
\end{align}
The basic operators are
\begin{equation}
W = e^\varphi + 2+ e^{- \varphi} \qquad \qquad M_{1,0} =  u_+ + u_- \qquad \qquad M_{1,n} = u_+ e^{n \varphi} + u_- e^{- n \varphi} 
\end{equation}
The Wilson loop should really be the full trace in the adjoint representation, $e^\varphi + 1+ e^{- \varphi}$, but we have simplified it for convenience. 
We have $M_{1,n} W= M_{1,n-1} + M_{1,n+1}$. 
We can compute 
\be
M_{1,0}^2 = u_+^2 - \frac{1}{(1-e^{-\varphi})(1-e^{\varphi-\epsilon})}- \frac{1}{(1-e^{-\varphi- \epsilon})(1-e^{\varphi})}+ u_-^2= u_+^2 - \frac{1 + e^{-\epsilon}}{(1-e^{\varphi-\epsilon})(1-e^{-\varphi- \epsilon})} + u_-^2
\ee
and identify this with $M_{2,0}$. 
 
We can compute some products of monopoles with charges. 
\begin{align}
M_{1,0}M_{1,1} &= u_+^2 e^\varphi - \frac{e^{- \varphi}}{(1-e^{-\varphi})(1-e^{\varphi-\epsilon})}- \frac{e^\varphi}{(1-e^{-\varphi- \epsilon})(1-e^{\varphi})}+ u_-^2 e^{-\varphi}= \cr
&=u_+^2 e^\varphi - \frac{-1 + e^{-\epsilon}(e^\varphi + 1 + e^{- \varphi})}{(1-e^{\varphi-\epsilon})(1-e^{-\varphi- \epsilon})} + u_-^2 e^{-\varphi}
\end{align}
which simplifies to
\begin{align}
M_{2,1} := M_{1,0}M_{1,1} -1 = u_+^2 e^\varphi - \frac{e^{-\epsilon}(1+ e^{-\epsilon})}{(1-e^{\varphi-\epsilon})(1-e^{-\varphi-\epsilon})} + u_-^2 e^{-\varphi}
\end{align}
We can also compute 
\begin{align}
M_{1,1}M_{1,0} &= u_+^2 e^{\varphi+\epsilon} - \frac{e^{\varphi-\epsilon}}{(1-e^{-\varphi})(1-e^{\varphi-\epsilon})}- \frac{e^{-\varphi-\epsilon}}{(1-e^{-\varphi-\epsilon})(1-e^{\varphi})}+ u_-^2 e^{-\varphi+\epsilon}= \cr
&=u_+^2 e^{\varphi+ \epsilon} - \frac{-e^{-2 \epsilon} + e^{-\epsilon}(e^\varphi + 1 + e^{- \varphi})}{(1-e^{\varphi-\epsilon})(1-e^{-\varphi-\epsilon})} + u_-^2 e^{-\varphi+\epsilon}
\end{align}
with $e^{-\epsilon} (M_{1,1}M_{1,0}-1) = M_{2,1}$ as well. 

It would be interesting to identify the geometric meaning of our tentative definition of $M_{2,1}$. 
Starting from $M_{2,0}$ and $M_{2,1}$ we can define general monopoles $M_{2,k}$ by the ``theta angle shift'', \ie\ applying the symmetry $u_\pm \to u_\pm e^{\pm\varphi + \frac{\epsilon}{2} }$.

We also have 
\begin{align}
M_{1,1}^2 &= u_+^2 e^{2 \varphi+\epsilon}  - \frac{e^{-\epsilon}}{(1-e^{-\varphi})(1-e^{\varphi-\epsilon})}- \frac{e^{-\epsilon}}{(1-e^{-\varphi-\epsilon})(1-e^{\varphi})}+ u_-^2 e^{-2\varphi+\epsilon} =\cr
&= u_+^2 e^{2 \varphi+\epsilon}  - \frac{e^{-\epsilon}(1 + e^{-\epsilon})}{(1-e^{\varphi-\epsilon})(1-e^{-\varphi- \epsilon})} + u_-^2 e^{-2\varphi+\epsilon}
\end{align}
and thus 
\be
e^{-\epsilon} M_{1,1}^2 + M_{2,0}+1 + e^{-\epsilon}= M_{2,1} \Phi
\ee

\subsection{Pure $U(N)$ theory}
\label{app:UN}

Define $h_i$ as the $N$-dimensional vectors with a single non-zero entry $(h_i)^i =1$. 
The abelian relations are
\begin{align}
[\varphi_a,\varphi_b]&=0 \cr
[\varphi_b,u^\pm_a] &= \pm \epsilon \delta_{ab} u^\pm_a \cr
u^+_a u^-_a &= \frac{1}{\prod_{b \neq a}(\varphi_b - \varphi_a)(\varphi_a - \varphi_b-\epsilon)} \cr
u^-_a u^+_a &= \frac{1}{\prod_{b \neq a}(\varphi_a - \varphi_b)(\varphi_b - \varphi_a-\epsilon)} \cr
u^+_a u^+_b &= -\frac{1}{(\varphi_a-\varphi_b)(\varphi_a-\varphi_b-\epsilon)} u^+_{h_a + h_b} \cr
u^+_a u^-_b  &= u^-_b u^+_a = u^+_{h_a - h_b}
\end{align}

We can start with $M_{\pm h_1,0} = \sum_a u^\pm_a$ and $M_{\pm(h_1+h_2),0} = \sum_{a<b} u^\pm_{(h_a + h_b)}$.
We get  
\begin{align}
M_{h_1,0} M_{h_1,0} &= \sum_a u^+_{2 h_a} - \sum_{a<b} \left[\frac{1}{(\varphi_a-\varphi_b)(\varphi_a-\varphi_b-\epsilon)} + \frac{1}{(\varphi_a-\varphi_b)(\varphi_a-\varphi_b+\epsilon)}\right] u^+_{h_a + h_b} =\cr
&= \sum_a u^+_{2 h_a} - \sum_{a<b} \frac{2}{(\varphi_a-\varphi_b-\epsilon)(\varphi_a-\varphi_b+\epsilon)} u^+_{h_a + h_b}
\end{align}
This is a natural candidates for $M_{2 h_1,0}$. The moduli space $\CM_{2 h_1}^{h_a + h_b}$ is again an $A_1$ singularity, resolved to $T^* \cp^1$ by splitting the charge $2 h_1$ Dirac singularities into two charge $h_1$ Dirac singularities. The two terms 
in the coefficient of $u^+_{h_a + h_b}$ correspond to these two fixed points. 

We can also compute 
\be
M_{h_1,0} M_{-h_1,0} = \sum_{a \neq b} u^+_{h_a - h_b} + \sum_a \frac{1}{\prod_{b \neq a}(\varphi_b - \varphi_a)(\varphi_a - \varphi_b-\epsilon)} 
\ee
Although it is far from obvious, the above expression coincides with 
\be
M_{-h_1,0} M_{h_1,0} = \sum_{a \neq b} u^+_{h_a - h_b} + \sum_a \frac{1}{\prod_{b \neq a}(\varphi_b - \varphi_a)(\varphi_a - \varphi_b+\epsilon)} 
\ee
They are the natural candidate for defining $M_{h_1-h_N,0}$. The moduli space $\CM_{h_1 - h_N}^0$ is the 
blowdown of $T^* \cp^{N-1}$. The latter is precisely the resolution corresponding to the factorization 
$M_{h_1,0} M_{-h_1,0}$ and the individual terms in the sum above are the equivariant fixed points 
of $T^* \cp^{N-1}$. The two factorizations are related by a flop. 

We can start adding scalar dressings as well. We encountered in the main text the polynomials $U_\pm(z)$ 
which capture the dressed monopoles of the form $M_{\pm h_1, \det(z - \varphi_{U(N-1)})}$,
as well as the ratios $U_\pm(z)Q(z)^{-1}$ which capture monopoles of the form $M_{\pm h_1, \varphi_{U(1)}^n}$.
Notice that our choice of signs in the algebra of $u^\pm_a$ is a bit different than in our example section in the main text. 

We also encountered a polynomial $\wt Q$, which should capture the monopoles of the form 
$M_{h_1-h_N, \det(z - \varphi_{U(N-2)})}$. This gives us our first example of a universal bundle on 
a monopole moduli space which cannot quite be mapped to a bundle over the resolution. 
We need to compute $U_+(z) U_-(z + \epsilon) Q^{-1}(z + \epsilon)$:
\begin{align}
&\sum_a u^+_a \prod_{b\neq a} (z-\varphi_b) \sum_c u^-_c \frac{1}{z - \varphi_c + \epsilon} = \cr &= \sum_{a \neq c} u^+_{h_a - h_c} \prod_{b\neq a\neq c} (z-\varphi_b) + \sum_a \frac{ \prod_{b\neq a} (z-\varphi_b)}{(z - \varphi_a + \epsilon)\prod_{b \neq a}(\varphi_b - \varphi_a)(\varphi_a - \varphi_b- \epsilon)} 
\end{align}
We can subtract $Q^{-1}(z + \epsilon)$, which is a power series in gauge-invariant polynomials of $\varphi$, 
to get a rank $N-2$ polynomial in $z$:
\begin{align}
\wt Q(z)&=  \sum_{a \neq c} u^+_{h_a - h_c} \prod_{b\neq a\neq c} (z-\varphi_b) + \cr &+\sum_a \frac{ \prod_{b\neq a} (z-\varphi_b)-\prod_{b\neq a}(\varphi_a - \varphi_b- \epsilon)}{(z - \varphi_a + \epsilon)}\frac{1}{\prod_{b \neq a}(\varphi_b - \varphi_a)(\varphi_a - \varphi_b- \epsilon)} 
\end{align}

The $U(N-2)$ universal bundle $E_{h_1 - h_N}$ on $\CM_{h_1 - h_N}^0$ associated to the Dirac singularity 
does {\it not} extend in a natural way to the resolution $T^* \cp^{N-1}$, which has only $U(1)$ and $U(N-1)$ bundles associated to 
either individual Dirac singularity in the factorization. It is not hard to disentangle the geometric description of these bundles:
if we describe the moduli space as a hyperk\"ahler quotient, in terms of two vectors $X$ and $Y$ with $X \cdot Y =0$,
the rank $N-2$ bundle can be thought of intuitively as vectors orthogonal to $X$, modulo $Y$. After the blow-up,
this bundle is ill-defined on the vanishing cycle. The answer above produces some equivariant characteristic class 
$\frac{ \prod_{b\neq a} (z-\varphi_b)-\prod_{b\neq a}(\varphi_a - \varphi_b- \epsilon)}{(z - \varphi_a + \epsilon)}$
which somehow behaves as the characteristic class of the non-existent rank $N-2$ bundle. It would be interesting to work this out in detail, perhaps using equivariant intersection cohomology.

\bibliographystyle{JHEP_TD}

\bibliography{coulomb}

\end{document}